\documentclass[a4paper,11pt]{article}
\usepackage{jheppub} 
\usepackage{xfrac}
\usepackage[numbers]{natbib}
\usepackage{hyperref} 
\usepackage{float}
\usepackage{subcaption}

\title{\boldmath Chiral Plasma under Strong Magnetic Fields: A Holographic Analysis of Transport Phenomena }

\author{Michael Lublinsky and Hadas Tzarfati}
\affiliation{Department of Physics, Ben-Gurion University, Beer-Sheva 8410501, Israel}

\emailAdd{lublinm@bgu.ac.il} \emailAdd{hadastz@post.bgu.ac.il}
\abstract{ Chiral plasma appears in several areas of physics,  historically starting from primordial plasma in the early Universe, 
then in quark-gluon plasma produced in heavy ion collisions, and, more recently, in Dirac and Weyl semimetals. The major signature  of the plasma is the non-conservation of the axial current due to the chiral anomaly and the emergence of new, anomaly-induced transport phenomena. In this paper, we study the plasma exposed to arbitrarily strong constant magnetic and weak electric fields. Employing all-order gradient resummation, we write down  constitutive relations for electric and axial currents parameterized by thirteen momentum- and magnetic- field-dependent transport coefficient functions. The latter are computed utilizing  a theoretical lab for a realistic plasma, namely a holographic $U(1)_V \times U(1)_A$ Maxwell--Chern--Simons theory in Schwarzschild--AdS$_5$, in the probe limit. 
As an application, we revisit the phenomena of negative magnetoresistance and chiral magnetic waves, beyond the naive hydrodynamic limit. 

}

\begin{document}
\maketitle
\flushbottom

\section{Introduction}
Hydrodynamics offers an effective description of strongly interacting QFTs at non-zero temperatures, capturing the dynamics of macroscopic currents that emerge from a locally near-equilibrium thermal state. Central to this framework are  \textbf{constitutive relations}, which connect the macroscopic currents to thermodynamic variables such as conserved charges, temperature and external fields. 

In a charged plasma, the constitutive relation for the vector current density\(\vec{J}\) 
takes the form
\begin{equation}
\vec{J} = \vec{J} \left[ \rho,  T, \vec{E}, \vec{B} \right].   
\end{equation}
Here the 3-current density depends on the $U(1)$ charge density $\rho$, the temperature $T$, and the external electromagnetic fields $\vec{E}$ and $\vec{B}$.

The simplest example of a constitutive relation is Fick's law
for the  vector current, 
\begin{equation} \label{fick}
  \vec{J} = - D^0\, \vec{\nabla} \rho,
\end{equation}
where $D^0$ is the diffusion constant. Likewise, Ohm's law relates the electric current
density to an external electric field,
\begin{equation}
  \vec{J} = \sigma_e^0\, \vec{E},
\end{equation}
where $\sigma_e^0$ is the electrical DC conductivity. 
$D^0$ and
$ \sigma_e^0$ are known as \textbf{transport coefficients} (TCs). 
Fick’s and Ohm’s laws are valid in the strict hydrodynamic long wave-length limit. Beyond this limit,  higher derivative terms
are expected to contribute.
The dynamics of the charges is governed by vector current conservation
\begin{align}
  \partial_\mu J^\mu &= 0. 
\label{4}
\end{align}

In addition to the usual electric charge, 
chiral media possess another charge, the axial charge. Its
density $\rho_5$ is  defined as the difference between the right- and left-handed charge densities ($\rho_5 \equiv \rho_R - \rho_L$).  
Due to chiral anomalies of relativistic QFTs  with massless fermions \cite{Adler:1969gk,Bell:1969ts}, 
the axial vector current $J_5^\mu$ is not conserved
 in the presence of external electromagnetic fields, 
\begin{align}
  \partial_\mu J_5^\mu &= 12 \kappa\, \vec{E}\cdot\vec{B}.
\label{5}
\end{align}
The anomaly coefficient is $\kappa =  N_c/(24\pi^2)$ 
for an $SU(N_c)$ gauge theory with a
massless Dirac fermion in the fundamental representation.

In a chiral plasma, the constitutive relations additionally involve the axial charge $\rho_5$, even though it is not conserved in general. 
\begin{equation}
\vec{J} = \vec{J} \left[ \rho,\rho_5,  T, \vec{E}, \vec{B} \right]  ; \quad\quad
\vec{J}_5 = \vec{J}_5 \left[ \rho, \rho_5, T, \vec{E}, \vec{B} \right].
\end{equation}
Chiral plasma features many new, anomaly-induced transport effects \cite{Grasso:2000wj,Kuzmin:1985mm,Vilenkin:1982pn,Rubakov:1996vz,Metlitski:2005pr,
Miransky:2015ava,
Kharzeev:2015kna,Kharzeev:2013ffa,Son:2009tf,Landsteiner:2016led}:
\begin{itemize}
  \item \textbf{Chiral Magnetic Effect (CME) \cite{Fukushima:2008xe,Kharzeev:2013ffa}:}
 A chiral imbalance (nonzero $\rho_5$)
 in the presence of an external magnetic field induces a vector current along the magnetic field, 
  \begin{align}
    \vec{J} = \sigma_\chi^0 \kappa\, \rho_5\, \vec{B} .
    \label{6}
  \end{align}
  The TC $\sigma_\chi^0$ is referred to as the chiral magnetic conductivity \cite{Kharzeev:2009pj,Yee:2009vw}. 
For recent reviews on the CME in heavy ion collisions, see e.g., \cite{Kharzeev:2024zzm,Liu:2020ymh}.
For a comprehensive review on chiral kinetic theory, see \cite{Hidaka:2022dmn}.
A recent review of chiral transport phenomena in astrophysical and cosmological systems can be found in \cite{Kamada:2022nyt}. 
For a review of holographic studies of anomalous transport and chiral phenomena, see \cite{Landsteiner:2016led}.

  
  \item \textbf{Chiral Separation Effect (CSE)  \cite{son2004quantum,metlitski2005anomalous}:}
The axial charge densities of left- and right-handed fermions become separated along the external magnetic field,
  \begin{align}
    \vec{J}_5 = \sigma_\chi^0 \kappa\, \rho\, \vec{B} ,
    \label{7}
  \end{align}

  \item \textbf{Chiral Electric Separation Effect (CESE) \cite{Huang:2013iia}:}
Chiral charge flow is induced  by an external electric field
when both the vector and axial charge densities are finite. 
  \begin{align}
    \vec{J}_5 \propto  \rho\, \vec{E}.
    \label{7B}
  \end{align}
\end{itemize}
We notice that all the effects mentioned above are valid in the
strict hydrodynamic limit, that is, ignoring higher 
gradients, and
for weak external electromagnetic fields (linear response). At stronger fields, one expects the emergence of non-linear effects.

$\bullet$ \textbf{Negative Magnetoresistance (MR)} refers to the unexpected decrease in electrical resistance when an external magnetic field is applied parallel to the electric current \cite{Li:2016cm,Son:2012bg,Armitage:2017cjs}. The effect
emerges when Ohm's law is combined with CME/CSE:
\begin{align}
 \vec{J} = \sigma_e^0\vec E+ \sigma_\chi^0 \kappa\, \rho_5\, \vec{B} ,\qquad\qquad
    \vec{J}_5 = \sigma_\chi^0 \kappa\, \rho\, \vec{B} .
    \label{71}
  \end{align}
Assuming the electric and magnetic fields to be parallel and eliminating the charge densities  using the conservation laws \eqref{4} and \eqref{5}, one
arrives at the linear response expression for the current density:
\begin{align}
\vec J=\sigma_L\,\vec E,\qquad\qquad
\sigma_L = \sigma_e^0 
+\frac{12 i \sigma_\chi^0 \kappa^2 B^2} {\omega}.
\label{nm}
\end{align}
The pole $1/\omega$  in the longitudinal conductivity $\sigma_L$ is problematic in the  limit of $\omega\rightarrow 0$. It is frequently 
regularized by a finite "chirality-changing scattering time" $\tau_5$ via $\omega \to \omega + i/\tau_5$ \cite{Jimenez-Alba:2015awa, Li:2016cm}: 
\begin{equation}
    \sigma_L(\omega=0) = \sigma_e^0 
+12 \tau_5 \sigma_\chi^0 \kappa^2 B^2.
\label{nm1}
\end{equation}
The anomaly-driven contribution enhances charge transport and effectively reduces the resistivity, leading to a negative MR response. This is in sharp contrast to conventional materials, in which  magnetic fields typically enhance resistance. 
 
 This effect has been extensively investigated in solid-state systems \cite{Li:2016cm,ong2021experimental,Zhang:2017,Yuan:2020,Balduini:2024etx}, where the observation of the {negative longitudinal MR} in Weyl semimetals is interpreted as a definitive hallmark of the chiral anomaly \cite{Son:2012bg}. Holographic Weyl semimetals have been investigated from various perspectives, recent works include Refs. \cite{Baggioli:2026hpo,Rai:2024bnr,Ji:2021aan}.
For a review of Dirac and Weyl semimetals in solid-state systems, see e.g. Ref. \cite{Armitage:2018}.

  $\bullet$ \textbf{Chiral Magnetic Wave} (CMW) \cite{Kharzeev:2010gd}
is an interplay between the CME and the CSE: a local fluctuation in the
electric charge density $\rho$, through the CSE, induces a fluctuation in the axial charge  density $\rho_5$, which, through the CME, feeds back to $\rho$. The net result is a  density wave. 
Experimental searches for CMW in relativistic heavy-ion collisions \cite{ALICE:2023weh,Nain:2026lnc} remain inconclusive due to large background contributions in the measured observables.
In order to determine the CMW dispersion relation, we parameterize the density fluctuations using a plane-wave ansatz:
\begin{align}
\rho(t,\vec{x}) = \epsilon \rho_0 e^{-i\omega t + i\vec{q}\cdot \vec{x}}, 
\qquad \rho_5(t,\vec{x}) = \epsilon \rho_5 e^{-i\omega t + i\vec{q}\cdot \vec{x}}.\label{pw}
\end{align}
Throughout this work, we will be interested in weak/linear effects only. In order to trace those,  we have introduced a formal parameter $\epsilon$. Keeping only first-order terms in $\epsilon$  will be  referred to as the \textbf{linear approximation}. In the absence of an external electric field, the axial current $J_5^\mu$ is conserved. Substituting the plane-wave ansatz \eqref{pw} into the CME \eqref{6} and the CSE \eqref{7}, then using the conservation laws \eqref{4} and \eqref{5}, one arrives at the dispersion relation 
\begin{equation}
     \omega_\pm = \pm \sigma_\chi^0 \kappa \,(\vec{q}\cdot\vec{B})
     =\pm \sigma_\chi^0 \kappa \,({q}\cdot{B})\cos(\alpha).
\end{equation}
This is a purely real mode that freely propagates in two opposite directions.
The frequency of the mode depends on the angle $\alpha$ between the wave vector and the magnetic field.

The CMW mode acquires an imaginary part (dissipation) when the charge diffusion process is included. Putting together Fick’s law \eqref{fick}, CME \eqref{6}, and CSE \eqref{7}, the constitutive relations for the currents read
\begin{align}
\vec{J} =- D^0 \vec{\nabla}\rho +\sigma_{\chi}^0 \rho_5 \kappa\vec{B} 
\qquad; \qquad \vec{J}_5 =- D^0 \vec{\nabla}\rho_5+ \sigma_{\chi}^0 \rho \kappa\vec{B} .
\label{cons1}
\end{align}
Substituting ~\eqref{cons1} into the continuity equations ~\eqref{4}--\eqref{5}  yields the  dispersion relation  \cite{Kharzeev:2010gd}:
\begin{align}
    \omega_\pm = \pm \sigma_\chi^0 \kappa \,(\vec{q}\cdot\vec{B}) - i D^0 q^2.
    \label{eq:CMW_dispersion}
\end{align}
This excitation corresponds to a long-wavelength  density wave  with a sound-like dissipative dispersion relation. 

So far, we have discussed the CMW in the absence of any electric fields. Including Ohmic currents is expected to strengthen the dissipation of the CMW. With the electric field turned on,
the constitutive relations read 
\begin{align}
  \vec{J} &= 
    - D^0\, \vec{\nabla}\rho
    + \sigma_\chi^0 \kappa\, \vec{B}\, \rho_5
     + \sigma_e^0\, \vec{E}  \qquad; \qquad   \vec{J}_5 =
    - D^0\, \vec{\nabla}\rho_5
    + \sigma_\chi^0 \kappa\, \vec{B}\, \rho
    \label{consthy}
\end{align}
Here we assume a weak electric field,  $E=O(\epsilon)$. Hence, CESE \eqref{7B} is discarded. 

As was mentioned above, the electromagnetic fields entering the constitutive relations are treated as external. 
Yet, this is not a self-consistent description of a charged medium. In a real plasma, electromagnetic fields are generated dynamically due to the presence of  fluctuating charges and currents. These induced fields then affect the dynamics of the charges. Moreover, once both the electric and magnetic fields 
are present, the axial current $J_5^\mu$ is no longer conserved, and the chiral anomaly starts to source the axial charge.  
Therefore, even in the absence of external sources, it is not self-consistent to neglect the effects of, say, the electric field,  as we discussed earlier. 


Ref. \cite{SHOV} explored the constitutive relations (\ref{consthy}) with an external magnetic field but with the electric field being created purely dynamically.  That is, due to Gauss's law, the electric field $E\sim \rho\sim O(\epsilon)$, 
\begin{align}
\label{gauss}
    \vec{E} = -\frac{i \vec{q}}{q^2} \rho.
\end{align}
Eq. \eqref{gauss} holds  for the longitudinal component of the electric field only ($\vec{E} \parallel \vec{q}$).
The CMW, which follows from (\ref{consthy}) with the dynamic electric field, has the following dispersion relation:
\begin{align}
\omega_\pm
= -\frac{1}{2} i\sigma_e^0 - iD^0q^2
\pm
(\frac{i}{2}\sigma_e^0)\sqrt{1  -\frac{48\kappa^2 B^2\sigma_\chi^0}{(\sigma_e^0)^2}(1+  \frac{q^2 \sigma_\chi^0}{12})} 
\label{eq:17}
\end{align}
At  zero magnetic field, the "$+$" mode reduces to the usual diffusive motion, while the "$-$" mode is gapped. 
The dispersion relation (\ref{eq:17}) was further explored in \cite{SHOV}. In particular, they found that 
the effect of the Ohmic conductivity $\sigma_e^0$
is quite dramatic: the CMW develops a dominant negative imaginary part and becomes overdamped, rather than remaining a propagating mode. 

The results discussed above are limited to the 
conventional hydrodynamic limit $\omega,q\rightarrow 0$. Moreover, they assumed 
 a weak, order-$\epsilon$, external magnetic field. In a parallel development, Refs. \cite{P5,P6} argued that,
beyond the hydrodynamic limit, once higher (all) 
order gradient effects are taken into account, the CMW could in fact become {\bf dissipationless}: the imaginary part of the CMW vanishes at finite wave-vectors and frequencies,  for magnetic fields stronger than a certain threshold.  The results of Refs. \cite{P5,P6} 
were obtained in a holographic model,  which serves as a realistic theoretical laboratory for studying transport. 
\subsection*{All-order gradient resummation}
The constitutive relations discussed above are valid in the long-wavelength or hydrodynamic limit, where the gradients of the thermodynamic variables and external fields are small. 
Beyond this limit, constitutive relations are
naturally organized as a \textbf{gradient expansion}. The gradient expansion is, in principle, an infinite series. Effective hydrodynamics  derived from  any finite-order truncation 
of the gradient expansion is known to violate relativistic causality. 
It was first argued in \cite{Lublinsky:2009kv} and further developed in \cite{Bu:2014ena,Bu:2014sia,P1} that causality is restored upon resummation of the gradient series to \textbf{all orders}. Hence, to a large extent, such resummation is mandatory.


For the linear in $\epsilon$ regime, Ref. \cite{Lublinsky:2009kv} introduced a rather compact procedure for resumming the gradient expansion: 
the TCs $D^0$, $\sigma_e^0$, $\sigma_\chi^0$, etc., are promoted to
scalar functionals depending on spacetime derivative operators
\begin{equation}
  D^0 \rightarrow D[\partial_t,\vec{\nabla}], 
  \qquad
  \sigma_e^0 \rightarrow \sigma_e[\partial_t,\vec{\nabla}],
  \qquad
  \sigma_\chi^0 \rightarrow \sigma_\chi[\partial_t,\vec{\nabla}].
\end{equation}
The derivatives act on the charge densities and fields. 
The functionals are assumed to be Taylor expandable and their expansion is supposed to coincide with the gradient expansion.

In momentum space, the derivatives are mapped under
the Fourier transform as $[\partial_t, \vec{\nabla}] \to [-i\omega, i\vec{q}]$. As a result, 
the functionals of the derivative operators become scalar functions of frequency and 3-momentum \footnote{For homogeneous backgrounds, the dependence is on $q^2$ and not on $\vec q$.} 
\begin{equation}
D^0 \rightarrow D(\omega,q^2), \qquad \sigma_e^0 \rightarrow \sigma_e(\omega,q^2),  \qquad \sigma_\chi^0 \rightarrow \sigma_\chi(\omega,q^2).
\end{equation}
In \cite{Lublinsky:2009kv},  these functions were referred to as \textbf{transport coefficient functions} (TCFs). When Fourier transformed back into  real time, the TCFs become memory functions. 
By promoting TCs to TCFs, we effectively resum infinitely many higher-order gradient terms. The familiar constant hydrodynamic TCs  $D^0$, $\sigma_e^0$, $\sigma_\chi^0$ 
correspond to the lowest order in the
expansion of the corresponding TCFs in  small $\omega$ and $q^2$. 

A big advantage of having TCFs is that one can study physics at finite momenta, beyond the strict hydrodynamic limit. The discovery in \cite{P5,P6} of the non-dissipative regime in CMW  relied entirely on the concept of TCFs.


\subsection*{Transport in the presence of  strong magnetic fields}

One of the important observations made in \cite{P5, P6}
is that the non-dissipative CMW emerges at strong external magnetic fields, above a certain threshold. While the charge densities in \cite{P5, P6} were treated as $O(\epsilon^1)$,
the magnetic field was taken as order one: $\vec{B}=\mathcal{O}(\epsilon^0)$ (as mentioned, $\vec{E}=0$ in \cite{P5, P6}). 

The constitutive relations become non-linear in $\vec B$,
which in \cite{P5, P6} was taken as time-independent and uniform. The non-linearity in magnetic fields is manifested in two ways. First, the TCFs, in addition to being functions    of $\omega$ and  $q^2$, start to depend on the magnetic-field:
\begin{align}
\label{constitutive}
D(\omega,q^2) \rightarrow D(\omega,q^2,B^2,(\vec{q}\cdot\vec{B})^2), \quad \sigma_e(\omega,q^2) \rightarrow \sigma_e(\omega,q^2,B^2,(\vec{q}\cdot\vec{B})^2),\ \ldots
\end{align}
Second, the constitutive relations acquire new terms. In particular,
the constitutive relations explored in \cite{P6} take the form:
 \begin{align}
 \label{const2} 
\vec{J} =  - D\, \vec{\nabla}\rho
    + D_B \kappa^2 \vec{B} (\vec{B} \cdot \vec{\nabla}\rho)
    + \sigma_\chi \kappa\, \vec{B}\, \rho_5
    + D_\chi \kappa\, (\vec{B}\cdot\vec{\nabla})\vec{\nabla}\rho_5
\\ \notag \vec{J}_5 =     - D\, \vec{\nabla}\rho_5
    + D_B \kappa^2 \vec{B} (\vec{B} \cdot \vec{\nabla}\rho_5)
    + \sigma_\chi \kappa\, \vec{B}\, \rho
    + D_\chi \kappa\, (\vec{B}\cdot\vec{\nabla})\vec{\nabla}\rho.
\end{align}
Here $D_\chi$ is an anomaly induced magnetic field dependent TCF, which emerges at second order in the gradient expansion. Similarly, $D_B$ captures an explicitly non-linear in $\vec B$ effect.  The constitutive relation (\ref{const2})
is the most general relation, which is linear in the densities. 



Substituting the constitutive
relations~\eqref{const2} into ~\eqref{4}--\eqref{5} yields the modified dispersion relation for the CMW ~\cite{P6}:
\begin{equation}
\omega_\pm =\omega_\pm^R +i \omega_\pm^I  = \pm \left({\sigma}_{\chi} - q^{2} D_{\chi}\right)\kappa\, \vec{B}\!\cdot\!\vec{q}
- i \left( q^{2} D - D_{B} (\kappa\, \vec{B}\!\cdot\!\vec{q})^{2} \right).
\label{eq:modified_dispersion}
\end{equation}
The dispersion relation  (\ref{eq:modified_dispersion}) is quite complicated
because its RHS  depends on $\omega_\pm$ through the dependence of the TCFs on $\omega$.
It was shown in ~\cite{P6} that the imaginary part $\omega_{\pm}^I$  might vanish at finite, non-hydrodynamic values of $\omega_\pm$ and $q^2$, once the external magnetic field exceeds a critical threshold\footnote{It is important to keep in mind that all the TCFs are complex functions.}. 

The phenomenon of having a dissipationless wave is by itself 
 extremely interesting. 
Yet, Refs. \cite{P5,P6} did not take into account any  effects of the dynamical electric fields. The latter could be dramatic as was argued above, and, in principle, has the potential to entirely destroy the effect. It is thus our main motivation to revise and extend the work of \cite{P5,P6} incorporating the dynamical electric field in their analysis, with the  goal of exploring whether a long-lived propagating mode  still exists at finite momenta. 
A quite similar question  has been 
raised recently in \cite{Ahn:2024ozz}, although it was addressed in a different holographic model. 


\subsection*{Constitutive relations with electric and magnetic fields.}

The major limitation of ~\cite{P5, P6} is that their analysis was limited  to the case of vanishing  electric field ($\vec{E}=0$). 
However, as noted above, the electric field cannot be consistently decoupled from the dynamics and should be taken into account.
One of the main objectives of our work is  to fill this gap. 
We  consider the following scaling
\begin{align}
\rho(x^\alpha) \sim \mathcal{O}(\epsilon) , \quad 
\rho_5(x^\alpha)  \sim\mathcal{O}(\epsilon) ,  \quad 
E(x^\alpha) \sim\mathcal{O}(\epsilon) , \quad 
B  \sim\mathcal{O}(\epsilon^0). \quad 
\label{linearschem}
\end{align}
The  general constitutive relations for the vector and axial currents consistent with the  scaling (\ref{linearschem})
have  the following  form:
\begin{align}
  \vec{J} &= 
    - D\, \vec{\nabla}\rho
    + D_B \kappa^2 \vec{B} (\vec{B} \cdot \vec{\nabla}\rho)
    + \sigma_\chi \kappa\, \vec{B}\, \rho_5
    + D_\chi \kappa\, (\vec{B}\cdot\vec{\nabla})\vec{\nabla}\rho_5
     + \sigma_e\, \vec{E} \label{11}\\\nonumber
    &\quad
    + \delta\sigma_\chi \kappa^2 (\vec{E}\cdot\vec{B}) \vec{B}
    + \gamma_D\, \vec{\nabla}(\vec{\nabla}\cdot \vec{E}) +\tau_D\kappa^2\vec{\nabla}(\vec{B}\cdot\vec{\nabla})(\vec{E}\cdot\vec{B}) +\tau_B\kappa^2\vec{B} (\vec{B}\cdot\vec{\nabla})(\vec{\nabla}\cdot\vec{E}),
\end{align}
\begin{align}
  \vec{J}_5 &=
    - D\, \vec{\nabla}\rho_5
    + D_B \kappa^2 \vec{B} (\vec{B} \cdot \vec{\nabla}\rho_5)
    + \sigma_\chi \kappa\, \vec{B}\, \rho
    + D_\chi \kappa\, (\vec{B}\cdot\vec{\nabla})\vec{\nabla}\rho
\label{12} \\    \nonumber
    &\quad + \gamma_\chi \kappa\, \vec{\nabla}(\vec{E}\cdot\vec{B})+\tau_\chi \kappa\vec{\nabla}(\vec{B}\cdot\vec{\nabla})(\vec{\nabla} \cdot \vec{E})+\sigma_B \kappa^3\vec{B}(\vec{B}\cdot\vec{\nabla})(\vec{E}\cdot\vec{B})+ \gamma_B \kappa\, \vec{B}(\vec{\nabla}\cdot \vec{E}).
\end{align}
Eqs. \eqref{11} and \eqref{12} introduce  thirteen distinct magnetic field dependent TCFs. We notice that the inclusion of the electric field is not limited to merely adding the Ohmic term, but also includes several additional terms  allowed by the  scaling (\ref{linearschem})\footnote{Within the holographic model,
which we are going to exploit for the computation of the TCFs, all the
non-linear effects are due to the chiral anomaly and hence proportional to $\kappa$. This is why we do not have terms that would be analogous to, say, the usual Hall effect.}.

 Our theoretical laboratory will be the very same holographic model that was 
explored in ~\cite{P5, P6}:   a $U(1)_V \times U(1)_A$ Maxwell--Chern--Simons theory in a Schwarzschild--AdS$_5$ background. 
The model will be reviewed in the next Section. Our calculation, while closely following those of Refs. \cite{P5, P6}, introduces  major new elements associated with the electric field. 

Within this holographic framework, we systematically derive the constitutive relations \eqref{11} and \eqref{12} and compute all the TCFs as functions of real frequency, 3-momentum, and constant magnetic field. Among the thirteen TCFs, $D$, $D_B$, $D_{\chi}$, and $\sigma_{\chi}$ have been previously studied in \cite{P5,P1,P2,P3,P4} including at a strong magnetic field ~\cite{P6}. $\delta\sigma_{\chi}$  has been analyzed  only in two limiting cases ($B=0$~\cite{P6} or $q^2=0$~\cite{P3}). 

The combination of the terms $\sigma_e$ and $\gamma_D$ 
 can be presented equivalently  
using the Bianchi identity:
\begin{equation}
    \sigma_e\, \vec{E} 
    + \gamma_D\, \vec{\nabla}(\vec{\nabla}\cdot \vec{E})
=\tilde \sigma_e\, \vec{E}+ \sigma_m\, \vec\nabla\times \vec B; ~~~~~~~~  {\sigma_m}=\gamma_D {i\omega}.;
~~~~~~~~\tilde\sigma_e=\sigma_e+q^2\gamma_D.
\end{equation}
It is in this second form that these two terms were introduced in Ref.~\cite{P1}. 
The remaining six TCFs are computed here for the first time.


Having computed all the TCFs, we focus on CMW with the constitutive relations  \eqref{11} and \eqref{12} being our starting point. Following
~\cite{SHOV}, we treat the electric field as dynamical, that is, by using (\ref{gauss}). The resulting dispersion relation for the CMW is given below, in (\ref{eq:dispersion_compact}). Implementing the same strategy as introduced in ~\cite{P5, P6}, we  
search for a dissipationless regime at finite momenta 
and at strong magnetic fields. Not that surprisingly but quite disappointingly, such a regime was not discovered: the CMW dispersion relation has a non-vanishing imaginary part in the explored range of parameters. More specifically,
we find two modes, 
one overdamped  and another one underdamped. 
To obtain this result it is not sufficient to know the TCFs as functions of real frequencies only. This is why we had to recompute the TCFs also 
as functions of complex frequencies, while keeping the 3-momentum real. There has been some interest in complex
frequencies (and momentum) related to the convergence 
of the hydrodynamic gradient expansion \cite{Grozdanov:2019kge} and the phenomenon of pole skipping \cite{Blake:2018leo,Grozdanov:2019uhi,Abbasi:2023myj}, but we will not touch upon these points here.

Returning to the effect of negative MR, we notice that the expression
(\ref{nm}) for the longitudinal conductivity $\sigma_L$ is valid for a homogeneous medium with homogeneous fields only ($q=0$). Furthermore, it is leading in $\omega$ and ignores any possible nonlinearities in the magnetic field. Finally, the regularization by $\tau_5$ is introduced largely by hand  even though it is obvious that some kind of a regulator must be present.  With the generalized constitutive relations  \eqref{11} and \eqref{12} we are fully equipped to recompute 
$\sigma_L$, or more generally a full response function, 
relaxing all the approximations mentioned above. Our findings confirm the effect of negative MR.
In particular, we find that a non-zero $q$ acts as a natural regulator of the $\omega$-pole in ({\ref{nm})}, which eliminates the need to introduce any phenomenological parameters such as $\tau_5$.
Put differently, we derive
an analytical expression for $\tau_5$ in terms of the TCFs.

In the next Section, we review the holographic model.
Sec. \ref{Sec3} presents our results for the TCFs focusing mainly on the new ones. Sec. \ref{Sec3} is supplemented by Appendix \ref{Num}, which  introduces the numerical setup,  and Appendix \ref{complex_tcfs}, which demonstrates how the TCFs extend into the complex $\omega$ plane. The results on MR and CMW are summarized in Sec. \ref{CMWMR}. We conclude with a short 
Summary (Sec. \ref{Summary}).




\section{The Holographic setup}

Anomaly-induced transport has been intensively studied in the past in various holographic models (see e.g., Refs. \cite{Ahn:2024ozz,Lin:2013sga,Gynther:2010ed,Amado:2011zx,Gursoy:2014boa,Gursoy:2014ela,Grozdanov:2016ala,Baggioli:2024zfq,Ammon:2020rvg,Demircik:2024bxd} -- a by far not inclusive list). For our calculations we use the   model  detailed in \cite{Landsteiner:2016led,P5,P6,P2,P3,P4}.

\subsection{The model}

A strongly coupled chiral plasma in 4D 
is modeled holographically by a
$U(1)_V \times U(1)_A$ gauge theory living in a Schwarzschild--AdS$_5$ black brane spacetime, in the probe limit.  
The vector and axial currents of the boundary theory are dual to the bulk gauge fields
$V_M$ and $A_M$, respectively. 
The chiral  
anomaly  is captured by a Chern--Simons term in the bulk.


The bulk geometry is described in ingoing Eddington--Finkelstein coordinates by the metric\footnote{We use capital Latin letters for 5D coordinate indices, Greek letters -- for 4D boundary space and time, and lowercase Latin letters -- for 3D space coordinates.}
\begin{equation}
ds^2 = g_{MN} dx^M dx^N
= 2\, dt\, dr - r^2 f(r) dt^2 + r^2 \delta_{ij} dx^i dx^j ,
\qquad
f(r) = 1 - \frac{1}{r^4} .
\end{equation}
The event horizon is located at $r=1$. The Hawking temperature, identified with the
temperature of the boundary theory, 
is normalized to
\begin{equation}
\pi T = 1 .
\end{equation}
In this work, all dimensionful variables are expressed in units of $\pi T$ and are  dimensionless.

On a constant-$r$ hypersurface $\Sigma$, the induced metric reads
\begin{equation}
ds^2\big|_{\Sigma}
= \gamma_{\mu\nu} dx^\mu dx^\nu
= - r^2 f(r) dt^2 + r^2 \delta_{ij} dx^i dx^j .
\end{equation}
The bulk action is given by
\begin{equation}
S = \int d^5x \sqrt{-g}\, \mathcal{L} + S_{\text{c.t.}},
\end{equation}
where the Lagrangian density takes the form
\begin{align}
\mathcal{L}
&=
-\frac14 (F_V)_{MN}(F_V)^{MN}
-\frac14 (F_A)_{MN}(F_A)^{MN}
\nonumber \\
& \hspace{1.6em} +
\frac{\kappa}{2\sqrt{-g}}\,
\epsilon^{MNPQR}
\left[
3 A_M (F_V)_{NP}(F_V)_{QR}
+
A_M (F_A)_{NP}(F_A)_{QR}
\right] .
\end{align}
Here, $(F_V)_{MN}$ and $(F_A)_{MN}$ denote the field strengths of the vector and axial gauge
fields, respectively, and $\kappa$ is a Chern--Simons coupling. The Chern--Simons term captures the Adler–Bell–Jackiw anomaly of the boundary axial current.

To render the theory holographically renormalizable, the following
counter-term is introduced \cite{Matsuo:2009xn,Taylor:2000xw,Sahoo:2010sp}:
\begin{equation}
S_{\text{c.t.}}
=
\frac14 \log r
\int d^4x \sqrt{-\gamma}
\left[
(F_V)_{\mu\nu}(F_V)^{\mu\nu}
+
(F_A)_{\mu\nu}(F_A)^{\mu\nu}
\right] .
\end{equation}
Varying the action with respect to $V_M$ and $A_M$ yields the bulk equations of motion:
\begin{equation}
\nabla_N (F_V)^{NM}
+
\frac{3\kappa}{\sqrt{-g}}
\epsilon^{MNPQR}
(F_A)_{NP}(F_V)_{QR}
= 0 ,
\end{equation}
\begin{equation}
\nabla_N (F_A)^{NM}
+
\frac{3\kappa}{2\sqrt{-g}}
\epsilon^{MNPQR}
\left[
(F_V)_{NP}(F_V)_{QR}
+
(F_A)_{NP}(F_A)_{QR}
\right]
= 0 .
\end{equation}
These equations  split into the $M=\mu$ and $M=r$ components. We impose the
radial gauge
\begin{equation}
V_r = A_r = 0.
\end{equation}
The equations of motion with $M=\mu$ are the dynamical equations for the bulk gauge fields. The radial components ($M=r$) impose constraint relations. 

The boundary currents are defined as functional derivatives of the 
bulk action with respect to the boundary values of the gauge fields:
\begin{equation}
J^\mu
=
\lim_{r\to\infty}
\frac{\delta S}{\delta V_\mu},
\qquad\qquad
J_5^\mu
=
\lim_{r\to\infty}
\frac{\delta S}{\delta A_\mu}.
\label{Jlim}
\end{equation}
From (\ref{Jlim}), the boundary vector and axial currents can be expressed in terms of the bulk gauge fields \cite{P2}:
\begin{align}
J^\mu
&=
\lim_{r\to\infty}
\sqrt{-\gamma}\, 
\Bigg[
(F_V)^{\mu M} n_M
+
\frac{6\kappa}{\sqrt{-g}}
\epsilon^{M\mu NQR}
n_M A_N (F_V)_{QR}
-
\widetilde{\nabla}_\nu (F_V)^{\nu\mu} \log r
\Bigg], \label{eq:boundary-currentsJ}
\\
J_5^\mu
&=
\lim_{r\to\infty}
\sqrt{-\gamma}\,
\Bigg[
(F_A)^{\mu M} n_M
+
\frac{2\kappa}{\sqrt{-g}}
\epsilon^{M\mu NQR}
n_M A_N (F_A)_{QR}
-
\widetilde{\nabla}_\nu (F_A)^{\nu\mu} \log r
\Bigg],
\label{eq:boundary-currentsJ5}
\end{align}
where $n_M$ is the outward-pointing unit normal vector to the constant-$r$ hypersurface
$\Sigma$ and $\widetilde{\nabla}_\mu$ denotes the covariant derivative compatible with the induced
metric $\gamma_{\mu\nu}$.


To evaluate the currents explicitly, it is convenient to express them in terms of the
near-boundary asymptotic expansion of the bulk gauge fields. As $r\to\infty$, the vector
and axial fields admit the expansions
\begin{align}
V_\mu
&=
\mathcal{V}_\mu
+
\frac{V^{(1)}_\mu}{r}
+
\frac{V^{(2)}_\mu}{r^2}
-
\frac{2 V^L_\mu}{r^2} \log r
+
\mathcal{O}\!\left(\frac{\log r}{r^3}\right), \label{27}
\\
A_\mu
&=
\mathcal{A}_\mu
+
\frac{A^{(1)}_\mu}{r}
+
\frac{A^{(2)}_\mu}{r^2}
-
\frac{2 A^L_\mu}{r^2} \log r
+
\mathcal{O}\!\left(\frac{\log r}{r^3}\right). \label{28}
\end{align}
The boundary values $\mathcal{V}_\mu$ and $\mathcal{A}_\mu$ are the 
 gauge potentials of the electromagnetic/axial fields in the boundary theory:
\begin{align}
E_i &= (F_V)_{it} = \partial_i \mathcal{V}_t - \partial_t \mathcal{V}_i,
&
B_i &= \frac12 \epsilon_{ijk} (F_V)_{jk} = \epsilon_{ijk} \partial_j \mathcal{V}_k,
\\ \label{axial_fields}
E^a_i &= (F_A)_{it} = \partial_i \mathcal{A}_t - \partial_t \mathcal{A}_i,
&
B^a_i &= \frac12 \epsilon_{ijk} (F_A)_{jk} = \epsilon_{ijk} \partial_j \mathcal{A}_k .
\end{align}
Some coefficients appearing in (\ref{27}) and (\ref{28})
are  fixed entirely
by the asymptotic analysis 
\begin{equation}
V^{(1)}_\mu = (F_V)_{t\mu},
\qquad
4 V^L_\mu = \partial^\nu (F_V)_{\mu\nu},
\end{equation}
\begin{equation}
A^{(1)}_\mu = (F_A)_{t\mu},
\qquad
4 A^L_\mu = \partial^\nu (F_A)_{\mu\nu}.
\end{equation}
The coefficients $V^{(2)}_\mu$ and $A^{(2)}_\mu$ cannot  be fixed by the asymptotic analysis alone -- they must be determined by integrating the dynamical bulk equations from the horizon to
the boundary. Throughout this work, $\mathcal{A}_\mu$ is set to zero, that is, no  axial external fields are present in the boundary theory. 

Upon substituting the near-boundary expansions \eqref{27}--\eqref{28} into
\eqref{eq:boundary-currentsJ}--\eqref{eq:boundary-currentsJ5}, we find that the boundary currents take the form
\begin{align}
J^\mu
=
\eta^{\mu\nu}\Big(2V_\nu^{(2)}+2V_\nu^{L}+\eta^{\rho t}\partial_\rho (F_V)_{t\nu}\Big),
\qquad
J_5^\mu
=
\eta^{\mu\nu}\,2A_\nu^{(2)}.
\label{eq:currents-preasymp}
\end{align}


As has been extensively discussed in Refs. \cite{P5,P6,P2,P3,P4},  solving only the dynamical equations for the 
bulk fields $V_\mu$ and $A_\mu$  determines all the TCFs entering the constitutive relations uniquely. 
The  constraint equations translate into the continuity equations (\ref{4}) and (\ref{5}) --
the resulting vector current remains conserved, while the axial current is anomalous due to the Chern--Simons term.

 
To solve the dynamical equations, we adopt the following ansatz for the bulk  fields:
\begin{align}
V_\mu(r,x^\alpha)
&=
\mathcal{V}_\mu(x^\alpha)
-
\frac{\rho(x^\alpha)}{2 r^2} \delta_{\mu t}
+
\mathbb{V}_\mu(r,x^\alpha),
\\
A_\mu(r,x^\alpha)
&=
-
\frac{\rho_5(x^\alpha)}{2 r^2} \delta_{\mu t}
+
\mathbb{A}_\mu(r,x^\alpha),
\end{align}
where the  gauge potential $\mathcal{V}_\mu$ of the external electromagnetic field is assumed to parametrize both an order-one uniform and time-independent magnetic field and a weak electric field.
Within this setup, we recover
the constitutive relations exactly in the form (\ref{11}) and (\ref{12}). The fields $\mathbb{V}_\mu$ and $\mathbb{A}_\mu$ will be determined by solving the dynamical bulk equations with the boundary conditions  $\mathbb{V}_\mu \to 0$ and $\mathbb{A}_\mu \to 0$ at the conformal boundary. 
We further require regularity of all the bulk fields on the interval $r\in[1,\infty)$.
Finally, the remaining integration constants are fixed by the Landau frame convention \cite{landau1987fluid}:
\begin{equation}
J^t = \rho(x^\alpha),
\qquad
J^t_5 = \rho_5(x^\alpha).
\end{equation}

\subsection{Equations of motion in the bulk}

In order to extract the TCFs to all orders in the boundary
derivative expansion, we adopt a linearization scheme in which the background magnetic
field is kept non-perturbative, while the electric fields and charge densities are treated as
small perturbations, that is, 
we assume the scaling
\eqref{linearschem}. In the linear approximation (first order in $\epsilon$), the dynamical bulk equations reduce to the following
system:
\begin{align}
0 &=
r^{3}\,\partial_{r}^{2}\mathbb{V}_{t}
+ 3r^{2}\,\partial_{r}\mathbb{V}_{t}
+ r\,\partial_{r}\partial_{k}\mathbb{V}_{k}
+ 12\kappa\,\partial_{r}\mathbb{A}_{i}\,B_{i},
\label{eq:lin-Vt}
\end{align}
\begin{align}
0 &=
(r^{5}-r)\,\partial_{r}^{2}\mathbb{V}_{i}
+ (3r^{4}+1)\,\partial_{r}\mathbb{V}_{i}
+ 2r^{3}\,\partial_{r}\partial_{t}\mathbb{V}_{i}
- r^{3}\,\partial_{r}\partial_{i}\mathbb{V}_{t}
+ r^{2}\big(\partial_{t}\mathbb{V}_{i}-\partial_{i}\mathbb{V}_{t}\big)
\\ \nonumber
&\quad
+ r\big(\partial^{2}\mathbb{V}_{i}-\partial_{i}\partial_{k}\mathbb{V}_{k}\big)
-\frac{r}{i \omega}\big(\partial_{i}\partial_{k} E_{k}-\partial^{2} E_{i}\big)
- \frac{1}{2}\,\partial_{i}\rho
- r^{2}E_{i}
+ 12\kappa r^{2} B_{i}\,\partial_{r}\mathbb{A}_{t}
+ \frac{12\kappa}{r}\,B_{i}\rho_{5},
\end{align}
\begin{align}
0 &=
r^{3}\,\partial_{r}^{2}\mathbb{A}_{t}
+ 3r^{2}\,\partial_{r}\mathbb{A}_{t}
+ r\,\partial_{r}\partial_{k}\mathbb{A}_{k}
+ 12\kappa\,\partial_{r}\mathbb{V}_{i}\,B_{i},
\end{align}
\begin{align}
0 &=
(r^{5}-r)\,\partial_{r}^{2}\mathbb{A}_{i}
+ (3r^{4}+1)\,\partial_{r}\mathbb{A}_{i}
+ 2r^{3}\,\partial_{r}\partial_{t}\mathbb{A}_{i}
- r^{3}\,\partial_{r}\partial_{i}\mathbb{A}_{t}
+ r^{2}\big(\partial_{t}\mathbb{A}_{i}-\partial_{i}\mathbb{A}_{t}\big)
\nonumber\\
&\quad
+ r\big(\partial^{2}\mathbb{A}_{i}-\partial_{i}\partial_{k}\mathbb{A}_{k}\big)
- \frac{1}{2}\,\partial_{i}\rho_{5}
+ 12\kappa r^{2} B_{i}\,\partial_{r}\mathbb{V}_{t}
+ \frac{12\kappa}{r}\,B_{i}\rho .
\label{eq:lin-Ai}
\end{align}
In the presence of the background magnetic field $\vec{B}$, the Chern--Simons term couples the dynamical equations for the vector and axial bulk fields.

In order to solve \eqref{eq:lin-Vt}--\eqref{eq:lin-Ai} systematically, we decompose the
bulk gauge fields into a basis of all allowed linear structures in $\epsilon$  built from the boundary fields  $\{\rho,\rho_5,\vec E,\vec B\}$ and their boundary derivatives. We employ the ansatz:
\begin{align}
\mathbb{V}_{t} &=
S_1\,\rho
+ S_2\,\kappa\,B_k\partial_k\rho_5
+ S_3\,\partial_k E_k
+ S_4\,\kappa^{2} B_i B_k \partial_k E_i,
\label{eq:ansatz-Vt}
\\
\mathbb{V}_{i} &=
V_1\,\partial_i\rho
+ V_2\,\kappa^{2} B_i B_k \partial_k\rho
+ V_3\,\kappa B_i\rho_5
+ V_4\,\kappa B_k\partial_k\partial_i\rho_5
+ V_5\,E_i
+ V_6\,\kappa^{2} E_j B_j B_i
\nonumber\\
&\hspace{1.6em}
+ V_7\,\partial_i\partial_k E_k + V_8\,\kappa^{2} B_k B_j \partial_i\partial_k E_j +V_9\,\kappa^{2} B_i B_j \partial_j\partial_k E_k ,
\label{eq:ansatz-Vi}
\\
\mathbb{A}_{t}&=
\bar S_1\,\rho_5
+ \bar S_2\,\kappa\,B_k\partial_k\rho
+ \bar S_5\,\kappa\,E_k B_k
+ \bar S_6\,\kappa\,B_i\partial_i\partial_k E_k,
\label{eq:ansatz-At}
\\
\mathbb{A}_{i} &=
\bar V_1\,\partial_i\rho_5
+ \bar V_2\,\kappa^{2} B_i B_k \partial_k\rho_5
+ \bar V_3\,\kappa B_i\rho
+ \bar V_4\,\kappa B_k\partial_k\partial_i\rho
+ \bar {V}_{10}\,\kappa\,\partial_i E_k B_k
\nonumber\\
&\hspace{1.6em}
+ \bar {V}_{11}\,\kappa B_j\partial_i\partial_j\partial_k E_k + \bar {V}_{12}\,\kappa^{3} B_i B_j B_k \partial_j E_k + \bar {V}_{13}\,\kappa B_i \partial_k E_k .
\label{eq:ansatz-Ai}
\end{align}
The coefficient functions \(S_i,\bar S_i,V_i,\bar V_i\) are the components of the Green's function matrix. They depend on the radial coordinate and  boundary derivative operators,
becoming scalar functions of $(r,\omega,q^2)$ upon Fourier transformation with respect to the boundary coordinates.
In Fourier space, Eqs. 
\eqref{eq:lin-Vt}--\eqref{eq:lin-Ai} 
form a system of ODEs,
which can be split into two 
decoupled sub-sectors.

\textbf{Sub-sector (i):} \quad $\{S_1, \bar{S_2}, V_1, V_2, \bar{V}_3, \bar{V}_4\}$
\begin{align}\notag
&0= r^3 \partial_r^2 S_1 + 3r^2 \partial_r S_1 - q^2 r \partial_r V_1 - r(\kappa \vec{B} \cdot \vec{q})^2  \partial_r V_2 + 12 (\kappa B)^2 \partial_r \bar{V}_3 - 12 (\kappa \vec{B} \cdot \vec{q})^2 \partial_r \bar{V}_4, \quad \\[6pt]  \label{61}
&0= r^3 \partial_r^2 \bar{S}_2 + 3r^2 \partial_r \bar{S}_2 + r \partial_r \bar{V}_3  - q^2 r \partial_r \bar{V}_4 + 12\partial_r V_1 + 12 (\kappa B)^2 \partial_r V_2 , \quad \\[6pt] \notag
&1/2= (r^5 - r) \partial_r^2 V_1 + (3r^4 + 1 - 2i\omega r^3) \partial_r V_1 - i\omega r^2 V_1 - r^2 (S_1 + r \partial_r S_1) + r(\kappa \vec{B} \cdot \vec{q})^2  V_2 ,\quad \\[6pt] \notag
&0= (r^5 - r) \partial_r^2 V_2 + (3r^4 + 1 - 2i\omega r^3) \partial_r V_2 - i\omega r^2 V_2 - q^2 r V_2 + 12 r^2 \partial_r \bar{S}_2, \quad \\[6pt] \notag
&0= (r^5 - r) \partial_r^2 \bar{V}_3 + (3r^4 + 1 - 2i\omega r^3) \partial_r \bar{V}_3 - i\omega r^2 \bar{V}_3 - q^2 r \bar{V}_3 + 12 r^2 \partial_r S_1 + 12/r  , \quad \\[6pt] \notag
&0= (r^5 - r) \partial_r^2 \bar{V}_4 + (3r^4 + 1 - 2i\omega r^3) \partial_r \bar{V}_4 - i\omega r^2 \bar{V}_4 - r^2 (\bar{S}_2 + r \partial_r \bar{S}_2) - r \bar{V}_3.\quad \notag
\end{align}

\textbf{Sub-sector (ii):} \quad $\{S_3, S_4, \bar{S_5}, \bar{S_6}, V_5, V_6, V_7,V_8,V_9, \bar{V}_{10},\bar{V}_{11},\bar{V}_{12},\bar{V}_{13}\}$
\begin{align}
\notag
0 &= r^3 \partial_r^2 S_3+ 3r^2 \partial_r S_3+r\partial_rV_5 -q^2 r \partial_r V_7- r(\kappa \vec{B}\cdot\vec{q})^2\partial_rV_9 -12(\kappa \vec{B}\cdot\vec{q})^2\partial_r\bar{V}_{11}\notag  \\ \notag & +12(\kappa B)^2\partial_r\bar{V}_{13} , \quad \\[6pt] \notag
0 &= r^3 \partial_r^2 S_4+ 3r^2 \partial_r S_4 + r \partial_r V_6 -q^2r\partial_rV_8+ 12\partial_r \bar{V}_{10}+12(\kappa B)^2\partial_r \bar{V}_{12}, \quad \\[6pt] \notag
0&=r^3 \partial_r^2 \bar{S_5}+ 3r^2 \partial_r \bar{S_5}-q^2 r \partial_r \bar{V}_{10}-r(\kappa \vec{B}\cdot\vec{q})^2\partial_r \bar{V}_{12} +12\partial_r V_5+12(\kappa B)^2\partial_r V_6\notag  \\ \notag & -12(\kappa \vec{B}\cdot\vec{q})^2\partial_r V_8 ,\quad \\[6pt] \notag
0& =r^3 \partial_r^2 \bar{S_6}+ 3r^2 \partial_r \bar{S_6}-q^2r\partial_r \bar{V}_{11} +r \partial_r\bar{V}_{13} +12\partial_r V_7+12(\kappa B)^2\partial_rV_{9},
\quad \\[6pt]  \label{62}
0 &= (r^5 - r) \partial_r^2 V_5 + (3r^4 + 1 - 2i\omega r^3) \partial_r V_5 - i\omega r^2 V_5 - q^2 r V_5-r^2- \frac{rq^2}{i \omega} ,
\quad \\[6pt] \notag
0 &= (r^5 - r) \partial_r^2 V_6 + (3r^4 + 1 - 2i\omega r^3) \partial_r V_6 - i\omega r^2 V_6 -q^2rV_6+12 r^2 \partial_r \bar{S_5} , \quad \\[6pt]   \notag
0 &= (r^5 - r) \partial_r^2 V_7 + (3r^4 + 1 - 2i\omega r^3) \partial_r V_7 - i\omega r^2 V_7 -r^2(S_3+r\partial_r S_3) -rV_5 \notag  \\ \notag &+r(\kappa \vec{B}\cdot\vec{q})^2\bar{V}_{9}- \frac{r}{i \omega} ,
\quad \\[6pt] \notag
0 &= (r^5 - r) \partial_r^2 V_8 + (3r^4 + 1 - 2i\omega r^3) \partial_r V_8 - i\omega r^2 V_8 -r^2(S_4+r\partial_r S_4) -rV_6,
\quad \\[6pt] \notag
0 &= (r^5 - r) \partial_r^2 V_9 + (3r^4 + 1 - 2i\omega r^3) \partial_r V_9 - i\omega r^2 V_9 - q^2 r V_9 + 12r^2\partial_r{\bar{S}}_6
\quad \\[6pt] \notag
0 &= (r^5 - r) \partial_r^2 \bar{V}_{10} + (3r^4 + 1 - 2i\omega r^3) \partial_r \bar{V}_{10} - i\omega r^2 \bar{V}_{10} -r^2(\bar{S}_5+r\partial_r \bar{S}_5) +r(\kappa \vec{B}\cdot\vec{q})^2\bar{V}_{12},
\quad \\[6pt] \notag
0 &= (r^5 - r) \partial_r^2 \bar{V}_{11} + (3r^4 + 1 - 2i\omega r^3) \partial_r \bar{V}_{11} - i\omega r^2 \bar{V}_{11} -r^2(\bar{S}_6+r\partial_r \bar{S}_6) -r \bar{V}_{13},
\quad \\[6pt] \notag
0 &= (r^5 - r) \partial_r^2 \bar{V}_{12} + (3r^4 + 1 - 2i\omega r^3) \partial_r \bar{V}_{12} - i\omega r^2 \bar{V}_{12} -q^2 r \bar{V}_{12} +12r^2\partial_r S_4,
\quad \\[6pt] \notag
0 &= (r^5 - r) \partial_r^2 \bar{V}_{13} + (3r^4 + 1 - 2i\omega r^3) \partial_r \bar{V}_{13} - i\omega r^2 \bar{V}_{13} -q^2 r \bar{V}_{13} +12r^2\partial_r S_3.
\end{align}
The sub-sector $\{\bar{S}_1, S_2, \bar{V}_1, \bar{V}_2, V_3, V_4\}$ satisfies equations identical to those of sub-sector (i).
Given that they also obey the same boundary conditions, we conclude that
\[
\{\bar{S}_1, S_2, \bar{V}_1, \bar{V}_2, V_3, V_4\} = \{S_1, \bar{S}_2, V_1, V_2, \bar{V}_3, \bar{V}_4\}.
\]

The near-boundary behaviour \eqref{27}--\eqref{28} translates into
the following asymptotics for the decomposition coefficients in
\eqref{eq:ansatz-Vt}--\eqref{eq:ansatz-Ai}:
\begin{align}
S_i &\to \frac{s^{(1)}_i}{r} + \frac{s_i}{r^{2}} + \frac{s_i^{L}}{r^{2}}\log r + \cdots,
&
V_i &\to \frac{v^{(1)}_i}{r} + \frac{v_i}{r^{2}} + \frac{v_i^{L}}{r^{2}}\log r + \cdots,
\label{expansion_sv}
\\
\bar S_i &\to \frac{\bar s^{(1)}_i}{r} + \frac{\bar s_i}{r^{2}} + \frac{\bar s_i^{L}}{r^{2}}\log r + \cdots,
&
\bar V_i &\to \frac{\bar v^{(1)}_i}{r} + \frac{\bar v_i}{r^{2}} + \frac{\bar v_i^{L}}{r^{2}}\log r + \cdots .
\label{expansion}
\end{align}
The coefficients $s_i^{(1)}, s_i^{L}, v_i^{(1)}$, and $v_i^{L}$ (and their barred analogues) are fixed uniquely by the near-boundary analysis. The coefficients $s_i$ and $\bar{s}_i$ are specified by the Landau frame convention, while the subleading coefficients $v_i$ and $\bar{v}_i$ are determined by solving the full set of radial ODEs.

Substituting \eqref{eq:ansatz-Vt}--\eqref{eq:ansatz-Ai} into \eqref{eq:currents-preasymp} and using the expansions \eqref{expansion_sv} and (\ref{expansion}), we can express the boundary currents as:
\begin{align}
J_t &=
(1-2s_1)\rho
-2s_2\,\kappa B_k\partial_k\rho_5
+\Big(\frac12-2s_3\Big)\partial_k E_k
-2s_4\,\kappa^{2} B_i B_k \partial_k E_i,
\end{align}
\begin{align} \label{eq:constitutive-currents J}
J_i &=
2v_1\,\partial_i\rho
+2v_2\,\kappa^{2} B_i B_k \partial_k\rho
+2v_3\,\kappa B_i\rho_5
+2v_4\,\kappa B_k\partial_k\partial_i\rho_5
+\Big(2v_5-\frac{i\omega}{2} +\frac{q^2}{2i\omega}\Big)E_i
\\ \nonumber
&\hspace{1.6em}
+2v_6\,\kappa^{2} (E_j B_j) B_i
+\Big(2v_7  +\frac{1}{2i\omega} ,\Big)\partial_i\partial_k E_k
+2v_8\,\kappa^{2} B_k B_j \partial_i\partial_k E_j+2v_9\,\kappa^{2} B_i B_j \partial_j\partial_k E_k,
\end{align}
\begin{align}
J_t^{5} &=
(1-2\bar s_1)\rho_5
-2\bar s_2\,\kappa B_k\partial_k\rho
-2\bar s_5\,\kappa E_k B_k
-2\bar s_6\,\kappa B_i\partial_i\partial_k E_k,
\end{align}
\begin{align}
J_i^{5} &=
2\bar v_1\,\partial_i\rho_5
+2\bar v_2\,\kappa^{2} B_i B_k \partial_k\rho_5
+2\bar v_3\,\kappa B_i\rho
+2\bar v_4\,\kappa B_k\partial_k\partial_i\rho
+2\bar {v}_{10}\,\kappa\,\partial_i E_k B_k
\nonumber\\
&\hspace{1.6em}
+2\bar {v}_{11}\,\kappa B_j\partial_i\partial_j\partial_k E_k
+2\bar {v}_{12}\,\kappa^{3} B_i B_j B_k \partial_j E_k + 2\bar {v}_{13}\,\kappa B_i \partial_k E_k .
\label{eq:constitutive-currents}
\end{align}
Finally, applying the Landau frame convention  fixes the coefficients $s_i$ and $\bar{s}_i$, yielding
the relations
\begin{equation}
s_1=s_2=s_4=\bar s_1=\bar s_2=\bar s_5=\bar s_6=0,
\qquad
s_3=\frac14 .
\end{equation}
The 3-currents \eqref{eq:constitutive-currents J} and \eqref{eq:constitutive-currents} can then be cast into the form
(\ref{11}) and (\ref{12})
with the TCFs expressed in terms of the near-boundary data $v_i$
and $\bar v_i$:
\begin{align} 
D = &-2v_1,
\qquad
D_B = 2v_2,
\qquad
\sigma_\chi = 2v_3,
\qquad
D_\chi = 2v_4,
\qquad
\sigma_e = 2v_5 - \frac{i\omega}{2}+\frac{q^2}{2i\omega},
\notag \qquad\\ 
&
\delta\sigma_\chi = 2v_6,
\qquad
\gamma_D = 2v_7+\frac{1}{2i\omega},
\qquad
\tau_D = 2v_8,
\qquad
\tau_B = 2v_9,
\qquad
\gamma_\chi = 2\bar {v}_{10},
 \notag \qquad\\ \label{13tcf}
& \qquad
\qquad
\tau_\chi = 2\bar {v}_{11},
\qquad
\sigma_B = 2\bar {v}_{12},
\qquad
\gamma_B = 2\bar {v}_{13}.
\end{align}


\section{Results for the TCFs}\label{Sec3}
\subsection{Analytical results: the hydrodynamic limit}
The TCFs $D$, $D_B$, $D_\chi$, and $\sigma_\chi$ have been extracted and analyzed  previously in ~\cite{P1,P2,P3,P6}.  $\sigma_e$ and $\delta\sigma_\chi$  were studied only in the limiting cases of $B=0$~\cite{P6} or $q^2=0$~\cite{P3}.
In the  zeroth-order limit ($\omega=q^2=0$), these TCFs yield the following analytical values, in agreement with the earlier results~\cite{P1,P2,P3,P6}
\begin{align}
&D^0 = \frac{1}{2}+18\big(1-2\log(2)\big)\kappa^2 B^2+\mathcal{O}((\kappa B)^4), \qquad D_B^0(B=0) = -9\big(\pi-2\log(2)\big), \label{eq:anal_13tcf}\\ \notag &\sigma^0_\chi =  6 + 216 \kappa^2 B^2  \big(1- \log(4)\big)+\mathcal{O}((\kappa B)^4), \qquad D_\chi^0(B=0) = \frac{1}{8}\big(6\pi-12\log(2)-\pi^2\big), \\ \notag  &\delta\sigma_\chi^0(B=0) = 18\big(\pi-2\log(2)\big), \quad \sigma_e^0(B=0) = 1, \quad 
\gamma_D^0(B=0) = \frac{1}{16}\big(2\pi-\pi^2+4\log(2)\big).
\end{align}
From the analytical expression of $D^0(B)$, we can  estimate the magnetic field at which its contribution becomes significant\footnote{Among all the TCs, $D^0(B)$ provides the most stringent constraint on $B$.}. The deviation from  $D^0(B=0)$  reaches, say, $10\%$ already at $\kappa B \approx 0.09$. At this value,  the weak magnetic field approximation breaks down, and the TCF transitions into the regime of strong magnetic fields. As will be shown below when evaluating other TCFs, strong magnetic field regime starts at even smaller values of $\kappa B$.

The TCFs $\tau_D$, $\tau_B$, $\gamma_\chi$, $\tau_\chi$, $\sigma_B$, and $\gamma_B$ are new and computed here for the first time: 
\begin{align}
\gamma_\chi^0(B=0) = -\frac{3}{2}\big(\pi-2\log(2)\big) \qquad  \label{eq:anal_13tcf_new}  \gamma_B^0(B=0) & = -\frac{3}{2} \big(\pi+2\log(2)\big).
\end{align}
We were  unable to obtain any analytical expressions for the remaining TCFs.  Numerical results are presented in the next subsection.

\subsection{Numerical results}

\subsection*{TCF's dependence on the magnetic field}

\begin{figure}[htbp] 
    \centering
    \includegraphics[width=0.8\textwidth]{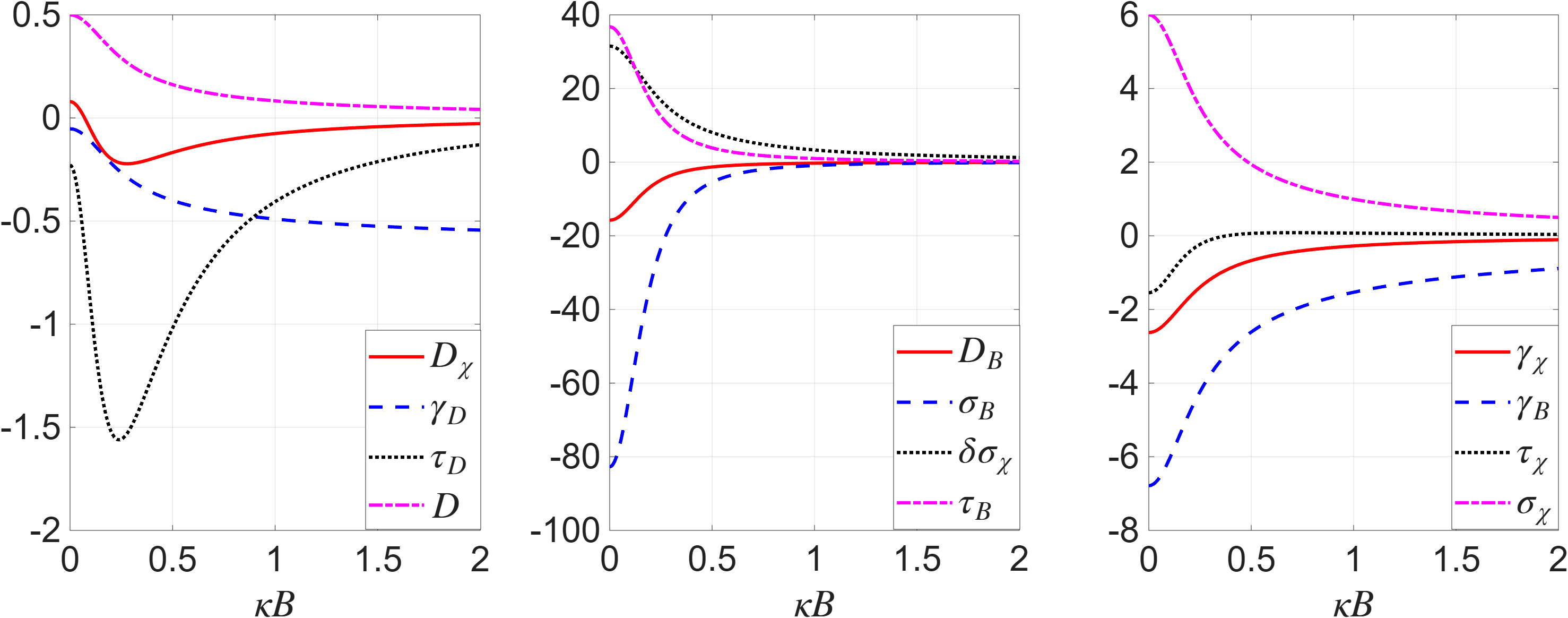}
    \caption{ TCFs as functions of $\kappa B$ at $\omega = q^2 = \alpha = 0$.}
    \label{fig:multi_dep}
\end{figure}

We extend our analysis beyond the weak magnetic field limit and explore the TCFs as functions of $\kappa B$ while fixing $\omega = q^2 = \alpha=0$. The  results are presented in Fig.~\ref{fig:multi_dep}.

An interesting exception in Fig.~\ref{fig:multi_dep} is $\gamma_D$, which is the only TCF that does not vanish in the large-$\kappa B$ limit. All other TCFs are strongly suppressed and asymptotically approach zero as $\kappa B$ increases, whereas $\gamma_D$ approaches a finite nonzero asymptotic value.\\
$\sigma_e$ is independent of $\kappa B$ and 
is thus not displayed in  Fig.~\ref{fig:multi_dep}. 

\subsection*{TCFs beyond the Hydrodynamic Limit}
\begin{figure}[htbp]
    \centering
    \includegraphics[width=0.8\linewidth]{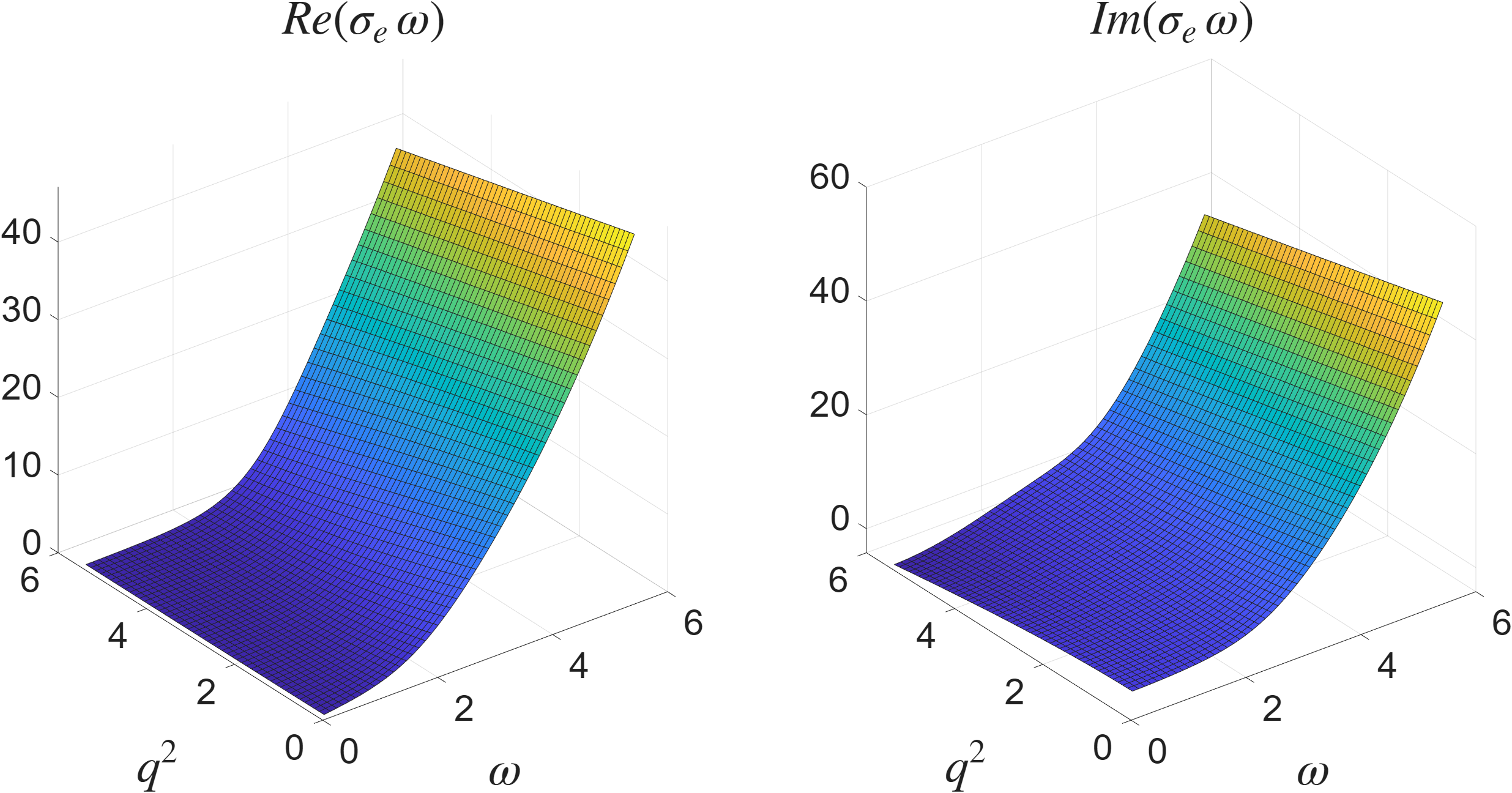}
    \caption{TCF $\sigma_e \omega$  as a function of  $\omega$  and  $q^2$.  $\kappa B = 0.25$ and $\alpha = 0$.}
    \label{1}
\end{figure}

\begin{figure}[htbp]
    \centering
    \includegraphics[width=0.8\linewidth]{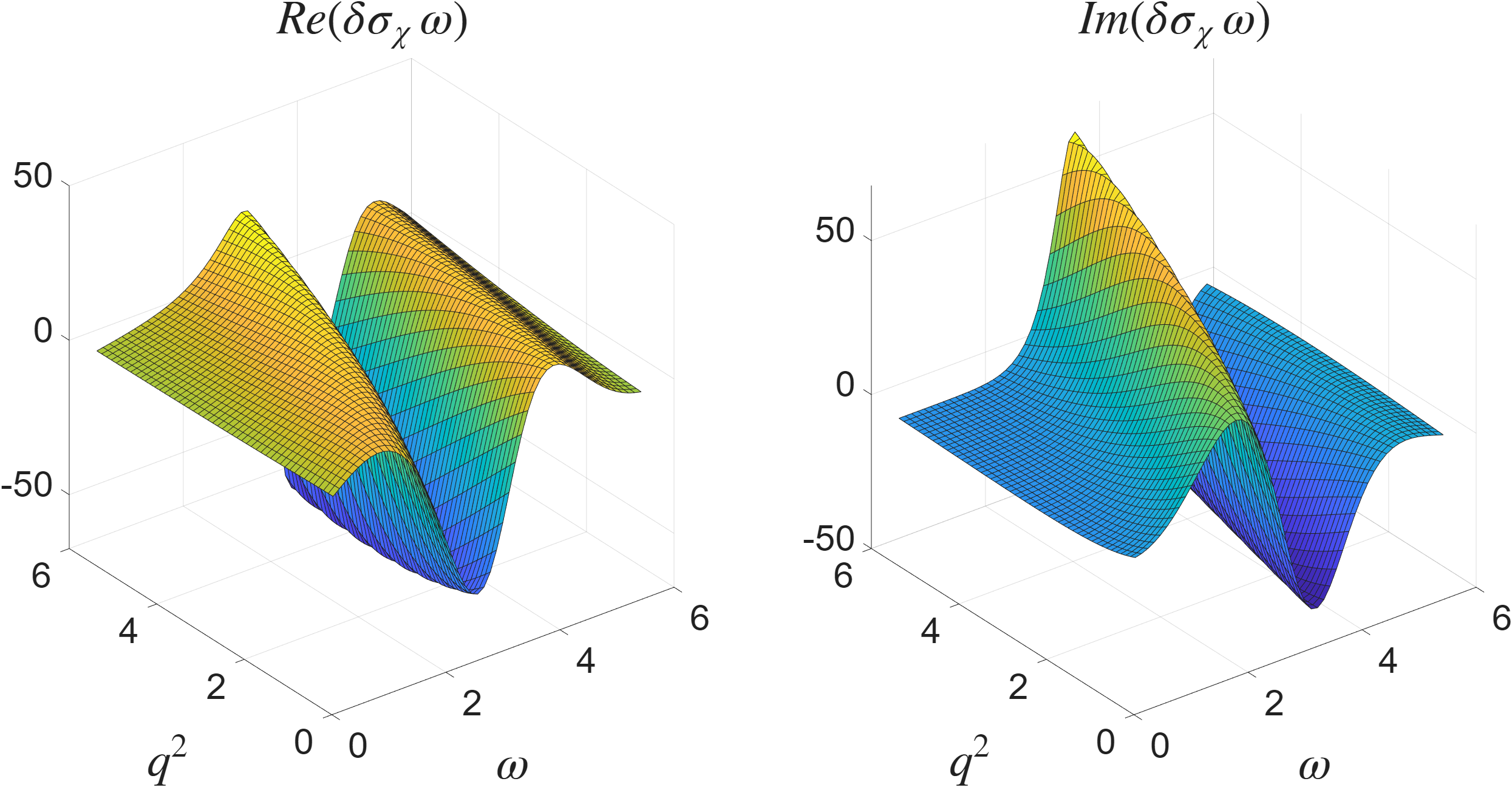}
    \caption{TCF $\delta\sigma_\chi \omega$  as a function of  $\omega$  and  $q^2$.  $\kappa B = 0.25$ and $\alpha = 0$.}
    \label{2}
\end{figure}

\begin{figure}[htbp]
    \centering
    \includegraphics[width=0.8\linewidth]{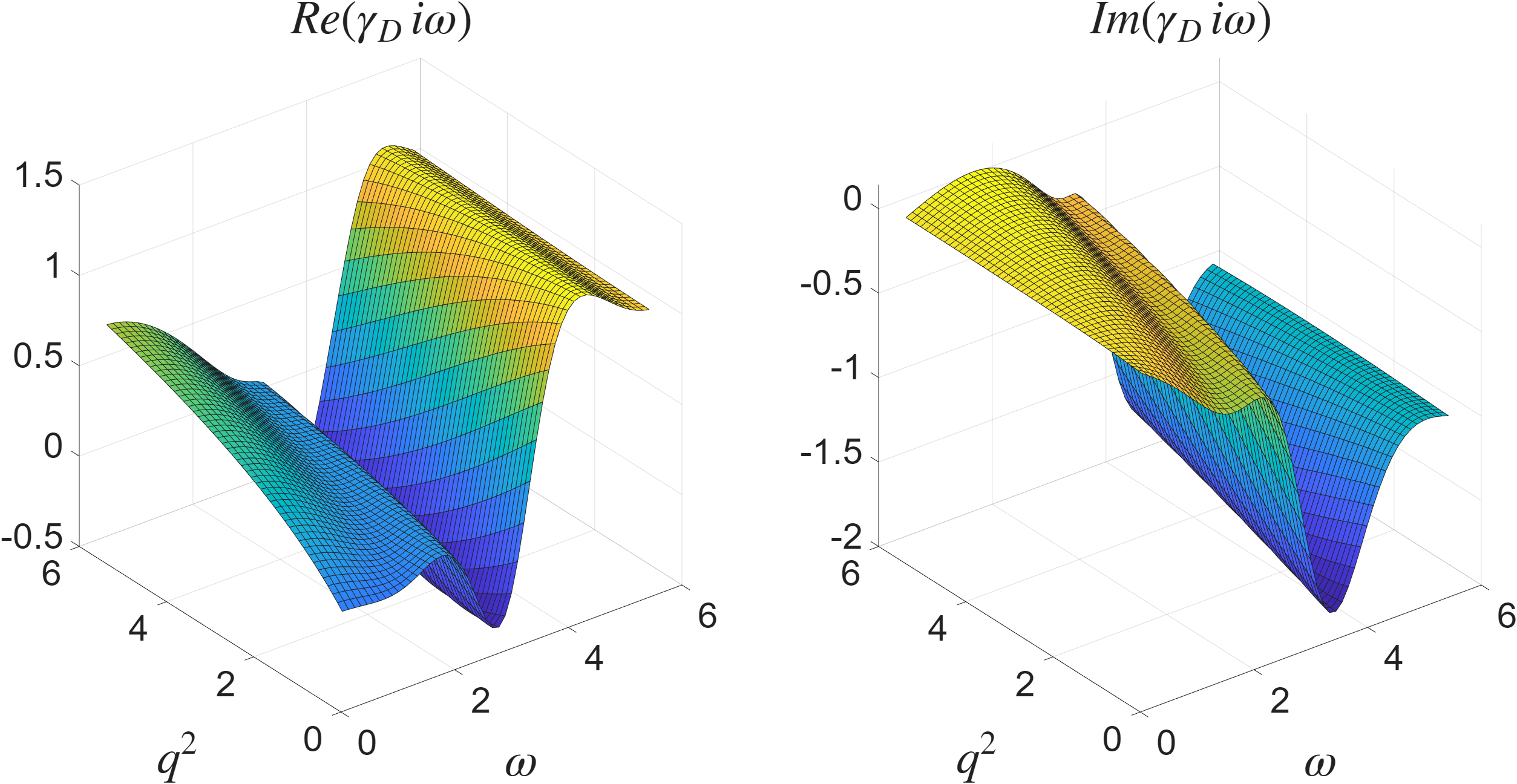}
    \caption{TCF $\gamma_D i \omega= \sigma_m$  as a function of  $\omega$  and  $q^2$.  $\kappa B = 0.25$ and $\alpha = 0$.}
    \label{3}
\end{figure}

\begin{figure}[htbp]
    \centering
    \includegraphics[width=0.8\linewidth]{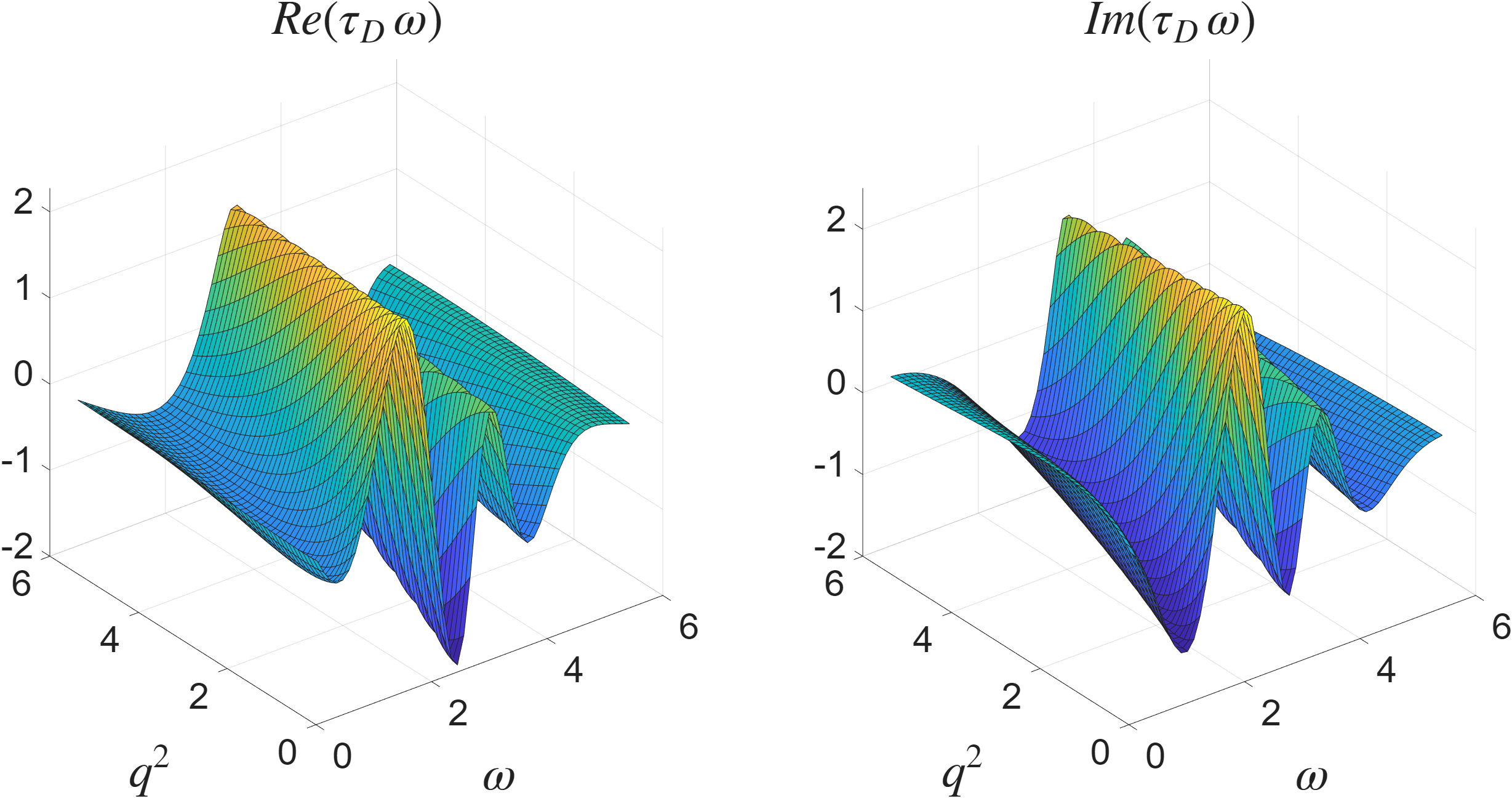}
    \caption{TCF $\tau_D \omega$  as a function of  $\omega$  and  $q^2$.  $\kappa B = 0.25$ and $\alpha = 0$.}
    \label{F4}
\end{figure}

\begin{figure}[htbp]
    \centering
    \includegraphics[width=0.8\linewidth]{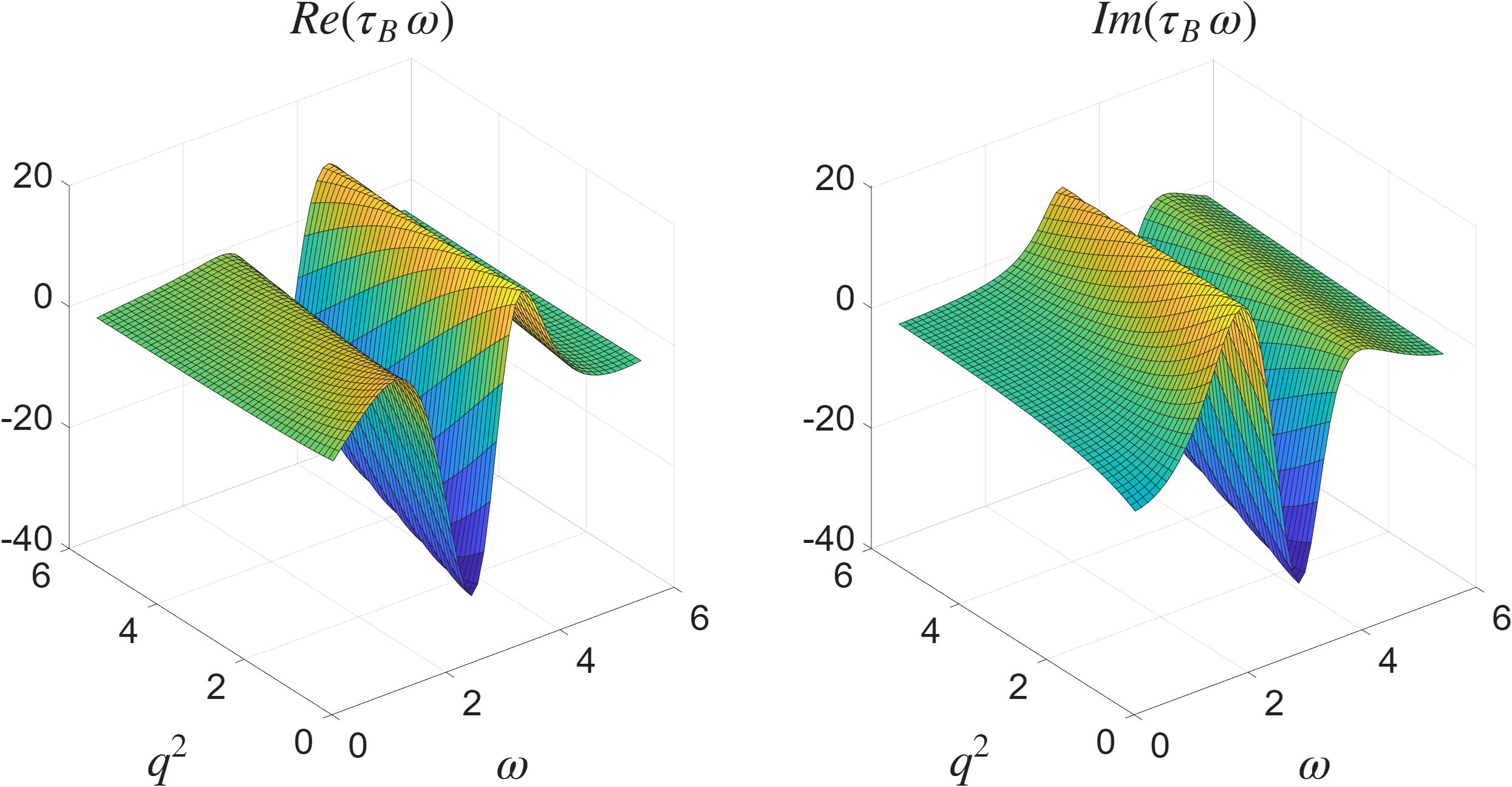}
    \caption{TCF $\tau_B \omega$  as a function of  $\omega$  and  $q^2$.   $\kappa B = 0.25$ and $\alpha = 0$.}
    \label{F5}
\end{figure}

\begin{figure}[htbp]
    \centering
    \includegraphics[width=0.8\linewidth]{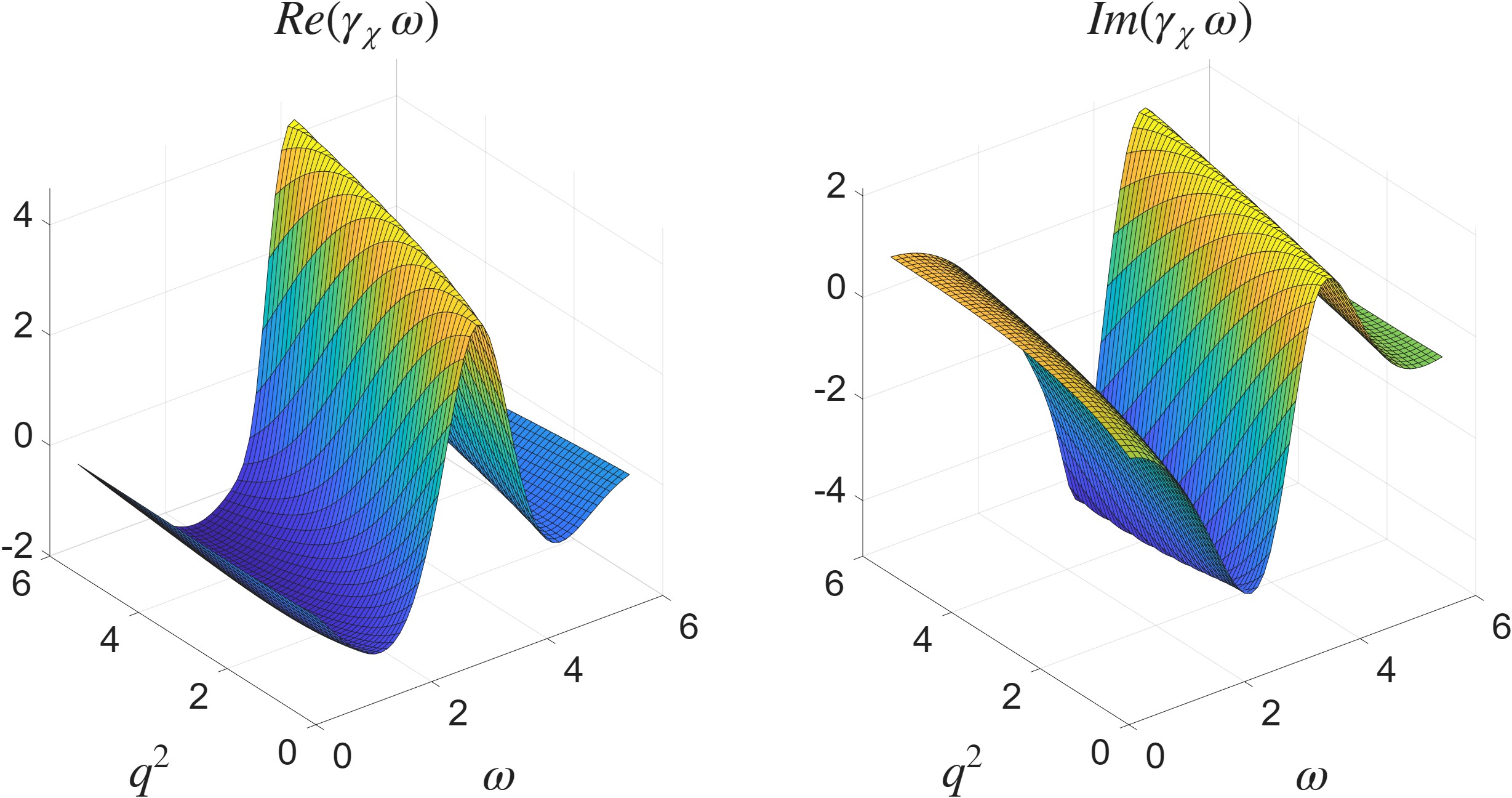}
    \caption{TCF $\gamma_\chi \omega$  as a function of  $\omega$  and  $q^2$.   $\kappa B = 0.25$ and $\alpha = 0$.}
    \label{F7}
\end{figure}

\begin{figure}[htbp]
    \centering
    \includegraphics[width=0.8\linewidth]{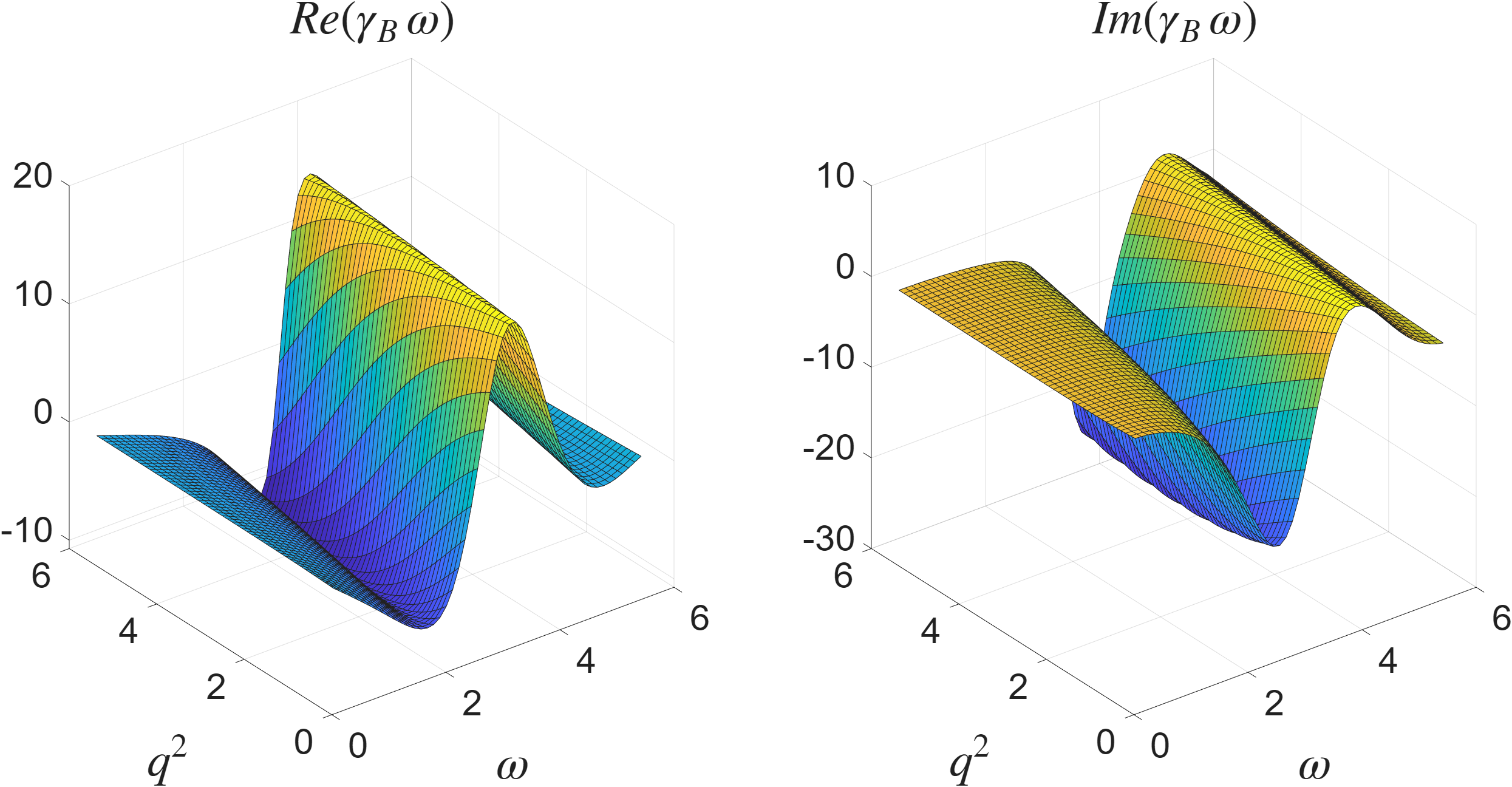}
    \caption{TCF $\gamma_B \omega$  as a function of  $\omega$  and  $q^2$. $\kappa B = 0.25$ and $\alpha = 0$.}
    \label{8}
\end{figure}

\begin{figure}[htbp]
    \centering
    \includegraphics[width=0.8\linewidth]{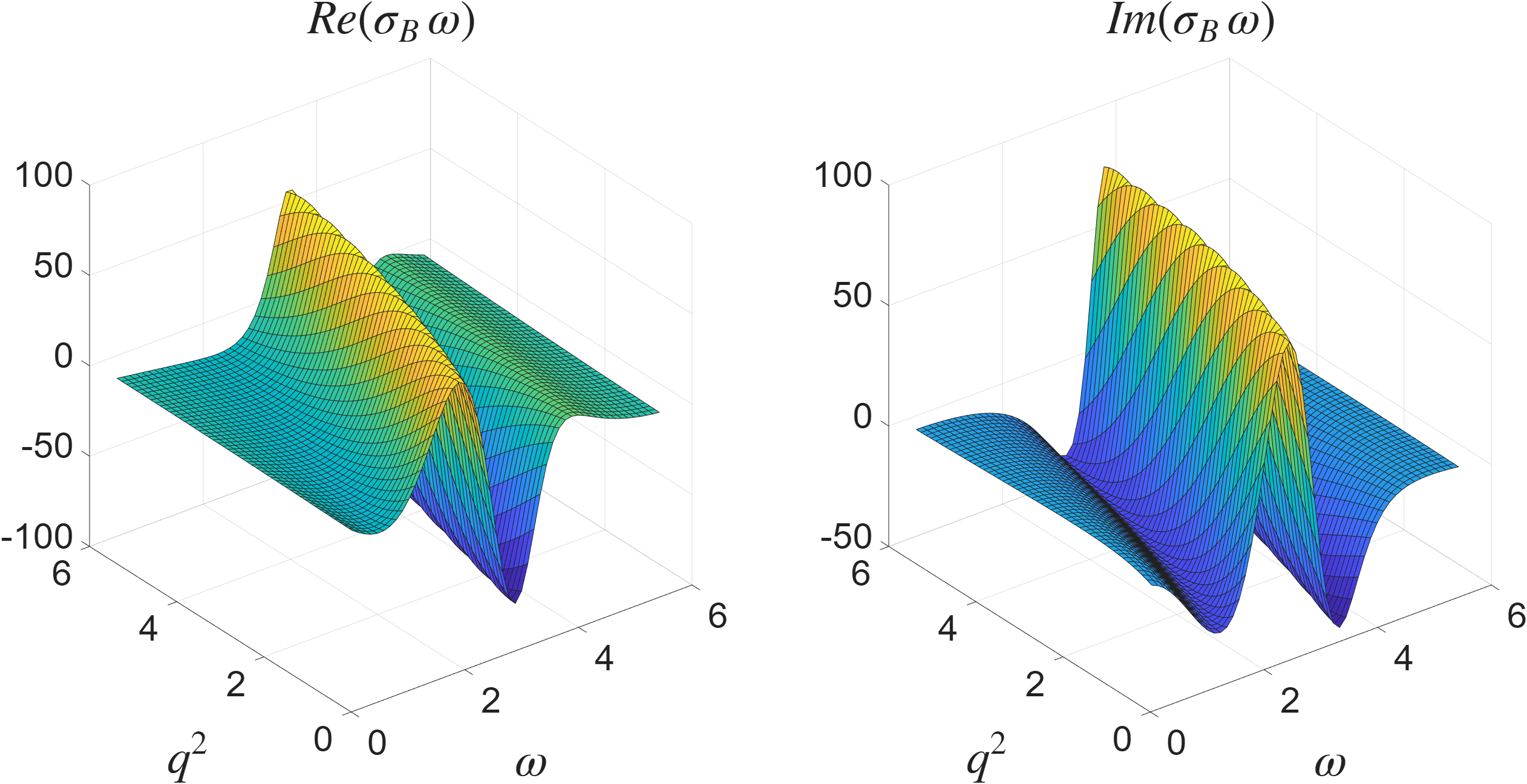}
    \caption{TCF $\sigma_B \omega$  as a function of  $\omega$  and  $q^2$.   $\kappa B = 0.25$ and $\alpha = 0$.}
    \label{9}
\end{figure}

\begin{figure}[htbp]
    \centering
    \includegraphics[width=0.8\linewidth]{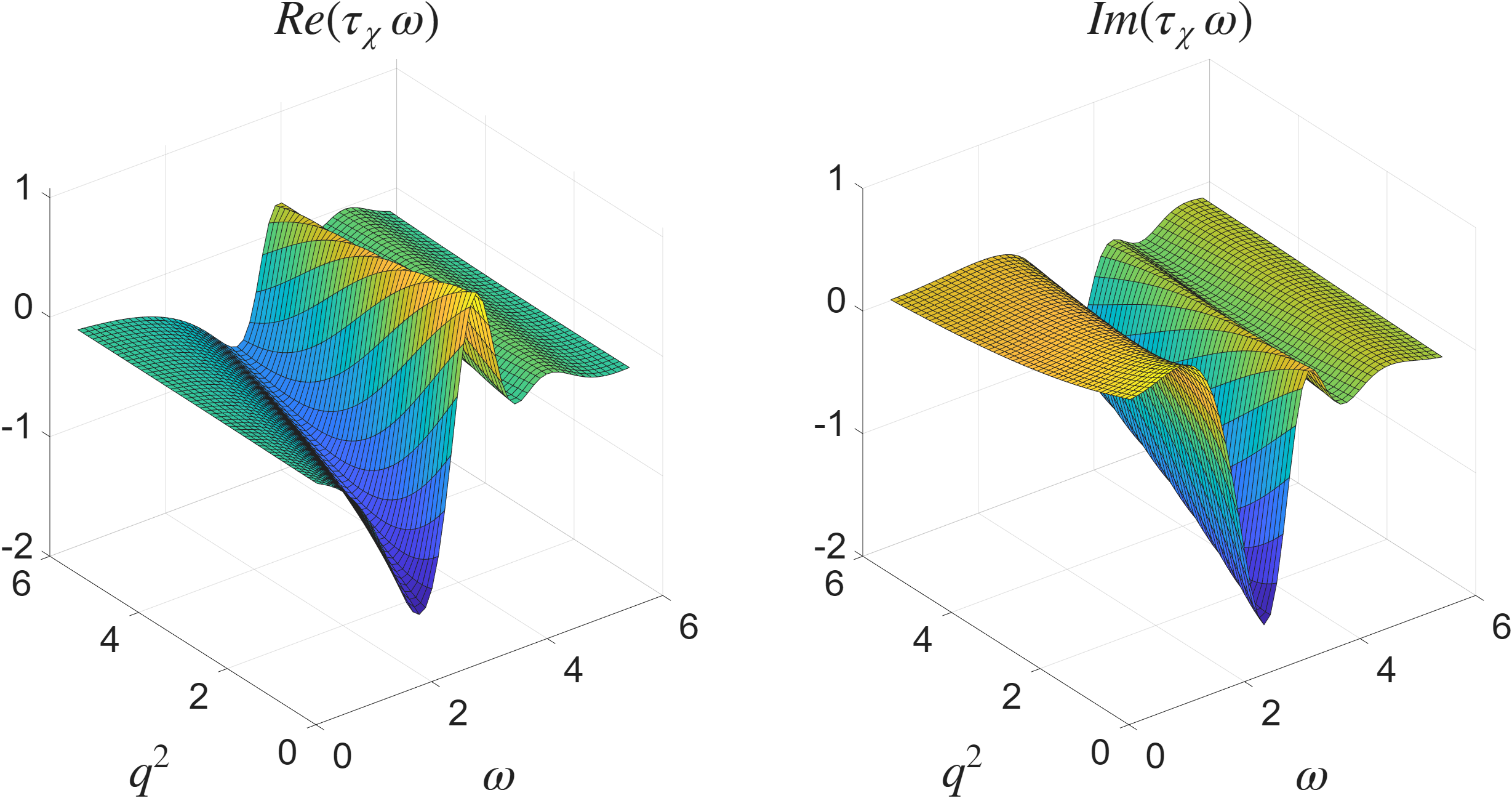}
    \caption{TCF $\tau_\chi \omega$  as a function of  $\omega$  and  $q^2$.  $\kappa B = 0.25$ and $\alpha = 0$.}
    \label{fig 11}
\end{figure}

We now extend our analysis beyond the hydrodynamic regime. In particular, we systematically compute 
all the TCFs as functions of $\omega$, $q^2$, $B$, and the angle $\alpha$ between $\vec q$ and $\vec B$. This is a 4D space of parameters and it would be too bulky to attempt to present all the dependencies. We thus choose to highlight  a few representative results, focusing primarily on new results (the TCFs $D$, $\sigma_\chi$, $D_\chi$, $D_B$  were featured in ~\cite{P6} and therefore are not shown here).
Figs.~\ref{1}--\ref{fig 11} display the TCFs as functions of $\omega$ and $q^2$
at $\kappa B = 0.25$, a representative value for the magnetic field within the strong-field regime. 
The numerical procedure for solving the bulk equations 
of motion is sketched  in Appendix \ref{Num}.

Since the ODEs \eqref{61}--\eqref{62} contain terms proportional to $1/\omega$, some of the TCFs become singular in the limit $\omega \to 0$ at finite $q$. 
We therefore choose to present the  TCFs  rescaled by $\omega$. In particular, we multiply $\gamma_D$ by $i\omega$, which precisely yields the TCF $\sigma_m$ introduced in~\cite{P1}.

All the TCFs share a similar qualitative behavior: they exhibit a relatively weak dependence on $q^2$ compared to strong variations with respect to $\omega$. This weak dependence on $q^2$ reflects a spatial quasi-locality of the relevant transport processes. 


All the TCFs, except $\sigma_e$, display damped oscillations with $\omega$. 
$\sigma_e$ behaves differently: both $\Re(\sigma_e)$ and $\Im(\sigma_e)$ increase with $\omega$, without developing any oscillatory behavior. This is consistent with earlier observations in Refs.~\cite{P1, Horowitz:2008bn} that at $q=0$, 
$\Re(\sigma_e)\sim \omega$ and
$\Im(\sigma_e)\sim \omega\log\omega$ when
 $\omega$ is asymptotically large.
 In Fig.~\ref{1}, we plot $\omega\,\Re(\sigma_e)$ and $\omega\,\Im(\sigma_e)$, where the additional factor of $\omega$ makes the frequency dependence even more pronounced.


An interesting phenomenon, previously noticed in ~\cite{P6}, is that at finite $\kappa B$  ($\kappa B \gtrsim 0.5$), all the TCFs (both the real and imaginary parts) exhibit singularities at the same value of $\omega$. 
While there might be some physics associated with this singularity \cite{Haack:2018ztx,Waeber:2024ilt}, we would not be exploring it here. 
Instead, we will assume that it sets the limit on our numerical analysis.  We have discovered that the larger the product $\omega q B$, the less accurate the numerical calculation. 
Below we will find even more  stringent constraint on the maximum magnetic field accessible numerically.

\label{sec:beyond-hydro-numerics}
\newpage
\subsection*{$\alpha$-dependence of the TCFs }
\begin{figure}[H]
    \centering
    \includegraphics[width=0.7\linewidth]{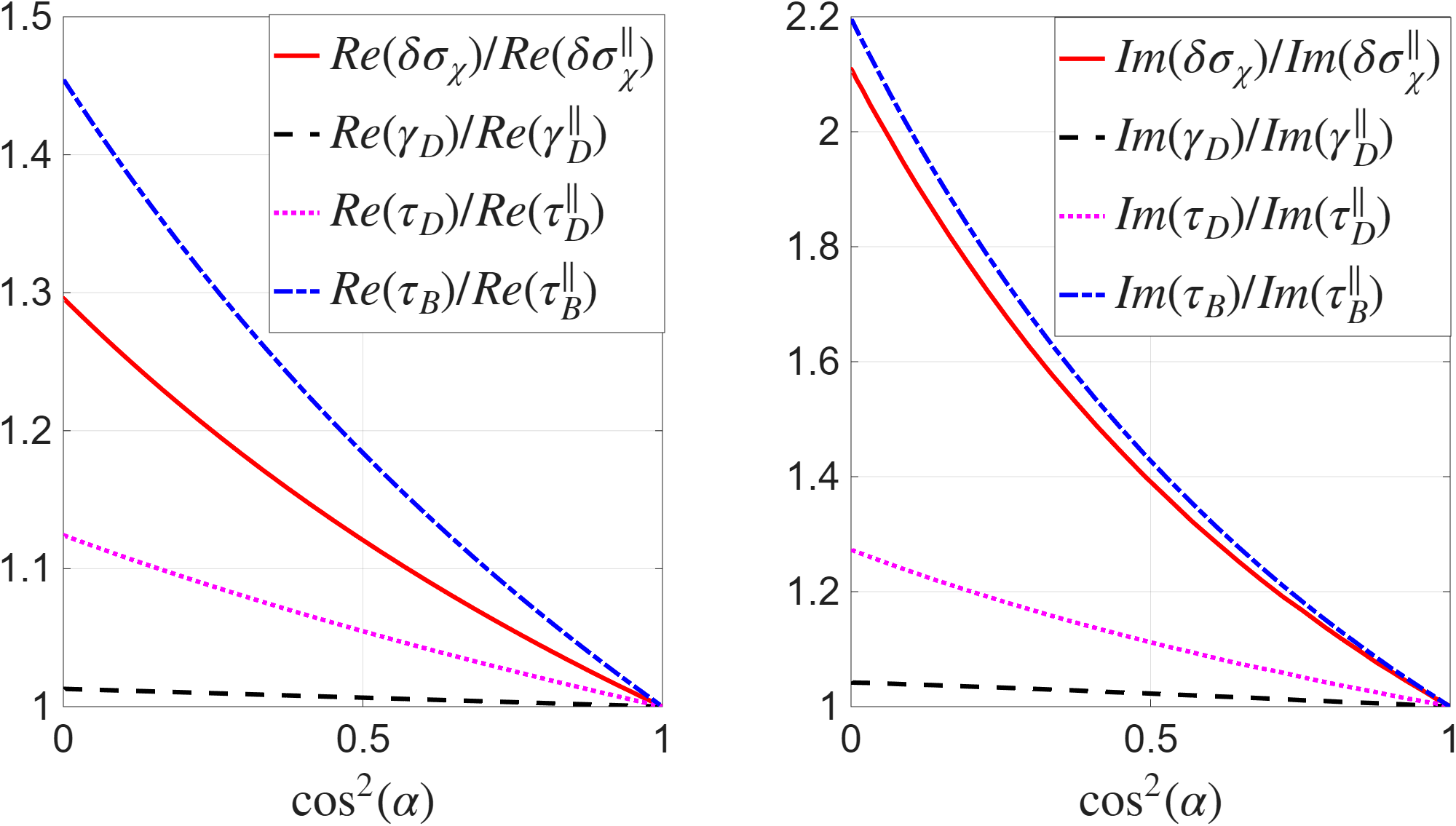}
    \caption{Normalized TCFs as functions of $\cos^2(\alpha)$. $\kappa B = 0.25$, $\omega =0.01$, and  $q^2 = 3$.}
    \label{fig:6p}
\end{figure}

\begin{figure}[H]
    \centering
    \includegraphics[width=0.7\linewidth]{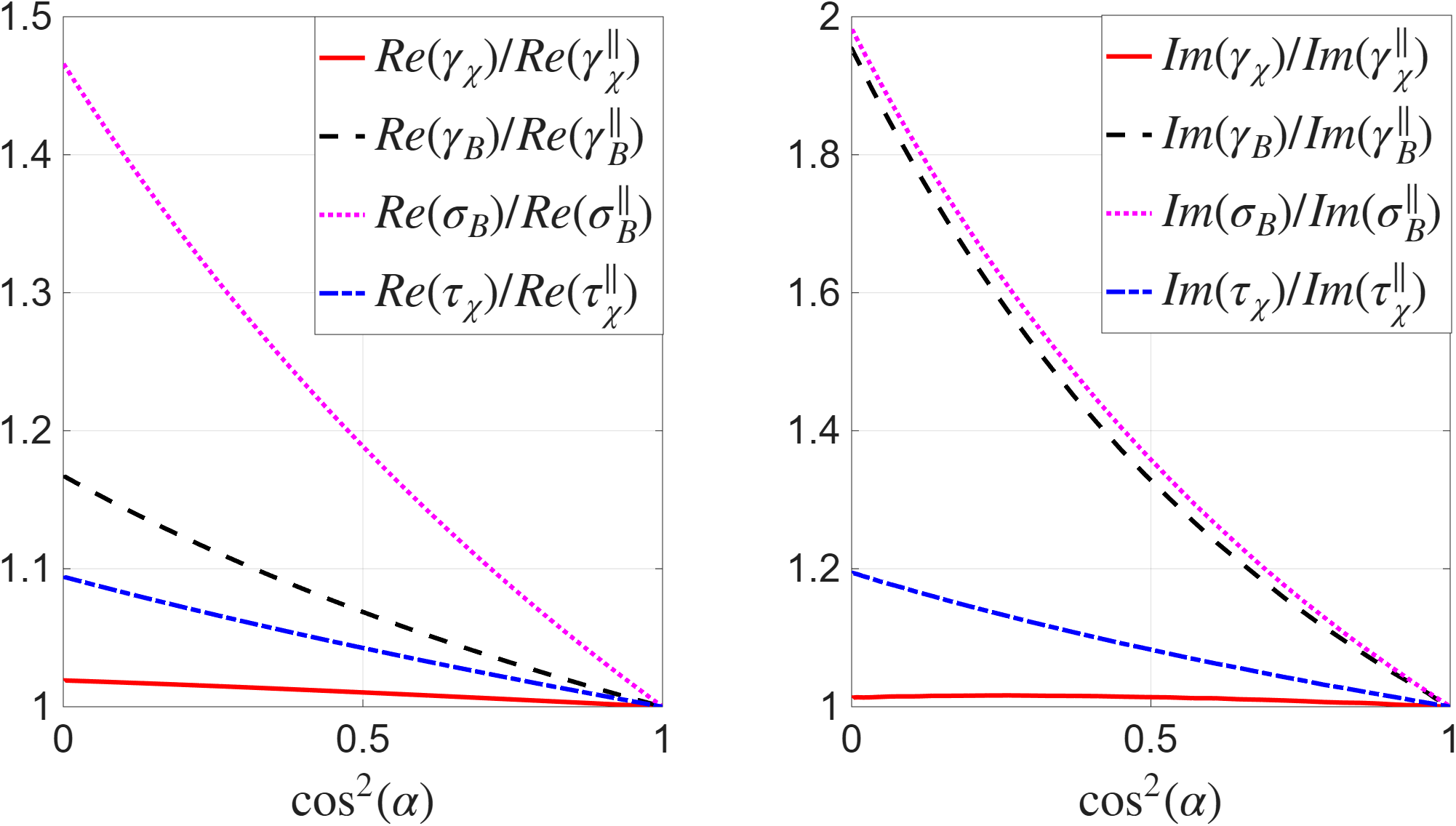}
    \caption{Normalized TCFs as functions of $\cos^2(\alpha)$. $\kappa B = 0.25$, $\omega =0.01$, and  $q^2 = 3$.}
    \label{fig:7p}
\end{figure}

\begin{figure}[H]
    \centering
    \includegraphics[width=0.7\linewidth]{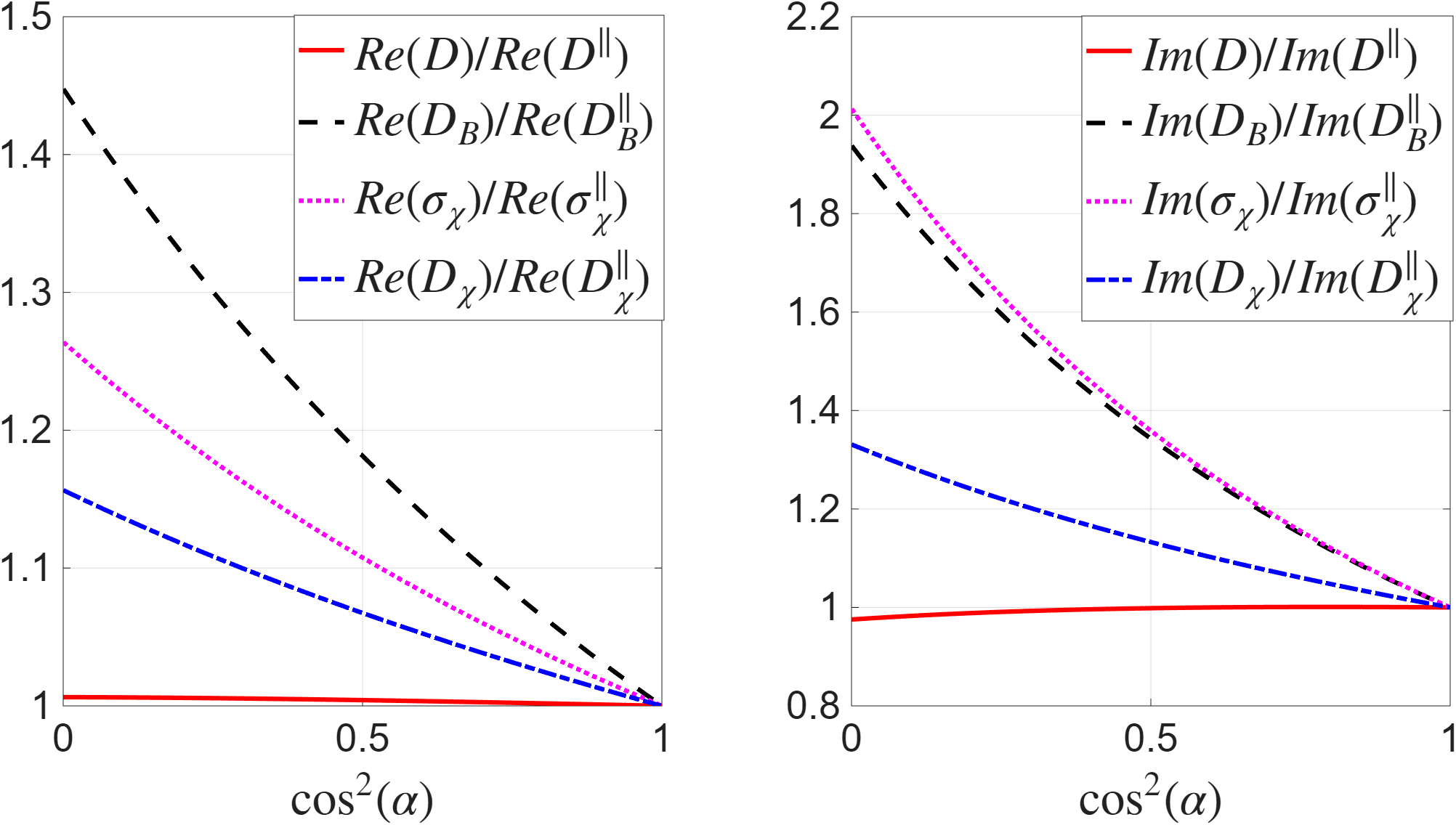}
    \caption{Normalized TCFs as functions of $\cos^2(\alpha)$. $\kappa B = 0.25$, $\omega =0.01$, and  $q^2 = 3$.}
    \label{fig:8p}
\end{figure}

Having explored the frequency and momentum dependence of the TCFs at the angle $\alpha=0$ (the angle between $\vec q$ and $\vec B$), we shift our focus to the $\alpha$  dependence of the TCFs. We limit our discussion to very small $\omega$ ($\omega=0.01$). However, to probe the dynamics beyond the hydrodynamic limit, the 3-momentum is fixed at an arbitrarily chosen value of $q^2=3$. The magnetic field is kept in the strong-field regime ($\kappa B=0.25$). Figs.~\ref{fig:6p}, \ref{fig:7p}, and \ref{fig:8p} reveal the effect of varying the angle $\alpha$ on the various TCFs. The results are normalized with respect to their values in the parallel  configuration, $\alpha=0$, which has been discussed above. 
The electric conductivity $\sigma_e$ was found to be independent of the magnetic field and hence does not depend on $\alpha$. Analyzing our numerical results, we  notice that some TCFs, and particularly the diffusion TCF $D$, exhibit only a very weak dependence on the orientation of $\vec q$ with respect to $\vec B$. In contrast, some others, such as the chiral magnetic conductivity TCF $\sigma_\chi$, have a very pronounced 
$\alpha$-dependence. A very interesting and quite unexpected observation is that all the TCFs (or rather, their absolute values) are minimized at $\alpha=0$ and maximized at $\alpha=\pi/2$. 
Given that several TCFs are highly sensitive to angular variations, mapping the entire angular parameter space is very bulky for presentation.
We therefore prefer to focus on the simplest scenario: the "all parallel" configuration \( \vec{q} \parallel \vec{B} \parallel \vec{E} \), as will be explored in the following subsection.

\subsection*{All-parallel case: \texorpdfstring{$\vec{q}\parallel\vec{B}\parallel\vec{E}$}{q || B || E} \label{section 3}}
The general constitutive relations \eqref{11} and \eqref{12}
 are quite complicated and involve a total
of thirteen distinct TCFs \eqref{13tcf}.  A significant simplification occurs when, in addition to $\alpha=0$, we impose  $\vec{q}\parallel\vec{E}$.
This case is of particular interest since it includes (\ref{gauss}).
Within the linearization scheme  \eqref{linearschem}, 
the constitutive relations for the "all parallel" case take the form:
\begin{align}
  \vec{J} &=
    - \tilde{D}\, \vec{\nabla}\rho
    + \tilde{\sigma}_\chi \kappa\, \vec{B}\, \rho_5
    + \tilde{\sigma}_e\, \vec{E} \, ,
\label{A.11}
\end{align}
\begin{align}
  \vec{J}_5 &=
    - \tilde{D}\, \vec{\nabla}\rho_5
    + \tilde{\sigma}_\chi \kappa\, \vec{B}\, \rho
    + \tilde{\sigma}_B \kappa\, \vec{\nabla}(\vec{E}\cdot\vec{B}) \, .
\label{A.12}
\end{align}
The TCFs appearing in (\ref{A.11}) and (\ref{A.12}) differ from the ones introduced in \eqref{11} and \eqref{12} (hence  we  denote them by a tilde). However, they are not independent. The relations between the two sets are given by:
\begin{align} \label{TCFT}
\tilde{D} = D - D_B \kappa^2 B^2, \quad 
\tilde{\sigma}_\chi = \sigma_\chi -&D_\chi q^2 , \quad  \\ \nonumber  \\ \nonumber
\tilde{\sigma}_e = \sigma_e + \delta\sigma_\chi \kappa^2 B^2  - \gamma_D  q^2  - \tau_D \kappa^2 B^2 q^2 -\tau_B \kappa^2 B^2 q^2,  \qquad
\tilde{\sigma}_B &= \gamma_B + \gamma_\chi -\tau_\chi q^2  +\sigma_B \kappa^2 B^2 .\nonumber
\end{align}
The resulting analytical expressions   for the TCFs at $\omega=q^2=0$ are derived from Eqs.~\eqref{eq:anal_13tcf}, \eqref{eq:anal_13tcf_new}, and \eqref{TCFT}:
\begin{align} \notag
&\tilde{D}^0 = \frac{1}{2} +9 \kappa^2 B^2 \big(2+\pi -6\log(2)\big) +\mathcal{O}((\kappa B)^4), \\ \label{eq71}
&\tilde{\sigma}^0_\chi = 6 + 216 \kappa^2 B^2  \big(1- \log(4)\big)+\mathcal{O}((\kappa B)^4) , \\ \notag
&\tilde{\sigma}^0_e = 1+ 18\kappa^2 B^2  \big(\pi- \log(4)\big)+\mathcal{O}((\kappa B)^4), \\ \notag
&\tilde{\sigma}_B^0(B=0) = -3\pi+\mathcal{O}((\kappa B)^2).
\end{align}
Requiring that the $B^2$ corrections are within 10\% of the leading contribution yields $\kappa B \lesssim 0.06$ (this constraint originates from $\tilde\sigma_e^0$).
This is our revised estimate for the weak-magnetic field limit.

\begin{figure}[htbp] 
    \centering
    \includegraphics[width=0.4\textwidth]{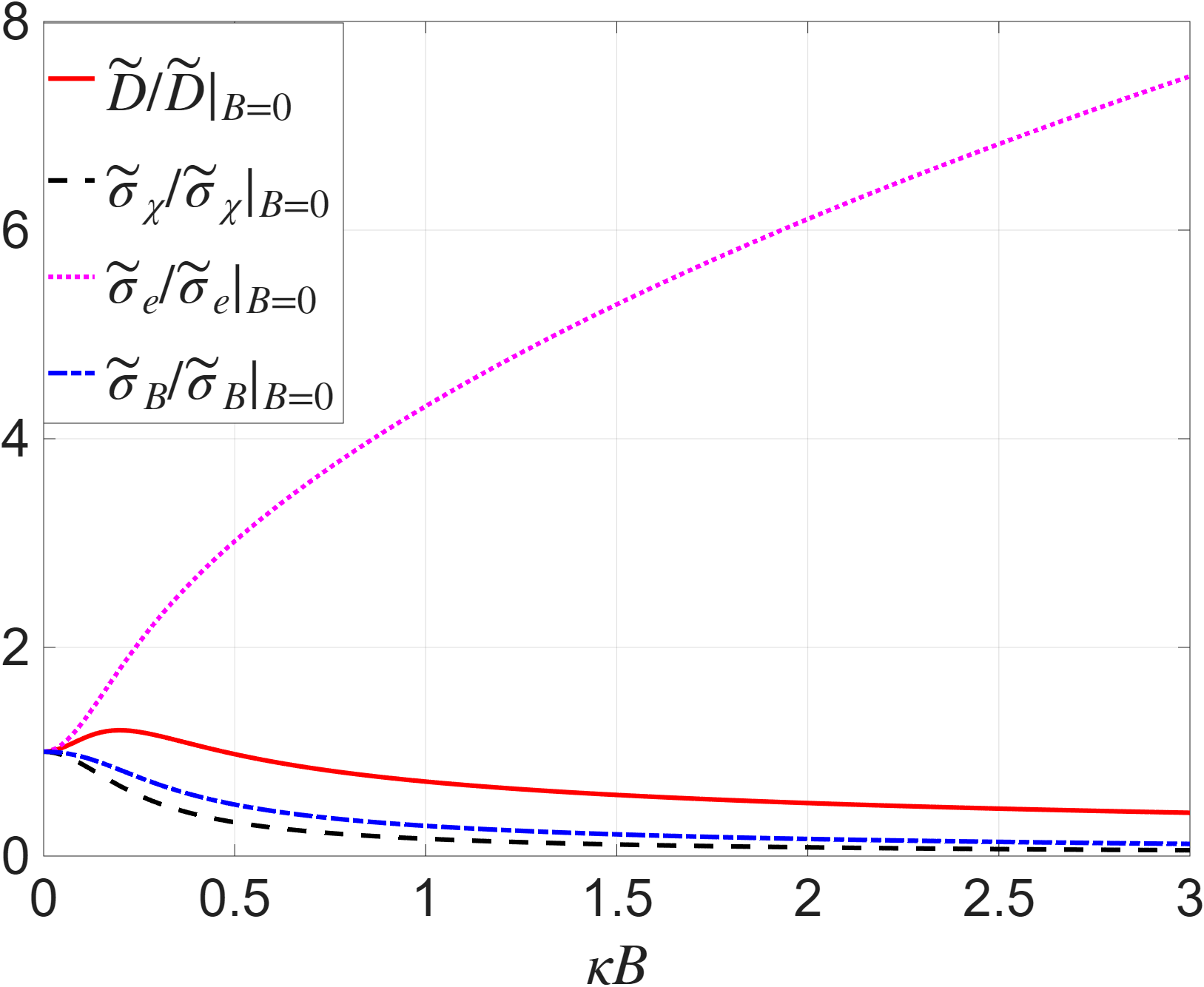}
    \caption{Normalized TCFs (all-parallel case) as functions of $\kappa B$ at $\omega = q^2 = 0$.}
    \label{b_dep_align}
        \label{fig1}
\end{figure}

To explore the regime of strong magnetic field, we examine the dependence of the TCFs on $B$ at $\omega = q^2 = 0$ numerically 
(Fig.~\ref{fig1}). 
Fig.~\ref{fig1} shows that all the TCFs are suppressed with  $\kappa B$ except $\tilde \sigma_e$, which  increases in the strong-field regime. 
This strong rise of 
$\tilde\sigma_e$ with the magnetic field is partially responsible for the effect of  negative MR.

As mentioned in the Introduction, we also study the TCFs as functions of complex $\omega$. Our results are presented in Appendix \ref{complex_tcfs}.

\section{  MR and CMW}\label{CMWMR}

\subsection{Negative MR}

In the presence of an external magnetic field $\vec{B}$, the linear-response electrical conductivity tensor decomposes into transverse ($\sigma_T$) and longitudinal ($\sigma_L$) components relative to $\vec{B}$,
\begin{align}\label{41}
J_i = \hat \sigma_{ij} E_j = \sigma_T \left( \delta_{ij} - \frac{B_i B_j}{B^2} \right) E_j + \sigma_L \frac{B_i B_j}{B^2} E_j.
\end{align}
The decomposition (\ref{41}) holds for  media exposed to an electric field whose gradients are aligned with the field itself ($\vec q \parallel \vec E$).\footnote{In principle, one could consider the most general situation of $\vec q \nparallel \vec E  \nparallel \vec B$, but we leave it beyond the scope of this paper.}  




Substituting the constitutive relations \eqref{11} and \eqref{12} into the continuity equations \eqref{4} and \eqref{5} yields:
\begin{align}
\label{sigma_l} \notag
&\sigma_L(\omega,q,B,\alpha) = \frac{-i\omega\left(\sigma_e+\delta \sigma_\chi \kappa^2 B^2 cos^2\alpha -\gamma_D q^2 -i \tau_D \kappa^2 B^2 q^2 \cos^2\alpha 
-i \tau_B \kappa^2 B^2 q^2 \cos^2\alpha \right) }{\left(-i\omega + D q^2 - D_B \kappa^2 B^2 q^2 \cos^2\alpha\right)^2+
(\sigma_\chi - D_\chi q^2)^2 \kappa^2 B^2 q^2 \cos^2\alpha }
\\\notag \\
&\times \left(-i\omega + D q^2 - D_B \kappa^2 B^2  q^2 \cos^2\alpha \right)
\\ \notag\\\notag
&+ \frac{-i\omega \kappa^2 B^2\cos^2\alpha\left(\sigma_\chi - D_\chi q^2\right)\left(12 + \gamma_\chi q^2 + \gamma_B q^2+ \sigma_B \kappa^2 B^2 q^2 \cos^2\alpha - \tau_\chi q^4\right)}
{\left(-i\omega + D q^2 - D_B \kappa^2 B^2 q^2 \cos^2\alpha\right)^2+ (\sigma_\chi - D_\chi q^2)^2 \kappa^2 B^2 q^2 \cos^2\alpha}
\end{align}
\begin{align}
 \notag
&\sigma_T(\omega,q,B,\alpha) = \frac{-i\omega}{-i\omega + D q^2} \, \left( \sigma_e -\gamma_Dq^2\right) .
\end{align}
The linear response conductivities $\sigma_{L,T}$ 
depend on the angle $\alpha$ (the angle between $\vec q\parallel\vec E$ and $\vec B$), both explicitly and through the $\alpha$-dependence of the TCFs. In principle, such dependence could be explored experimentally.

The simplest and commonly discussed case is that of spatially uniform fields ($q \to 0$), for which simpler expressions are obtained,
\begin{align}
\sigma_L(\omega, 0, B, \alpha) = \sigma_e + \delta\sigma_\chi \kappa^2 B^2cos^2\alpha + \frac{12 i \sigma_\chi \kappa^2 B^2 \cos^2\alpha}{\omega}, \qquad \sigma_T(\omega, 0, B, \alpha) =\sigma_e.
\label{q=0}
\end{align}
Notice that Eq. (\ref{q=0}) (taken at $\alpha=0$) has an extra term compared to (\ref{nm}).
As mentioned in the Introduction, the $1/\omega$ pole is problematic in the static limit of $\omega\rightarrow 0$. 
It is clear from  \eqref{sigma_l} that the physical regularization emerges due to finite $q$ rather than from a phenomenological regulator like the relaxation time $\tau_5$. 

\begin{figure}[H]
   \centering
   \begin{subfigure}{0.48\textwidth}
      \centering
      \includegraphics[width=\textwidth]{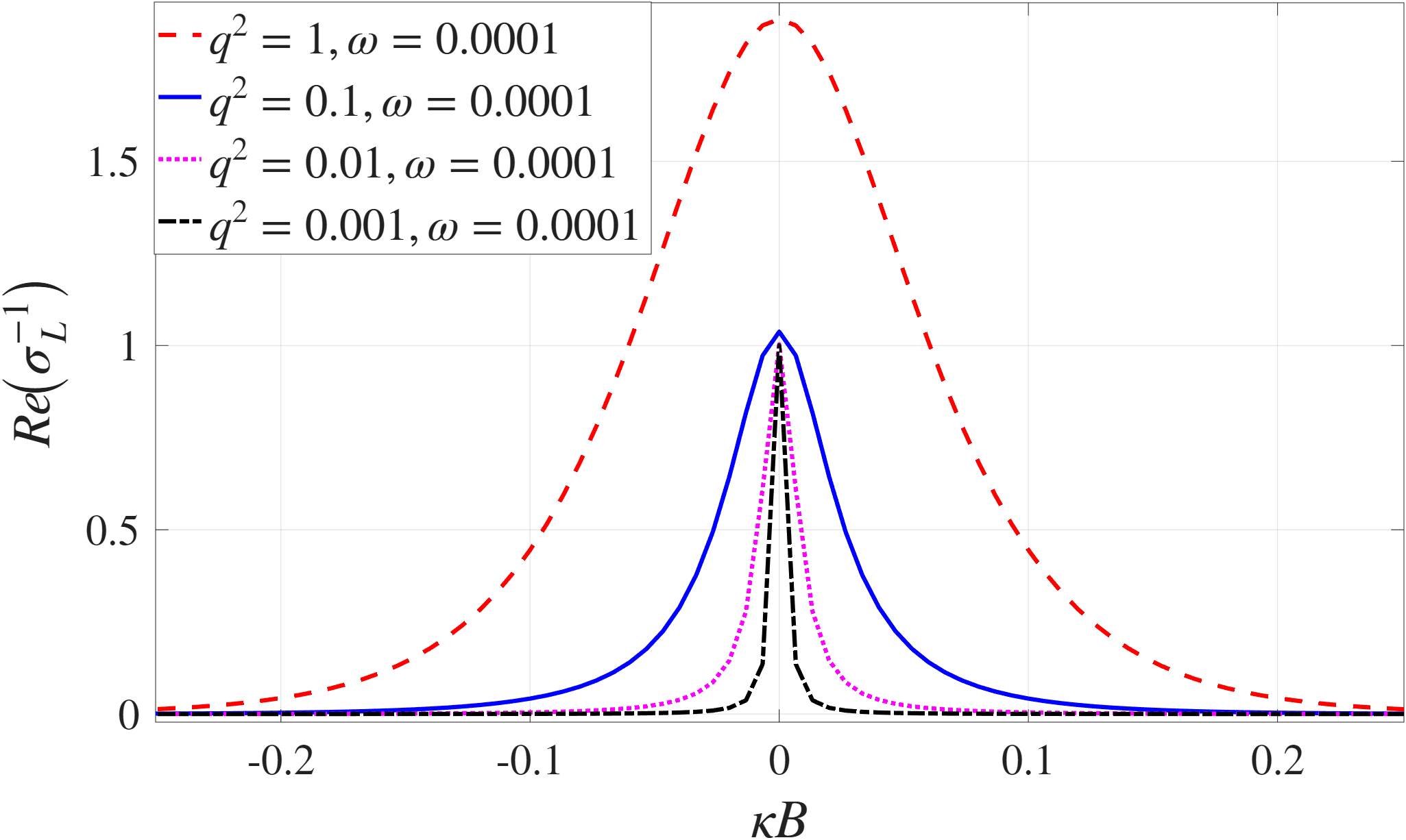}
      \caption{$\omega \ll q^2$}
      \label{fig:NMR1}
   \end{subfigure}
   \hfill
   \begin{subfigure}{0.48\textwidth}
      \centering
      \includegraphics[width=\textwidth]{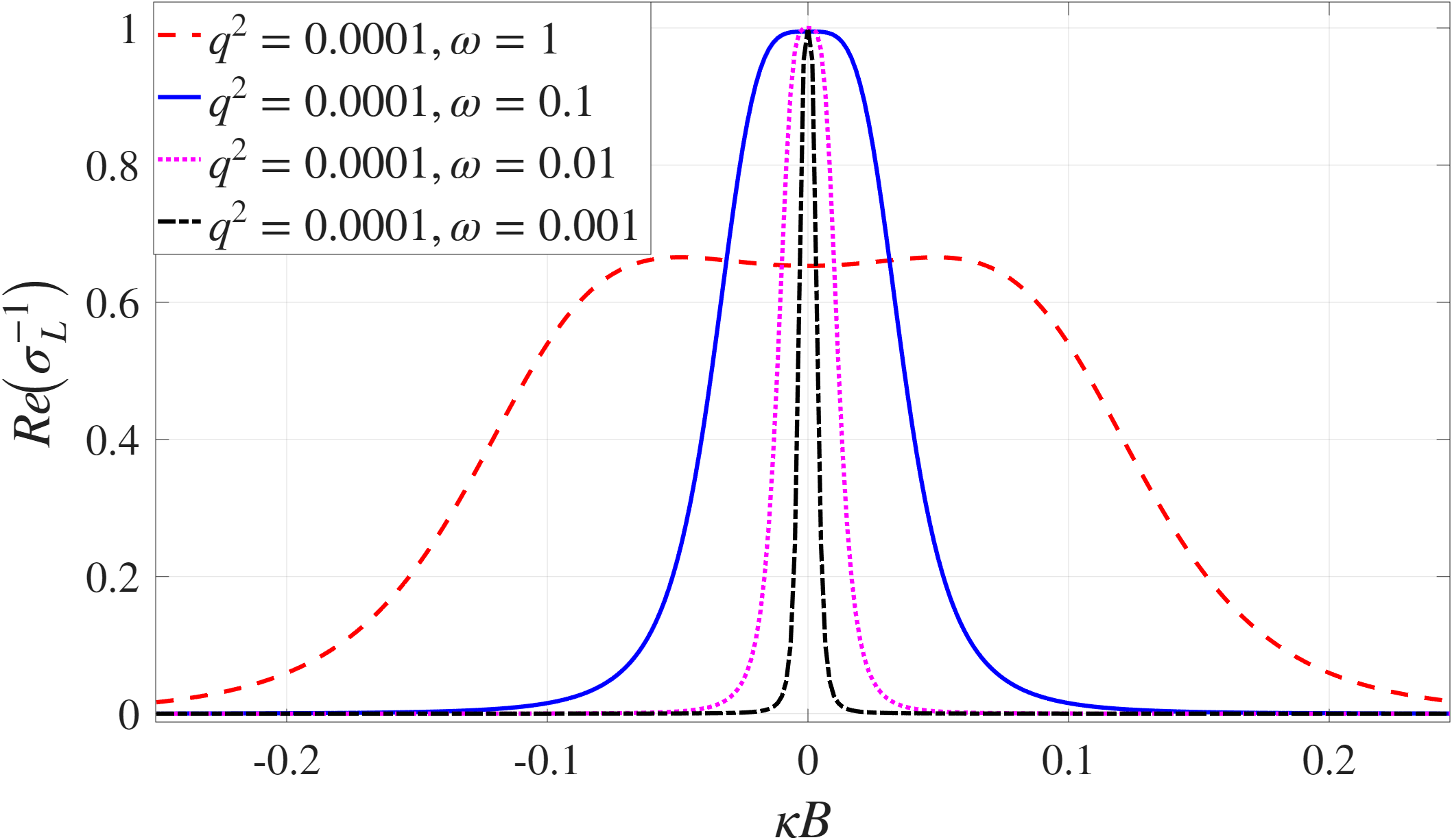}
      \caption{$q^2 \ll \omega$}
      \label{fig:NMR2}
   \end{subfigure}
   \caption{Resistivity $\Re(\sigma_L^{-1})$ as a function of $\kappa B$ for fixed $\omega$, $q^2$, and $\alpha=0$.}
   \label{fig:NMR_main}
\end{figure}

We are thus prompted to restore the  dependence of $\sigma_L$ on $q$. To simplify the analysis, we will only consider the all-parallel configuration of the fields  ($\alpha=0$).
Then $\sigma_L$ reads
\begin{align}
\label{sigma_lp} 
\sigma_L(\omega,q,B) = \frac{-i\omega\tilde\sigma_e \left(-i\omega + \tilde D q^2 \right) }{\left(-i\omega + \tilde D q^2 \right)^2+
\tilde\sigma_\chi^2 \kappa^2 B^2 q^2 }
+ \frac{-i\omega \kappa^2 B^2\tilde\sigma_\chi \left(12 + \tilde \sigma_B  q^2\right)}
{\left(-i\omega + \tilde D q^2 \right)^2+
\tilde\sigma_\chi^2 \kappa^2 B^2 q^2 }
\end{align}
In (\ref{sigma_lp}), the three limits $\omega\rightarrow 0$, $q\rightarrow 0$, and $B\rightarrow 0$  do not commute. If  the limit $\omega\rightarrow 0$ is taken first, 
the  longitudinal conductivity vanishes
(Debye screening). 
We proceed by computing the resistivity $\Re(\sigma_L^{-1})$, in the $\omega \to 0$ limit, assuming that all the TCFs are real\footnote{The imaginary part will vanish in the limit $\omega\rightarrow 0$.}:

\begin{equation}
    \Re(\sigma_L^{-1})(\omega\rightarrow 0)={\frac{1}{\tilde\sigma_e}}-
    \frac{\kappa^2 B^2 \tilde\sigma_\chi^2}{\tilde\sigma_e}\frac{
    (\kappa^2 B^2 (12 + q^2 \tilde\sigma_B)^2 + q^2 \tilde\sigma_e^2) }{ (\tilde D  \tilde\sigma_e q^2 + \kappa^2 B^2 (12 + q^2\tilde\sigma_B) \tilde\sigma_\chi)^2}.
\end{equation}
Compared with (\ref{nm1}), we
can identify the relaxation time $\tau_5$ with the following
$q$-dependent expression (also taking the small-$B$ limit as is implied in (\ref{nm1})):
\begin{equation}
\tau_5=\frac{\tilde\sigma_e^{1}}{12\tilde\sigma_\chi^0}+
    \frac{ \tilde\sigma_e^0\tilde\sigma_\chi^0}{ 12 (\tilde D^0)^2  q^2 }\label{tau5}
\end{equation}
Here, $\tilde{\sigma}_e^{1}$ is 
the coefficient of the $(\kappa B)^2$ term in the small-$B$ expansion of $\tilde\sigma_e(\omega=0,q,B)$.
This result for $\tau_5$ is quite generic. The holographic model provides explicit values for the relevant TCFs. In the limit $q\rightarrow 0$, the  analytical expressions are given
in (\ref{eq71}). In this limit, the first term in (\ref{tau5}) is subleading and our final result becomes very simple:
\begin{equation}
\tau_5(q\rightarrow 0)=
    \frac{ 2}{ q^2 }
\end{equation}
At small $q$,  one expects a negative contribution 
to the resistivity due to the anomaly-induced transport. 
Indeed, we verify this numerically without assuming any approximations. 
Fig. \ref{fig:NMR1} shows the resistivity  calculated from \eqref{sigma_lp} as a function of the magnetic field at small values of $\omega\ll q^2$. 
Fig. \ref{fig:NMR2} displays the resistivity in the opposite limit of $\omega\gg q^2$. In both scenarios,  we observe a clear signature of the effect of negative MR. 


\subsection{CMW}

In this section, we explore the CMW, extending previous studies \cite{P5,P6} by systematically incorporating the effects of the dynamical electric field. 
Our primary goal is  to determine whether the CMW can  still be rendered dissipationless despite
the inclusion of the dynamical electric field  given by \eqref{gauss}.


The dispersion relation is derived by substituting the constitutive relations \eqref{11}--\eqref{12} into the continuity equations \eqref{4}--\eqref{5}:
\begin{align} \label{83}
0 =& -i\omega\rho +Dq^2\rho - D_B \kappa ^2(\vec{B}\cdot\vec{q})^2\rho+\sigma_\chi i \kappa (\vec{B}\cdot \vec{q})\rho_5 -D_\chi i \kappa q^2(\vec{B}\cdot \vec{q})\rho_5 +\sigma_e \rho \quad \\[1ex] \notag &+\frac{\delta\sigma_\chi  \kappa ^2 (\vec{B}\cdot \vec{q})^2\rho}{q^2} - \gamma_D q^2 \rho-\tau_D\kappa^2 (\vec{B}\cdot \vec{q})^2\rho -\tau_B\kappa^2 (\vec{B}\cdot \vec{q})^2\rho 
\end{align}
\begin{align} \label{84}
    0 =& -i\omega\rho_5 +Dq^2\rho_5 - D_B \kappa ^2(\vec{B}\cdot\vec{q})^2\rho_5+\sigma_\chi i \kappa (\vec{B}\cdot \vec{q})\rho -D_\chi i \kappa q^2(\vec{B}\cdot \vec{q})\rho \quad \\[1ex] \notag & + i  \gamma_\chi  \kappa (\vec{B} \cdot \vec{q})\rho  - i \tau_\chi \kappa q^2(\vec{B}\cdot \vec{q}) \rho+ \frac { i \sigma_B \kappa^3(\vec{B}\cdot\vec{q})^3\rho}{q^2}
    +\frac{12 i  \kappa (\vec{q}\cdot \vec{B})\rho}{q^2} \\[1ex] \notag &+ i  \gamma_B  \kappa (\vec{B} \cdot \vec{q})\rho.
\end{align}
The resulting dispersion relation consists of two branches:
\begin{equation}
\omega_\pm
= \Omega(\omega_\pm,q^2,\kappa \vec B)
\pm
\sqrt{\zeta(\omega_\pm,q^2,\kappa \vec B)} ,
\label{eq:dispersion_compact}
\end{equation}
where
\begin{align}
\Omega
&= -\frac{1}{2} i\sigma_e - iDq^2
+ iD_B\kappa^2(\vec B\!\cdot\!\vec q)^2
- \frac{ i\,\delta\sigma_\chi\kappa^2(\vec B\!\cdot\!\vec q)^2}{2q^2}
\\\nonumber
&\quad
+\frac{1}{2} i\gamma_D q^2
+\frac{1}{2} i\tau_D\kappa^2(\vec B\!\cdot\!\vec q)^2+\frac{1}{2} i\tau_B\kappa^2(\vec B\!\cdot\!\vec q)^2
\end{align}
and 
\begin{align}
\zeta=&
-\Big(\frac{1}{2}\sigma_e
+ \frac{ \delta\sigma_\chi\kappa^2(\vec B\!\cdot\!\vec q)^2}{2q^2}
-\frac{1}{2}\gamma_D q^2
-\frac{1}{2}\tau_D\kappa^2(\vec B\!\cdot\!\vec q)^2  -\frac{1}{2} \tau_B\kappa^2(\vec B\!\cdot\!\vec q)^2 \Big)^2
\\\nonumber
&+ \kappa^2(\vec B\!\cdot\!\vec q)^2 (\sigma_\chi-D_\chi q^2)^2
+\kappa^2(\vec B\!\cdot\!\vec q)^2
\left(\frac{\sigma_\chi}{q^2}-D_\chi\right)\\\nonumber
& \times
\Big(\gamma_\chi q^2+\gamma_B q^2-\tau_\chi q^4
+\sigma_B\kappa^2(\vec B\!\cdot\!\vec q)^2+12\Big).
\label{eq:def_Z}
\end{align}
The obtained dispersion relation makes it possible to
study the CMW as a function of $q$ and the magnetic field $B$. Moreover, this general expression also depends on the angle $\alpha$ and, in principle, is valid for the case of $\vec{q} \nparallel \vec{B}$.

For our present analysis, however, we focus on the strictly parallel configuration, $\vec{q} \parallel \vec{B}$, where the simplified version of the constitutive relations
(\ref{A.11}) and (\ref{A.12}) can be used.
They lead to the following  dispersion relation:
\begin{equation}
\omega_\pm=\omega_\pm^R+i\omega_\pm^I
= -\frac{1}{2} i\tilde{\sigma}_e - i\tilde{D}q^2
\pm
\sqrt{-(\frac{1}{2}\tilde{\sigma}_e)^2  + \kappa^2 B^2  q^2 \tilde{\sigma}_\chi^2
+\kappa^2 B^2\tilde{\sigma}_\chi
(\tilde{\sigma}_B \cdot q^2+12)} .
\label{eq:92}
\end{equation}
The dispersion relation (\ref{eq:92}) looks very similar to the one in (\ref{eq:17}). The clearly visible difference is that in the latter, $\tilde\sigma_{B}=0$.
Yet, the major difference is that while (\ref{eq:17})
is meant to be correct in the hydrodynamic limit only,
 (\ref{eq:92}) is valid for arbitrary values of $\omega$ and $q$. Furthermore, through the $B$-dependence of the TCFs, (\ref{eq:92}) is non-linear in the magnetic field and also valid  
for strong magnetic fields.

The complex 
dispersion relation (\ref{eq:92}) can be
written as two equations for the real and imaginary parts to be satisfied simultaneously:
\begin{equation}
\Phi_\pm^R(\omega_\pm,q^2,\kappa  B)=
\Phi_\pm^I(\omega_\pm,q^2,\kappa  B)=0,
\label{dispersion}
\end{equation}
where
\begin{align}
\Phi^I_\pm(\omega,q^2,\kappa  B)
&\equiv -\Im[\omega]+
\Re\left[\frac{1}{2}\tilde{\sigma}_e  - \tilde{D}q^2\right]
\pm
\left(\Re[\zeta]^2+\Im[\zeta]^2\right)^{1/4}
\sin\!\frac{\theta}{2},
\label{eq:PhiI}
\\[1ex]
\Phi^R_\pm(\omega,q^2,\kappa B)
&\equiv
-\Re[\omega]
- \Im\!\left[\frac{1}{2}\tilde{\sigma}_e  + \tilde{D}q^2 \right]
\pm
\left(\Re[\zeta]^2+\Im[\zeta]^2\right)^{1/4}
\cos\!\frac{\theta}{2}.
\label{eq:PhiR}
\\[1ex]
\zeta(\omega,q^2,\kappa  B)&=-(\frac{1}{2}\tilde{\sigma}_e)^2  + \kappa^2 B^2  q^2 \tilde{\sigma}_\chi^2
+\kappa^2 B^2\tilde{\sigma}_\chi
(\tilde{\sigma}_B \cdot q^2+12)
\label{zeta}
\end{align}
\begin{equation}
\theta \equiv
\arctan\!\left(\frac{\Im[\zeta]}{\Re[\zeta]}\right).
\label{eq:theta_def}
\end{equation}
As indicated by $\pm$, the dispersion relation has two distinct branches. To explore both branches simultaneously, it is convenient to rewrite the dispersion relation somewhat differently. Let us introduce
\begin{align}
\phi(\omega,q^2,\kappa  B) =\phi_R+i\phi_I \equiv\left[-\omega - i\left(\frac{1}{2} \tilde{\sigma}_e + \tilde{D} q^2\right)\right]^2 -\zeta
\end{align}
and the dispersion relation is obtained from the
simultaneous vanishing of the real and imaginary parts of $\phi$,
\begin{align}
\phi_R(\omega_\pm,q^2,\kappa  B)=
\phi_I(\omega_\pm,q^2,\kappa  B)=0.
\label{phii_phir_0}
\end{align}
Despite the appearance of the "$\pm$" sign, the number of solutions is not known in advance and it can range from zero to infinity. This is due to the highly complicated dependence of $\phi$ on $\omega$. 

We have searched for a solution with $\omega_\pm^I=0$, but none was found in the range of explored parameters. Consequently, we turn to studying  the results in the complex plane.  To this end, the TCFs must be computed in the complex $\omega$ plane. 

\begin{figure}[htbp] 
    \centering
    \includegraphics[width=0.8\textwidth]{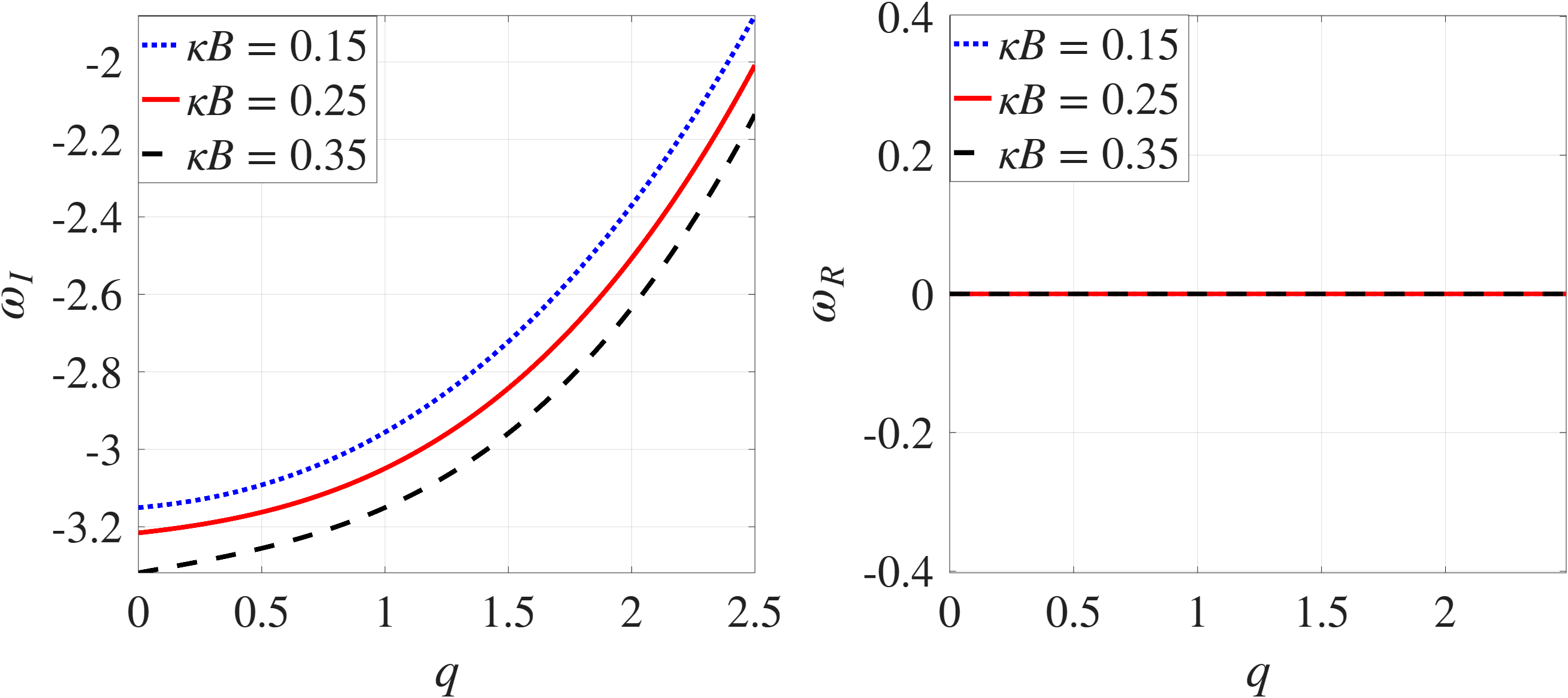}
    \caption{The overdamped mode of the "+" branch.}
    \label{fig:ov_dam}
\end{figure}

We identified two solutions, both corresponding to the "+" branch  in \eqref{eq:92}. The first is the  overdamped mode (see Fig.~\ref{fig:ov_dam}). 
This mode is characterized by $\Im(\omega_+) < 0$ and $\Re(\omega_+) = 0$.

\begin{figure}[H] 
    \centering
    \includegraphics[width=0.8\textwidth]{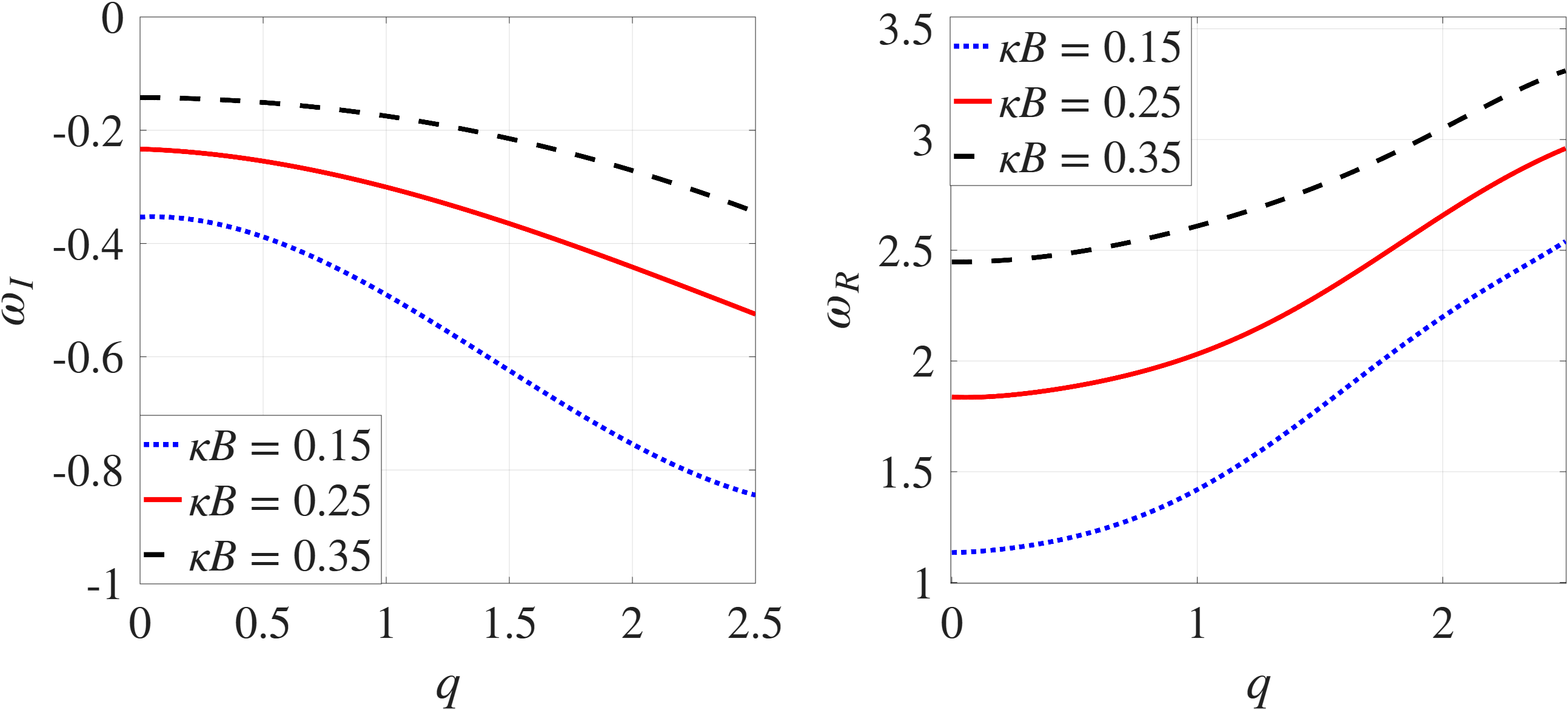}
    \caption{The underdamped mode of the "+" branch.}
    \label{fig:un_dam}
\end{figure}

The second solution represents an underdamped mode (Fig.~\ref{fig:un_dam}), where $\Im(\omega_+) < 0$ and $\Re(\omega_+) > 0$. 



\section{Summary}\label{Summary}




In this paper, we have introduced (Eqs. \eqref{11} and \eqref{12}) and analyzed thirteen TCFs with an emphasis on their anomaly-induced non-linear dependence on the external magnetic field. Within the
holographic model, all the TCFs are computable exactly, and we have reported on new results, including those for complex frequencies. 

As an application, we have considered two physical effects. The first one is a computation of linear-response conductivities and particularly a revision of the phenomenon of negative MR. We have demonstrated that there is no need to introduce any phenomenological axial charge relaxation time $\tau_5$ -- instead, its role is played by the non-uniformity of the external electric field.  We hope that our improved theoretical model can be found useful 
for the analysis of real  data on transport
measured in Dirac/Weyl semimetals.

As a second application, we have calculated a CMW dispersion relation,
taking into account effects that were missing in previous studies. 
Earlier results suggested the existence of non-dissipative modes above some threshold magnetic field. The new element we introduce here is a self-consistent  inclusion of the dynamical electric field. In line with expectations, we have not found any  dissipationless CMWs. 
Instead, we discovered two modes -- one overdamped and one propagating with dissipation. 

The medium described by the constitutive relations \eqref{11} and \eqref{12} could be studied further. 
A particularly  interesting  direction  would be 
to explore the propagation of e/m waves  in such a medium, in the spirit of \cite{Qiu:2016hzd}. The question of how 
the anomaly-induced transport  affects the refractive index is quite intriguing and should be answered in the future.

\section*{Acknowledgments}

The authors would like to thank  Tuna Demircik, Eran Maniv, and Muntaser Naamneh for useful discussions related to this work.   
The research was supported by  the  Binational Science Foundation grants \#2022132, \#2021789, \#2024818, and the Israel Science Foundation grant \#910/23.

\appendix

\section{Numerical setup}\label{Num}

In this Appendix, we detail the numerical methodology employed to solve the radial ODEs derived in \eqref{61}--\eqref{62}. As the resulting equations constitute a coupled system of linear, second-order ODEs with radially dependent coefficients, an analytical solution is generally intractable, necessitating a robust numerical approach. To facilitate a numerical solution, we utilize the compactified inverse radial coordinate $u = 1/r$. The system is then solved on the domain $u \in [0, 1]$ using a Chebyshev spectral collocation method \cite{trefethen2000spectral,driscoll2014chebfun,boyd2001chebyshev}.

Instead of dividing the space into a grid of discrete, local points (as in standard finite-difference schemes), the spectral method approximates the exact solution globally using a sum of continuous, smooth basis functions defined over the entire domain. Mathematically, each unknown radial function $V(u)$ is approximated by a truncated expansion of Chebyshev polynomials:
\begin{align}
V_N(u) &= \sum_{n=0}^{N} a_n T_n(\tilde{u}),
\end{align}
where $T_n$ are Chebyshev polynomials of the first kind. The variable $\tilde{u} = 2u - 1$ linearly maps the physical domain $u \in [0, 1]$ onto the standard Chebyshev interval $\tilde{u} \in[-1, 1]$. In doing so, the infinite-dimensional function space is projected onto a finite-dimensional polynomial subspace of dimension $N + 1$.
Chebyshev polynomials of degree $N$ exactly reproduce any polynomial of degree $\leq N$, and approximate perfectly smooth (analytic) functions with exponential accuracy \cite{trefethen2000spectral}.

In the limit of zero spatial momentum ($q = 0$) and vanishing magnetic field ($B = 0$), the conductivity $\tilde{\sigma}_e$ coincides with the current-current two-point correlator, which was computed analytically in \cite{Horowitz:2008bn}:
\begin{align}\label{analsigma}
\tilde{\sigma}_e(\omega, q=0)
&= i \omega \Bigg[
\frac{1}{2}\psi\!\left(\frac{1-i}{4}\omega + \frac{1}{2}\right)
+ \frac{1}{2}\psi\!\left(\frac{1+i}{4}(-\omega) + \frac{1}{2}\right)
 + \log 2 + \gamma_E - 1
\Bigg].
\end{align}
Here, $\psi(x) = \Gamma'(x)/\Gamma(x)$ is the digamma function and $\gamma_E$ is the Euler–Mascheroni constant. 
Note that 
in this specific limit of $B=0$ and $q=0$, the modified conductivity $\tilde{\sigma}_e$ reduces to $\sigma_e$. The accuracy of the numerical procedure is validated through a comparison of its results with the exact analytical solution (\ref{analsigma}). 
Our approach enables the extension of this comparison directly into the complex $\omega$-plane.

To robustly quantify the deviation between the numerical and analytical results across this domain, we compute the bounded relative error:
\begin{equation}
\varepsilon = \frac{|\tilde{\sigma}_e^{\text{num}} - \tilde{\sigma}_e^{(\ref{analsigma})}|}{|\tilde{\sigma}_e^{\text{num}}| + |\tilde{\sigma}_e^{(\ref{analsigma})}|}.
\end{equation}

\begin{figure}[H] 
    \centering
    \includegraphics[width=0.45\textwidth]{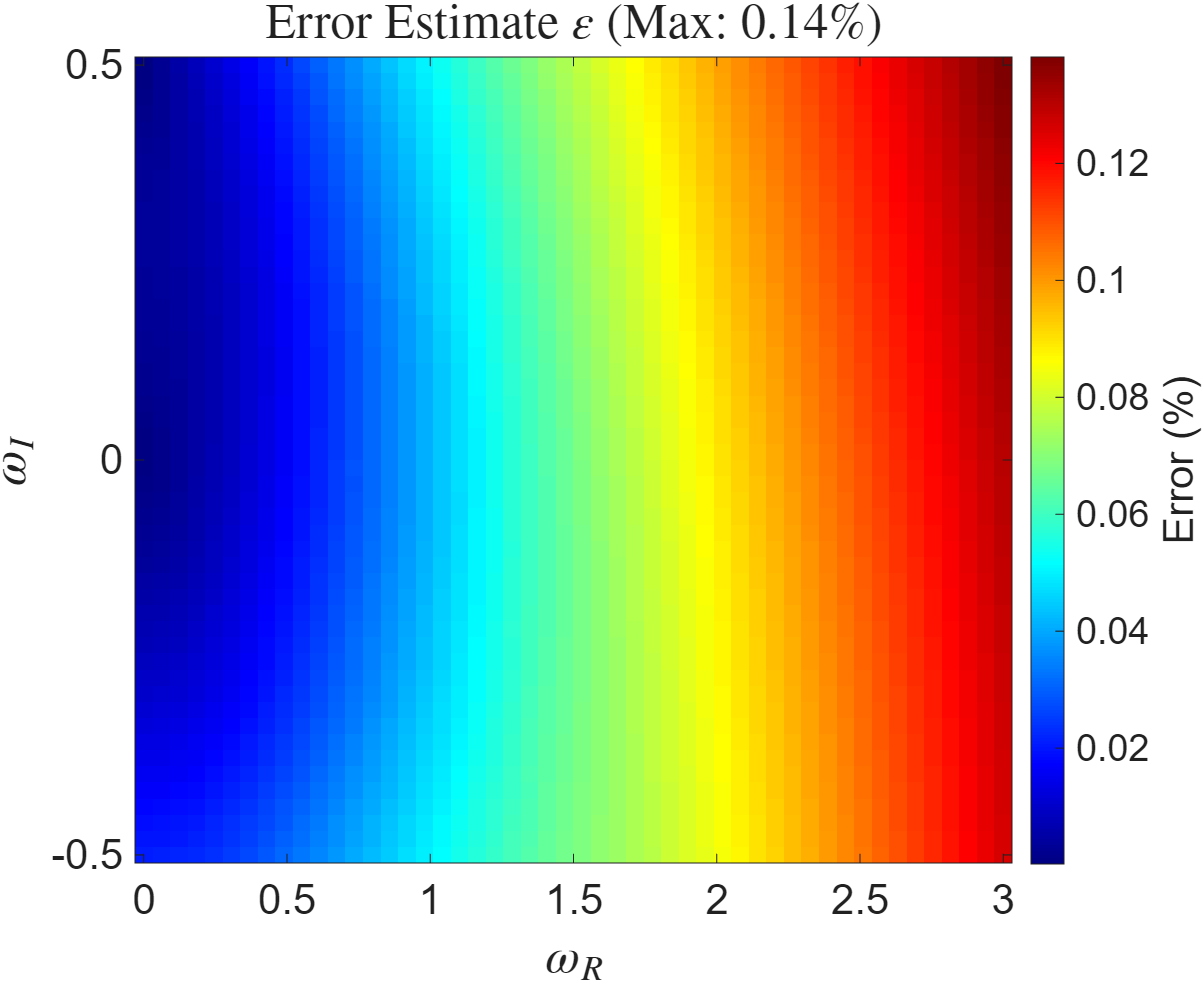}
    \caption{Relative error estimate $\varepsilon$ (in percent) over the complex $\omega$-plane. 
    }
    \label{error_est}
\end{figure}  
As illustrated in Fig. ~\ref{error_est}, the numerical procedure is highly accurate, with the bounded relative error $\varepsilon$ remaining below $0.14\%$ across the entire investigated domain of the complex $\omega$ plane. 
The error distribution reveals that while the deviation exhibits a marginal increase with the real part of the frequency ($\omega_R$), it remains notably stable along the imaginary axis ($\omega_I$). This difference can be traced to the fact that higher values of $\omega_R$ introduce rapid oscillations in the solution near the horizon ($u=1$),
while the dependence on $\omega_I$ is smooth. 

Having established the reliability of the method against the exact analytical solution, we extend our analysis to configurations involving finite momentum and magnetic fields, where analytical closed-form solutions are unavailable.

\section{\texorpdfstring{TCFs of complex $\omega$}{TCFs of complex omega}}
\label{complex_tcfs}
\begin{figure}[H]
    \centering
    \begin{subfigure}[b]{0.32\textwidth}
        \centering
        \includegraphics[width=\textwidth]{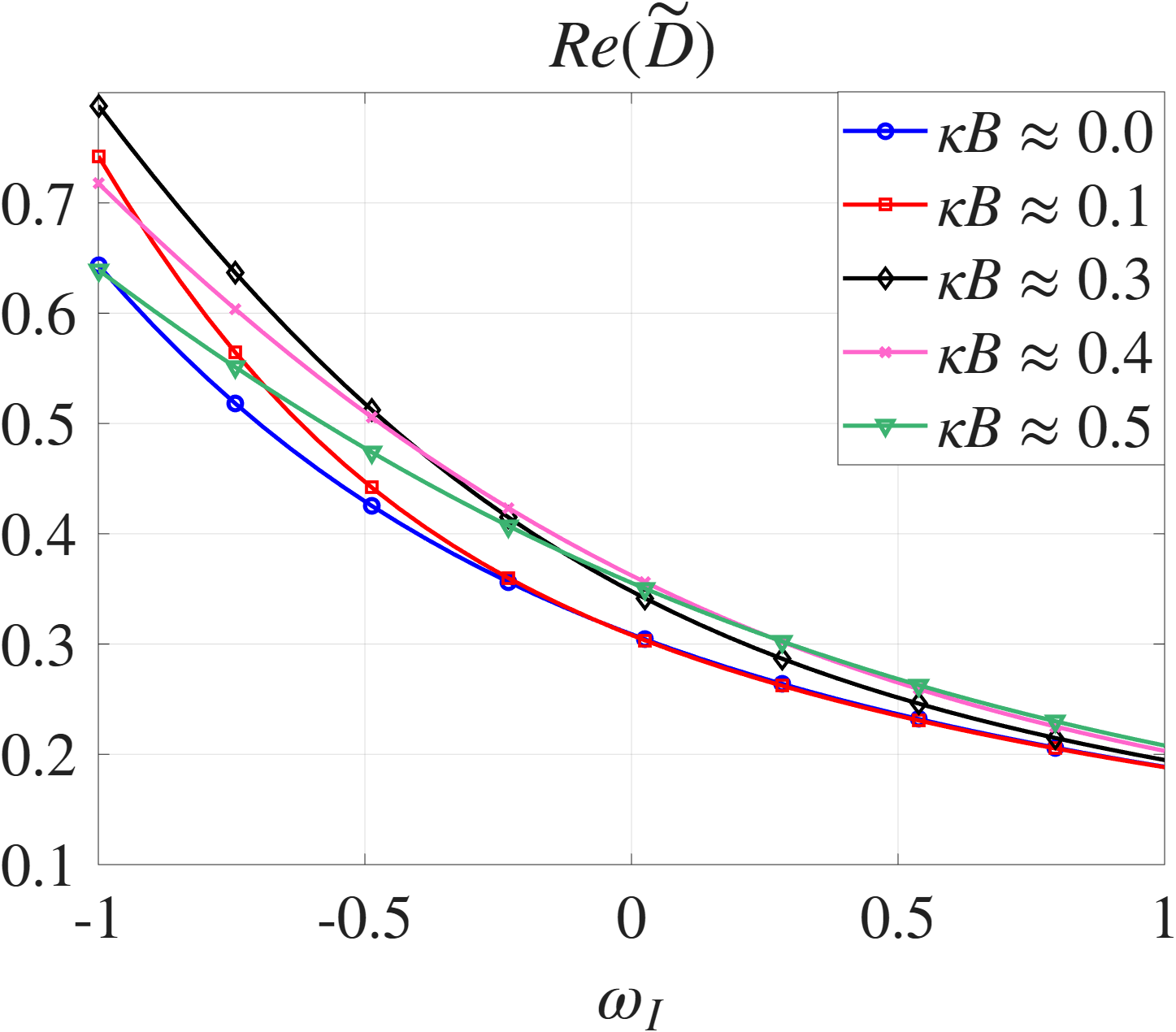}
        \caption{}
    \end{subfigure}
    \hfill
    \begin{subfigure}[b]{0.32\textwidth}
        \centering
        \includegraphics[width=\textwidth]{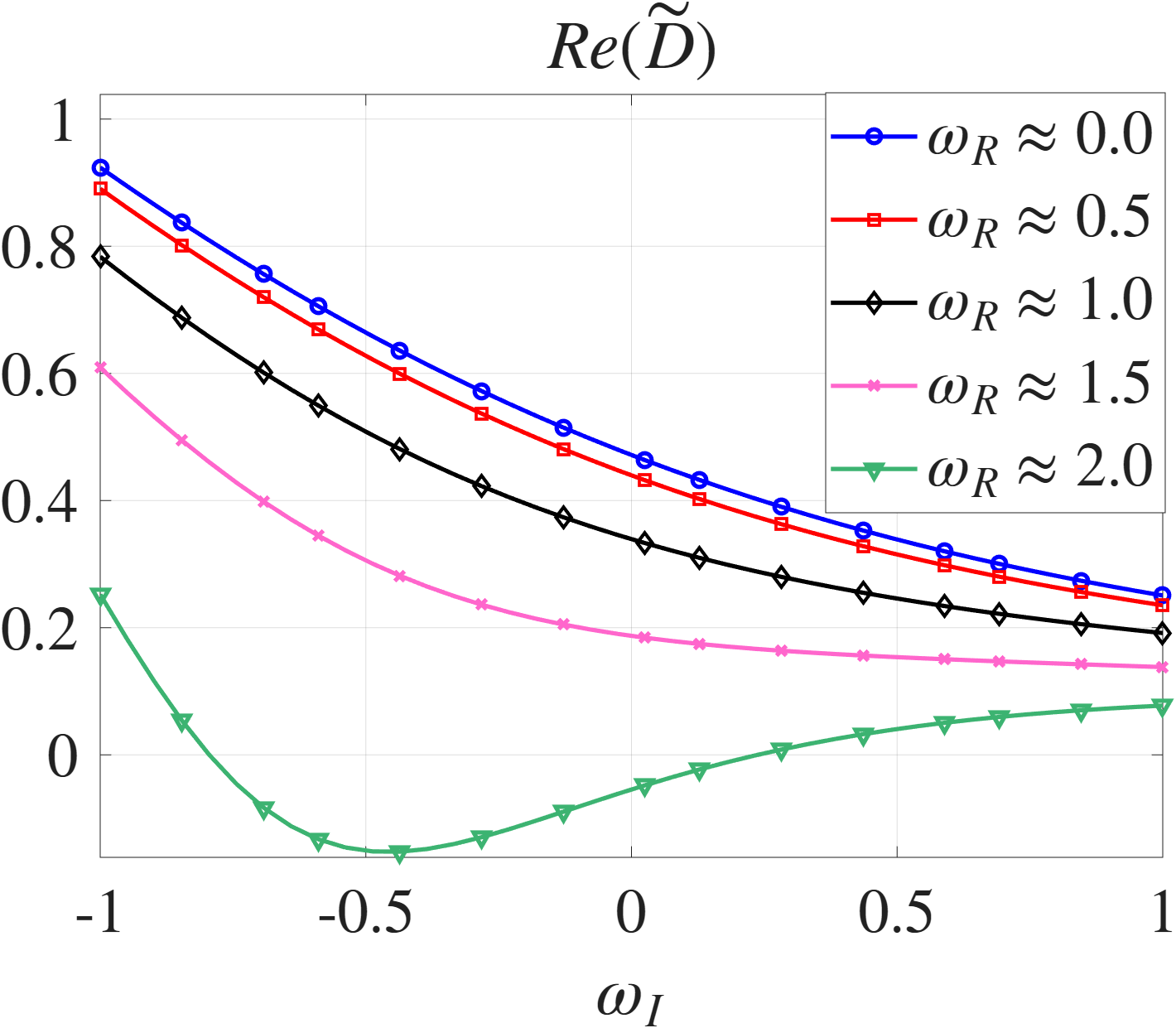}
        \caption{}
    \end{subfigure}
    \hfill
    \begin{subfigure}[b]{0.32\textwidth}
        \centering
        \includegraphics[width=\textwidth]{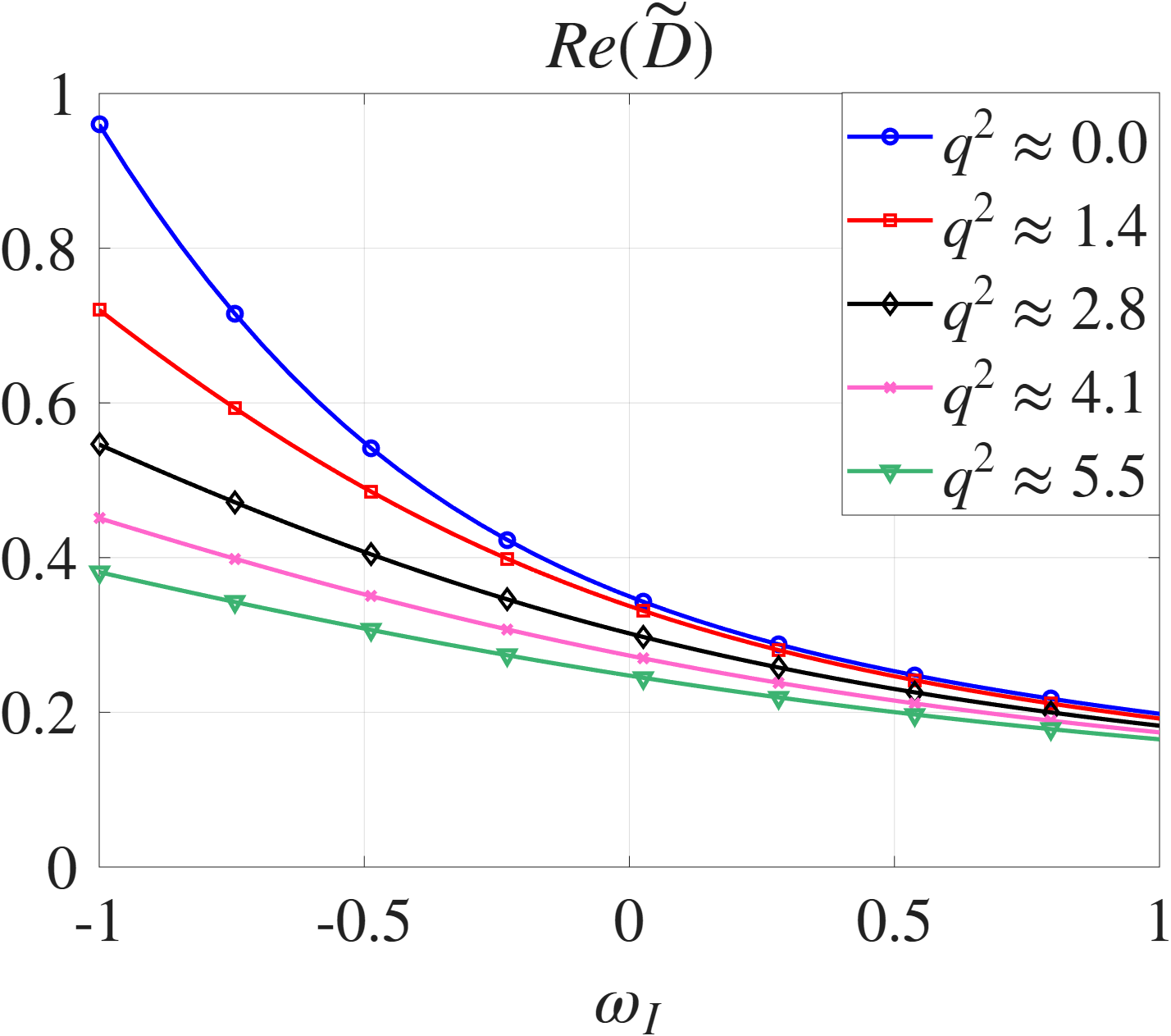}
        \caption{}
    \end{subfigure}
    \caption{Real part of $\tilde D$ as a function of $\omega_I$ for: (a) $\omega_R = q^2 = 1$, (b) $\kappa B = 0.25$, $q^2 = 1$, and (c) $\kappa B = 0.25$, $\omega_R = 1$.}
    \label{fig:re_diff_wi}
\end{figure}

\begin{figure}[htbp]
    \centering
    \begin{subfigure}[b]{0.32\textwidth}
        \centering
        \includegraphics[width=\textwidth]{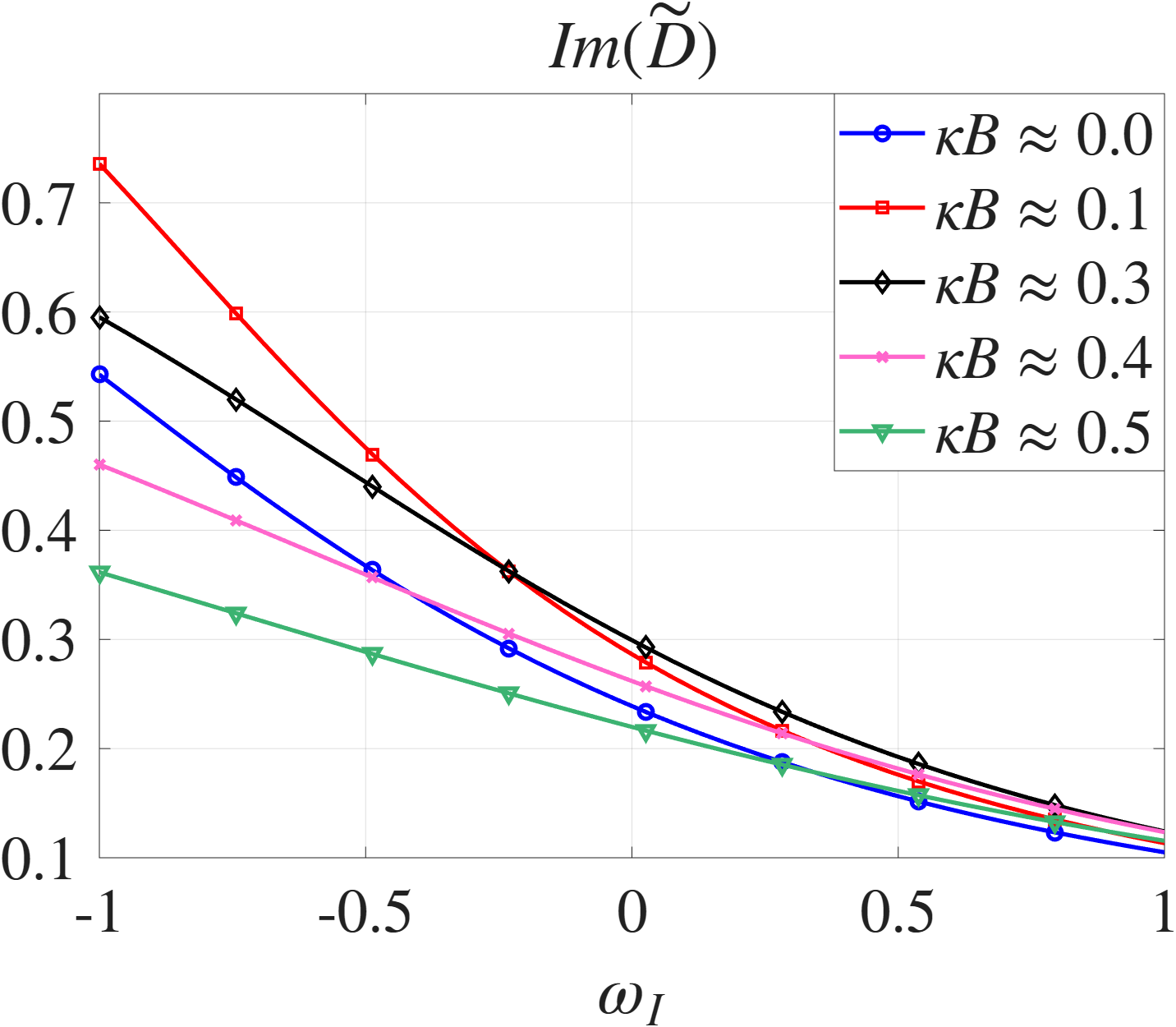}
        \caption{}
    \end{subfigure}
    \hfill
    \begin{subfigure}[b]{0.32\textwidth}
        \centering
        \includegraphics[width=\textwidth]{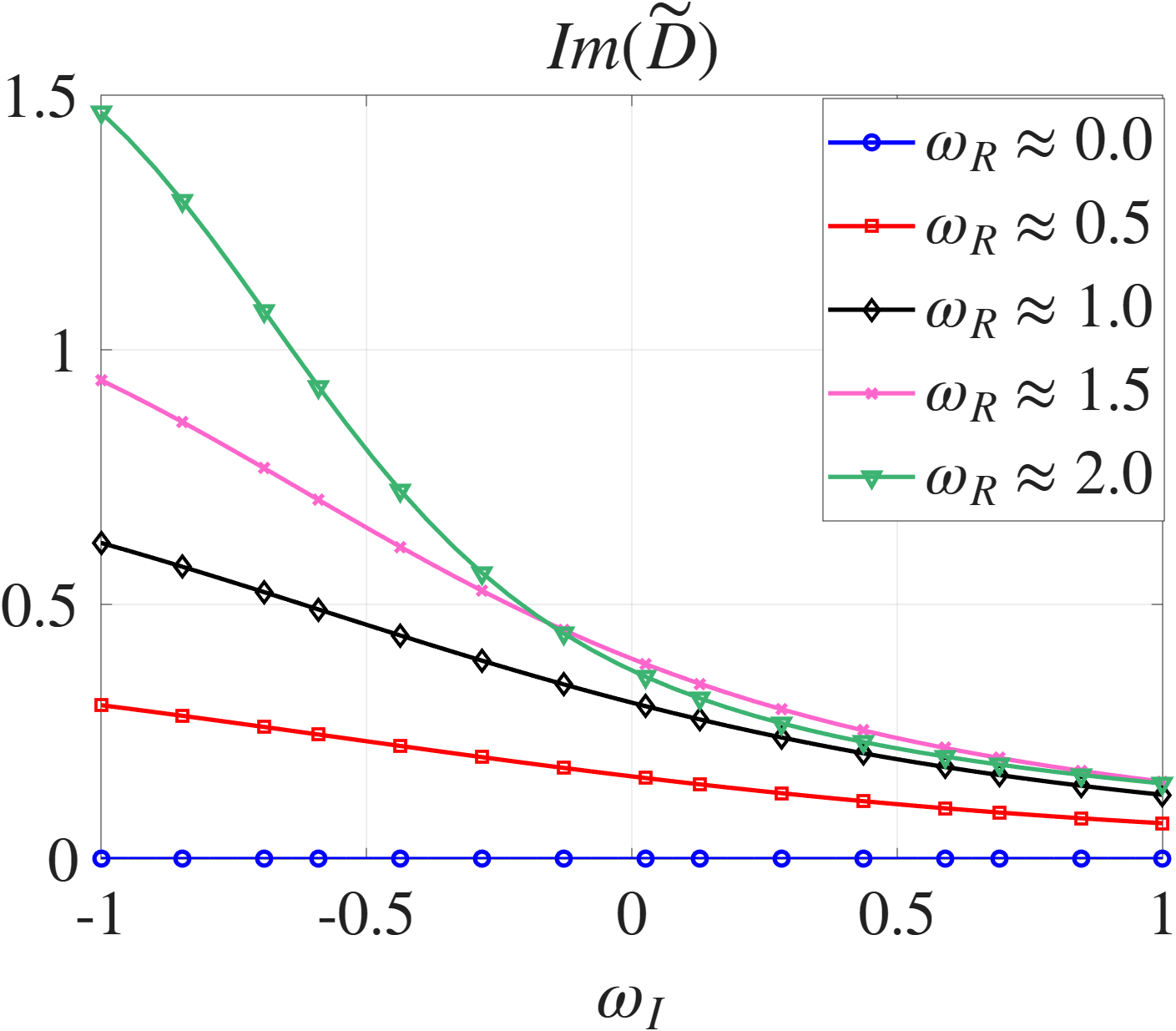}
        \caption{}
    \end{subfigure}
    \hfill
    \begin{subfigure}[b]{0.32\textwidth}
        \centering
        \includegraphics[width=\textwidth]{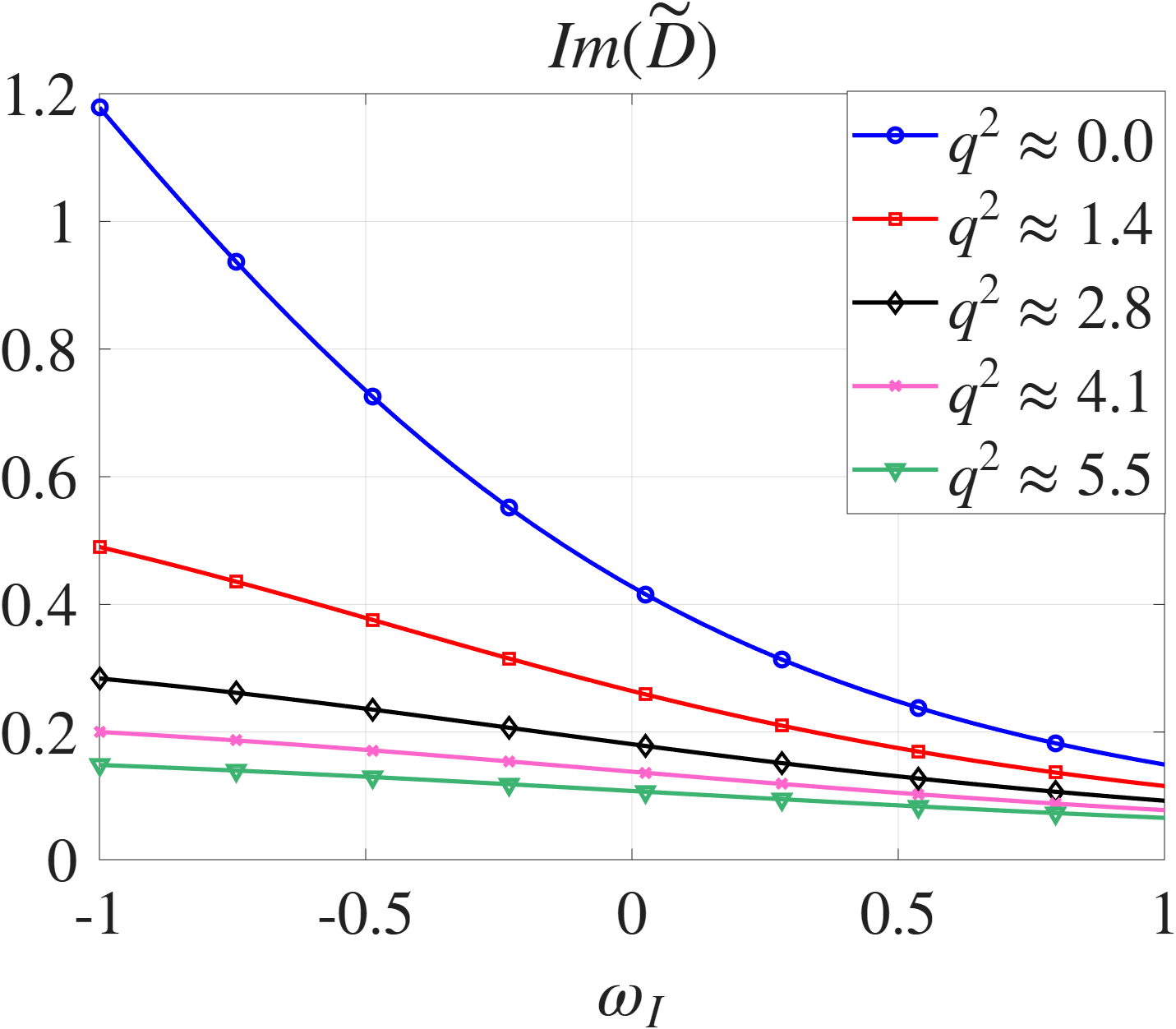}
        \caption{}
    \end{subfigure}
    \caption{Imaginary part of $\tilde D$ as a function of $\omega_I$ for: (a) $\omega_R = q^2 = 1$, (b) $\kappa B = 0.25$, $q^2 = 1$, and (c) $\kappa B = 0.25$, $\omega_R = 1$.}    \label{fig:im_diff_wi}
\end{figure}

\begin{figure}[htbp]
    \centering
    \begin{subfigure}[b]{0.32\textwidth}
        \centering
        \includegraphics[width=\textwidth]{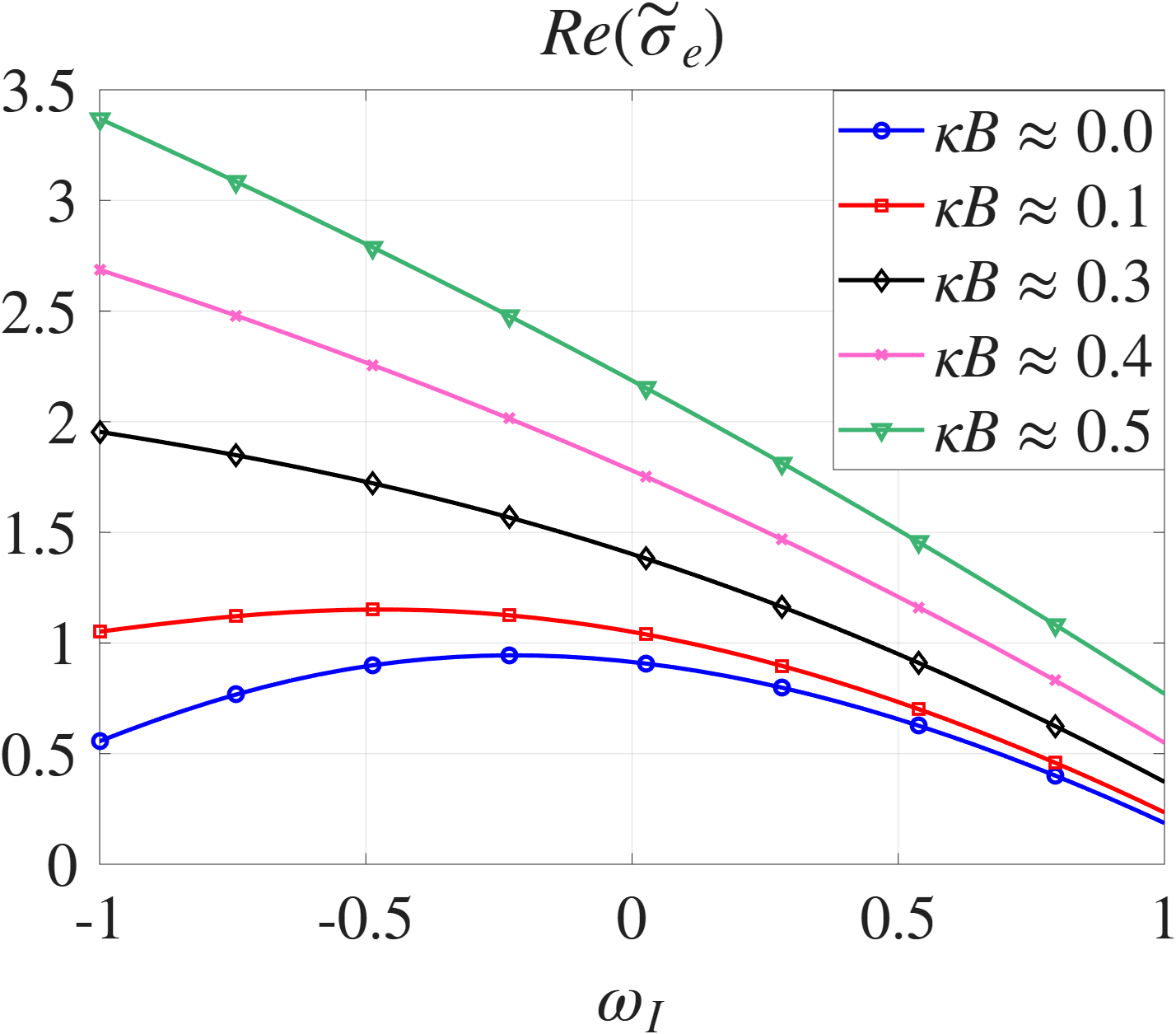}
        \caption{}
    \end{subfigure}
    \hfill
    \begin{subfigure}[b]{0.32\textwidth}
        \centering
        \includegraphics[width=\textwidth]{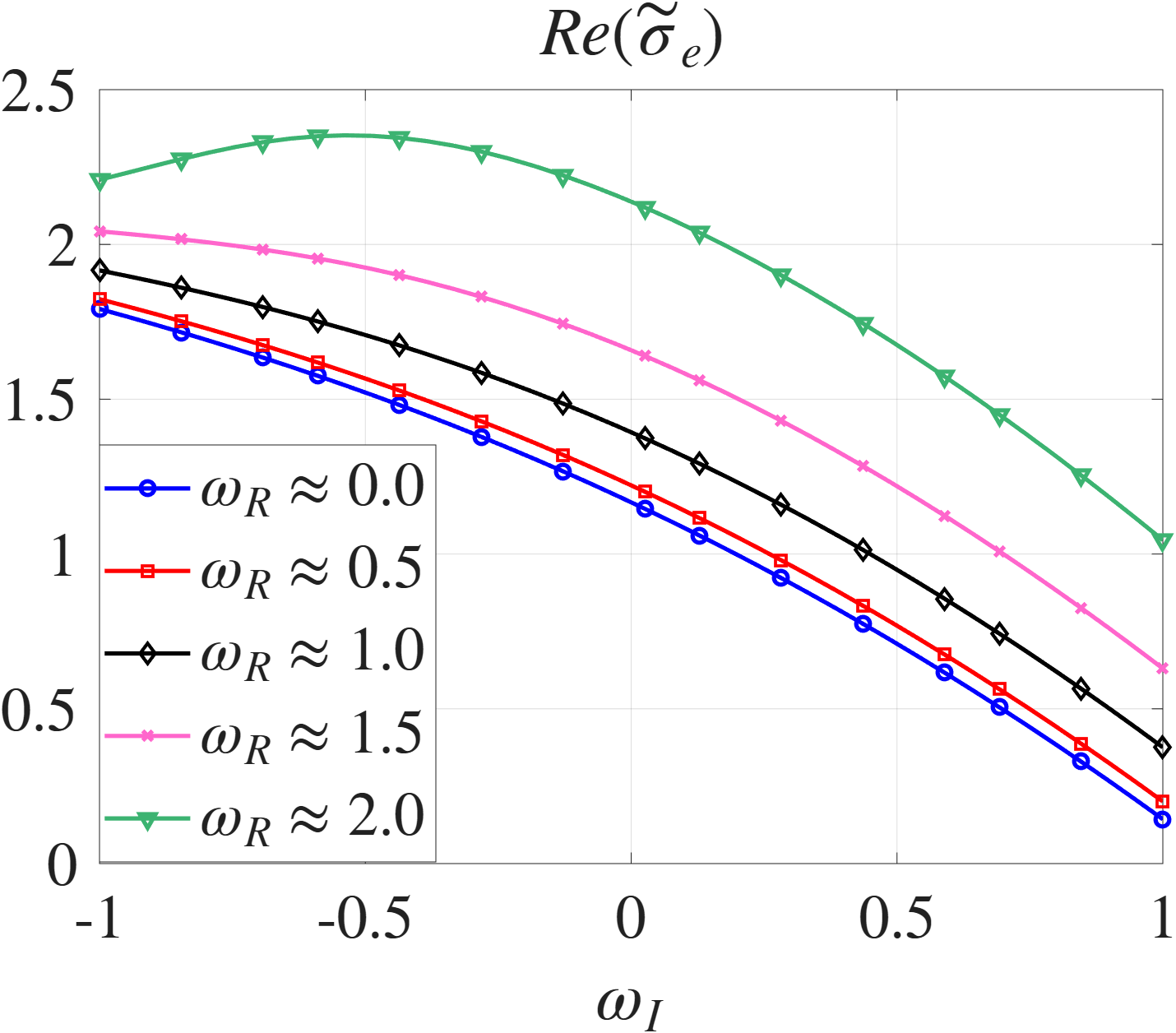}
        \caption{}
    \end{subfigure}
    \hfill
    \begin{subfigure}[b]{0.32\textwidth}
        \centering
        \includegraphics[width=\textwidth]{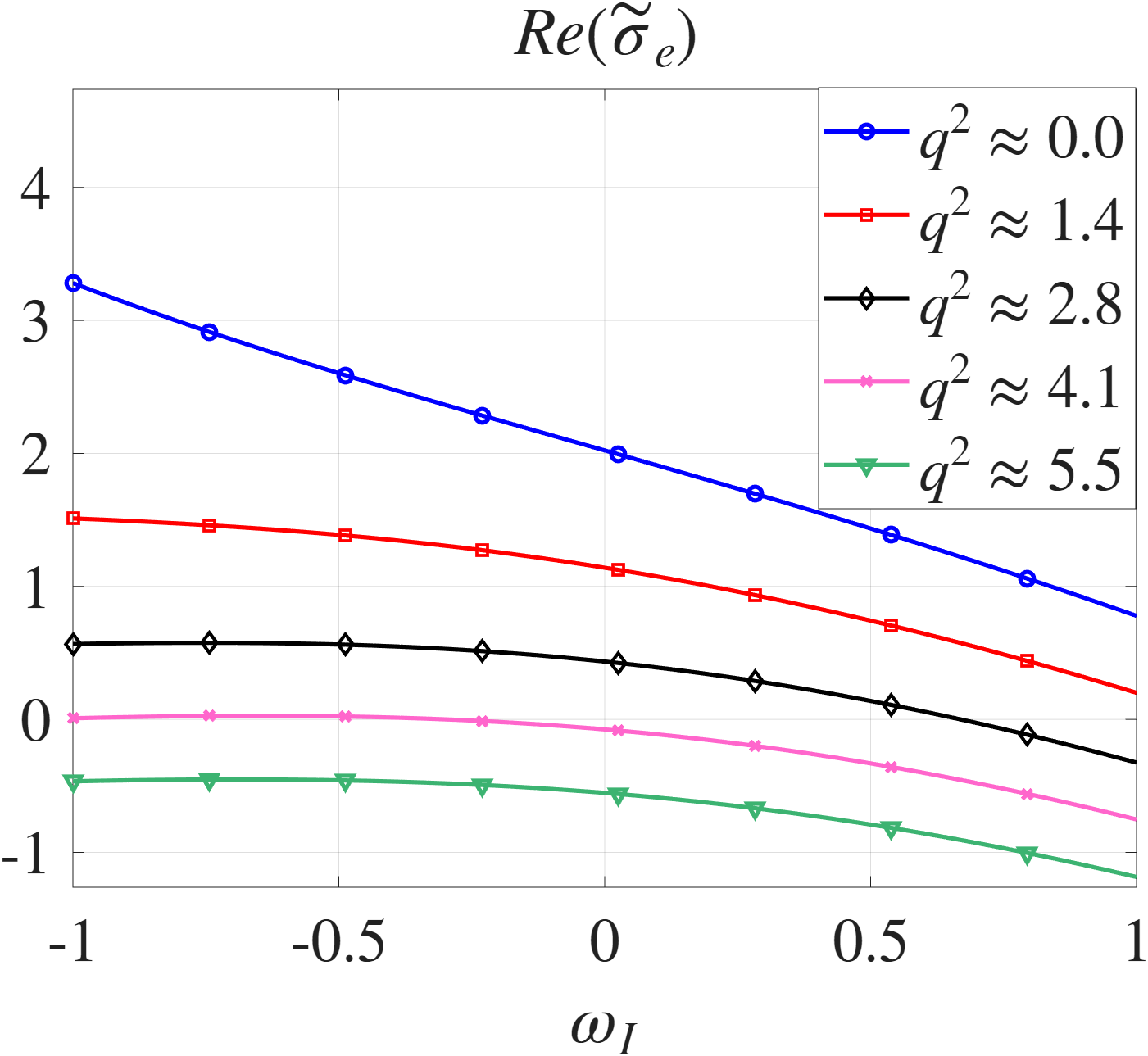}
        \caption{}
    \end{subfigure}
    \caption{Real part of $\tilde \sigma_e$ as a function of $\omega_I$ for: (a) $\omega_R = q^2 = 1$, (b) $\kappa B = 0.25$, $q^2 = 1$, and (c) $\kappa B = 0.25$, $\omega_R = 1$.}    \label{fig:re_sigmae_wi}
\end{figure}

\begin{figure}[htbp]
    \centering
    \begin{subfigure}[b]{0.32\textwidth}
        \centering
        \includegraphics[width=\textwidth]{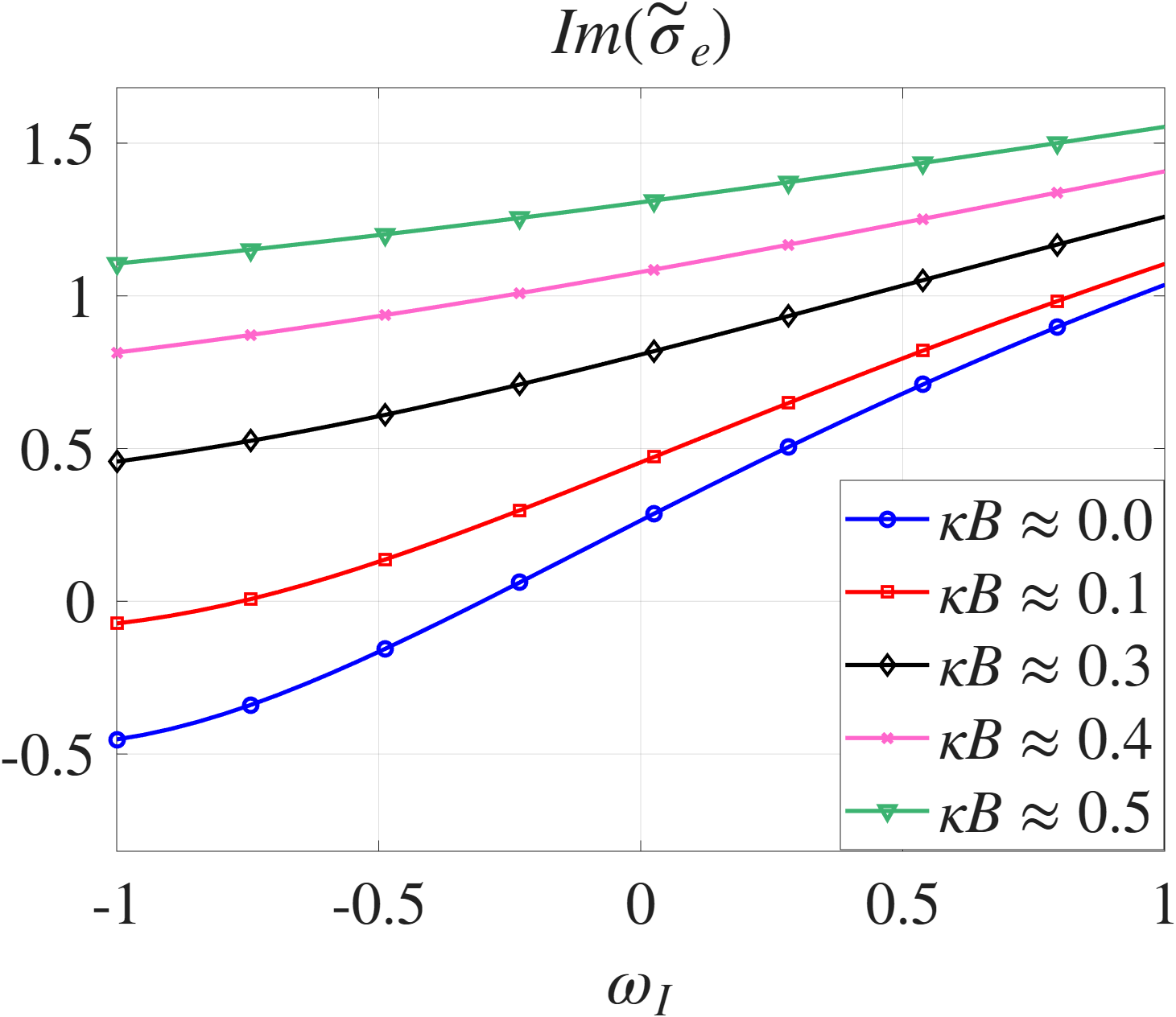}
        \caption{}
    \end{subfigure}
    \hfill
    \begin{subfigure}[b]{0.32\textwidth}
        \centering
        \includegraphics[width=\textwidth]{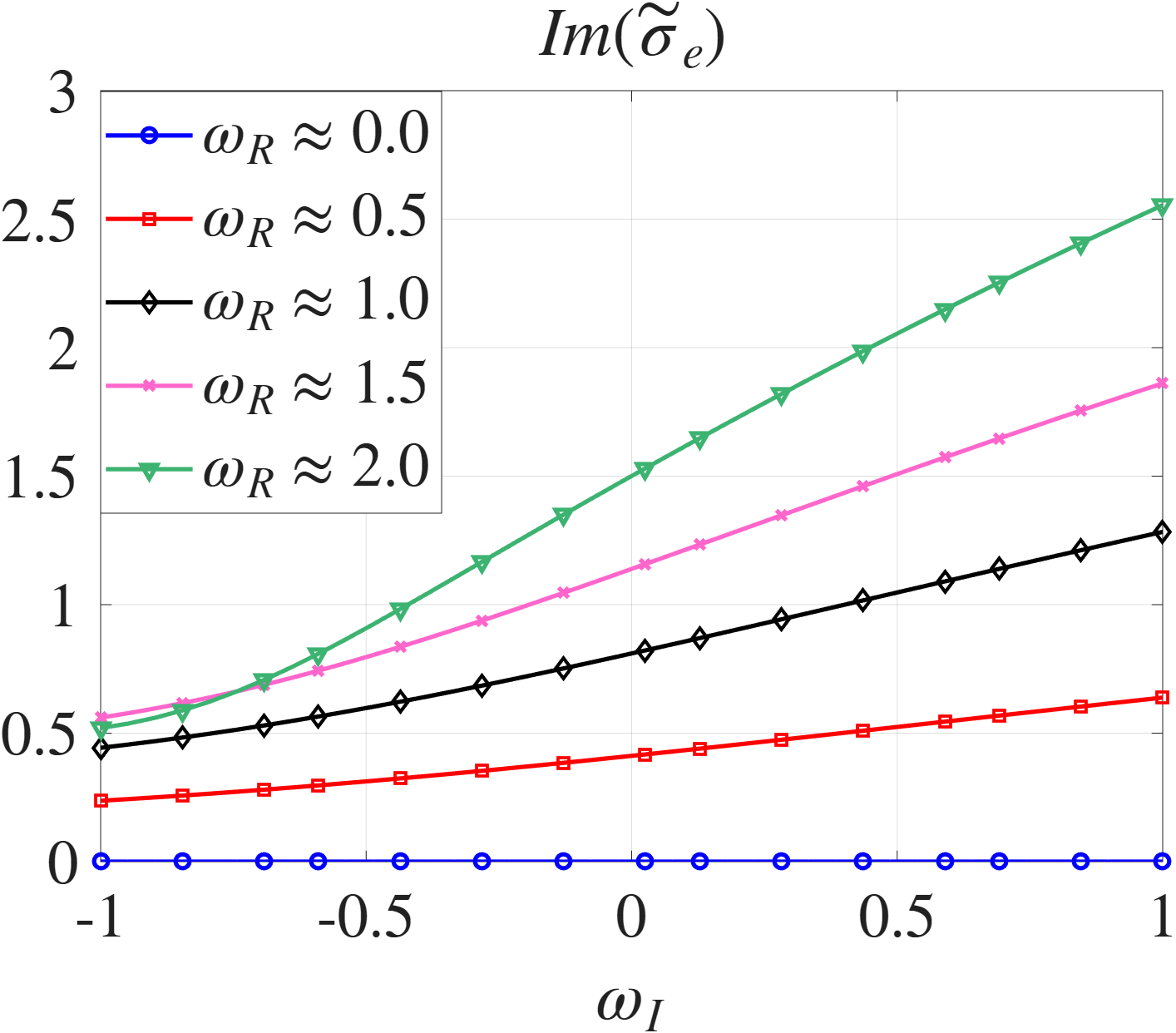}
        \caption{}
    \end{subfigure}
    \hfill
    \begin{subfigure}[b]{0.32\textwidth}
        \centering
        \includegraphics[width=\textwidth]{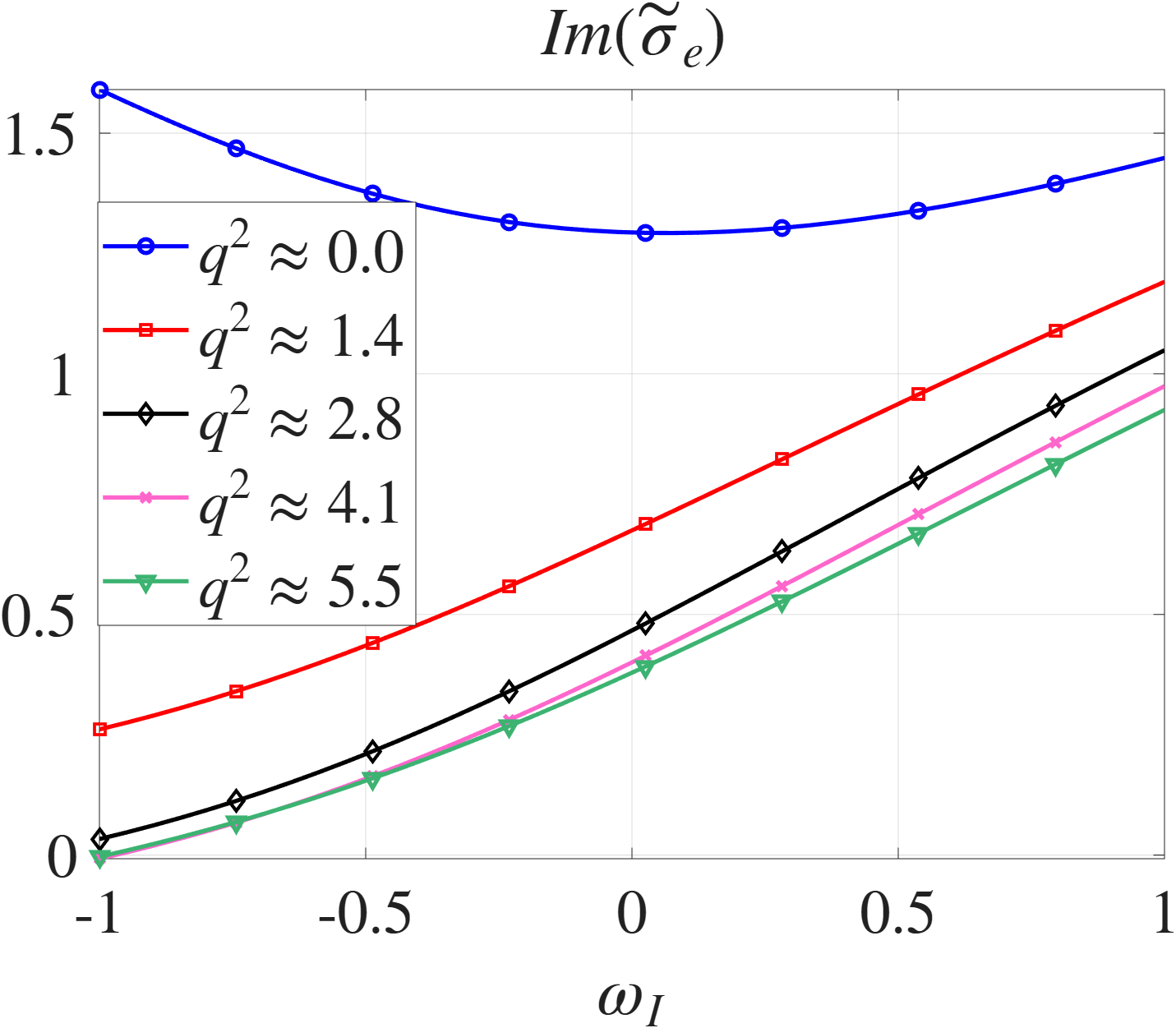}
        \caption{}
    \end{subfigure}
    \caption{Imaginary part of $\tilde \sigma_e$ as a function of $\omega_I$ for: (a) $\omega_R = q^2 = 1$, (b) $\kappa B = 0.25$, $q^2 = 1$, and (c) $\kappa B = 0.25$, $\omega_R = 1$.}        \label{fig:im_sigmae_wi}
\end{figure}

\begin{figure}[htbp]
    \centering
    \begin{subfigure}[b]{0.32\textwidth}
        \centering
        \includegraphics[width=\textwidth]{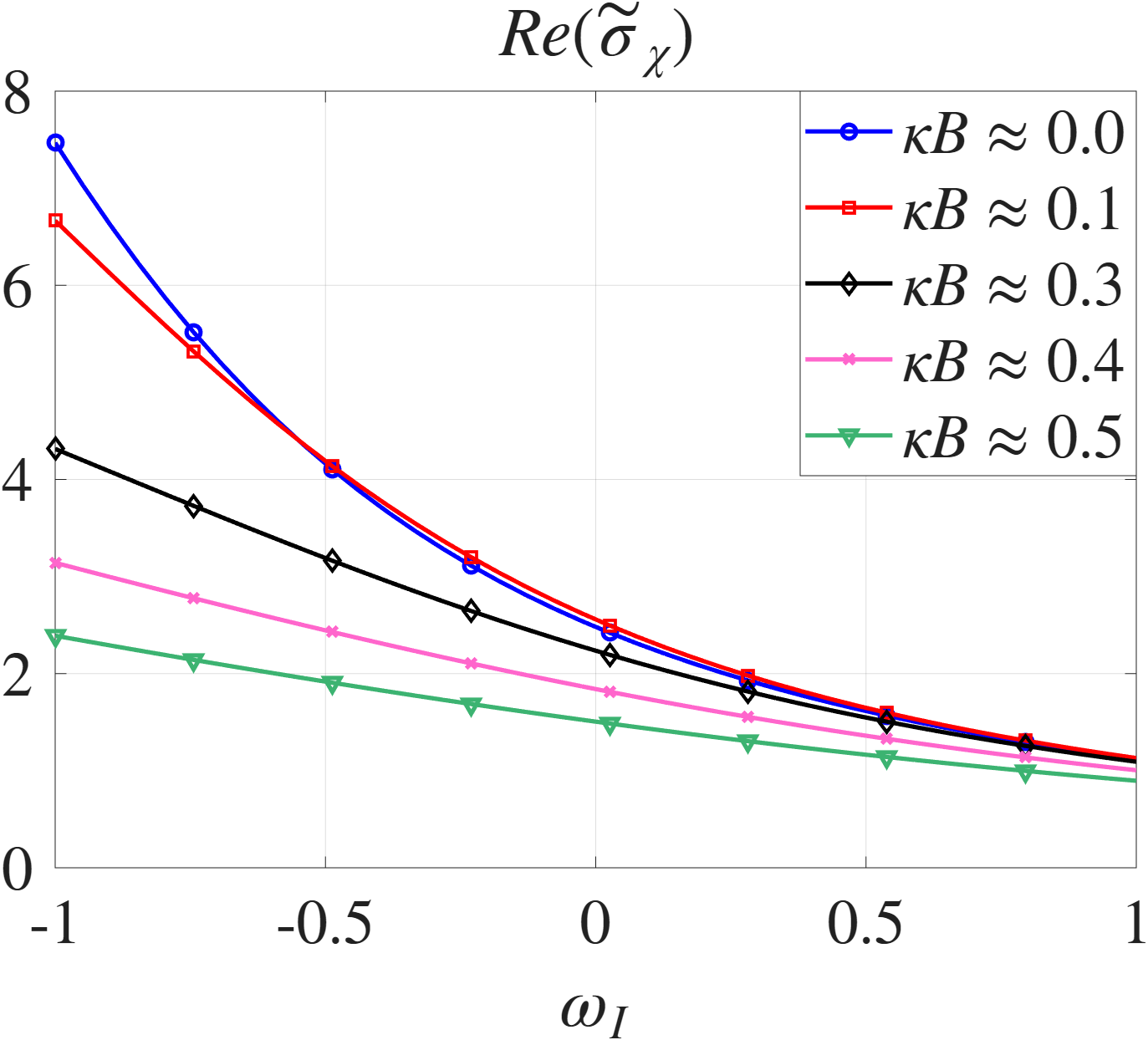}
        \caption{}
    \end{subfigure}
    \hfill
    \begin{subfigure}[b]{0.32\textwidth}
        \centering
        \includegraphics[width=\textwidth]{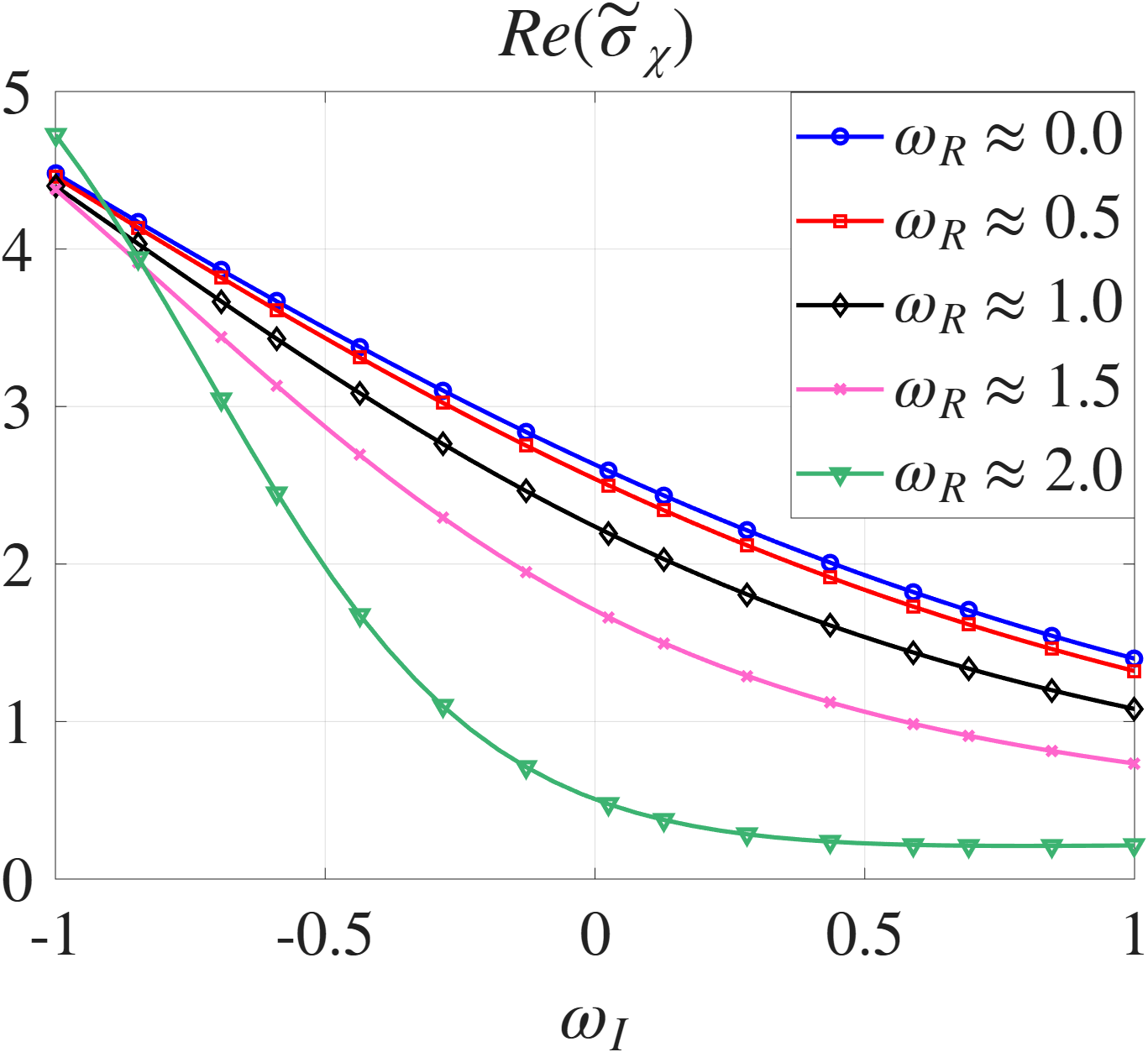}
        \caption{}
    \end{subfigure}
    \hfill
    \begin{subfigure}[b]{0.32\textwidth}
        \centering
        \includegraphics[width=\textwidth]{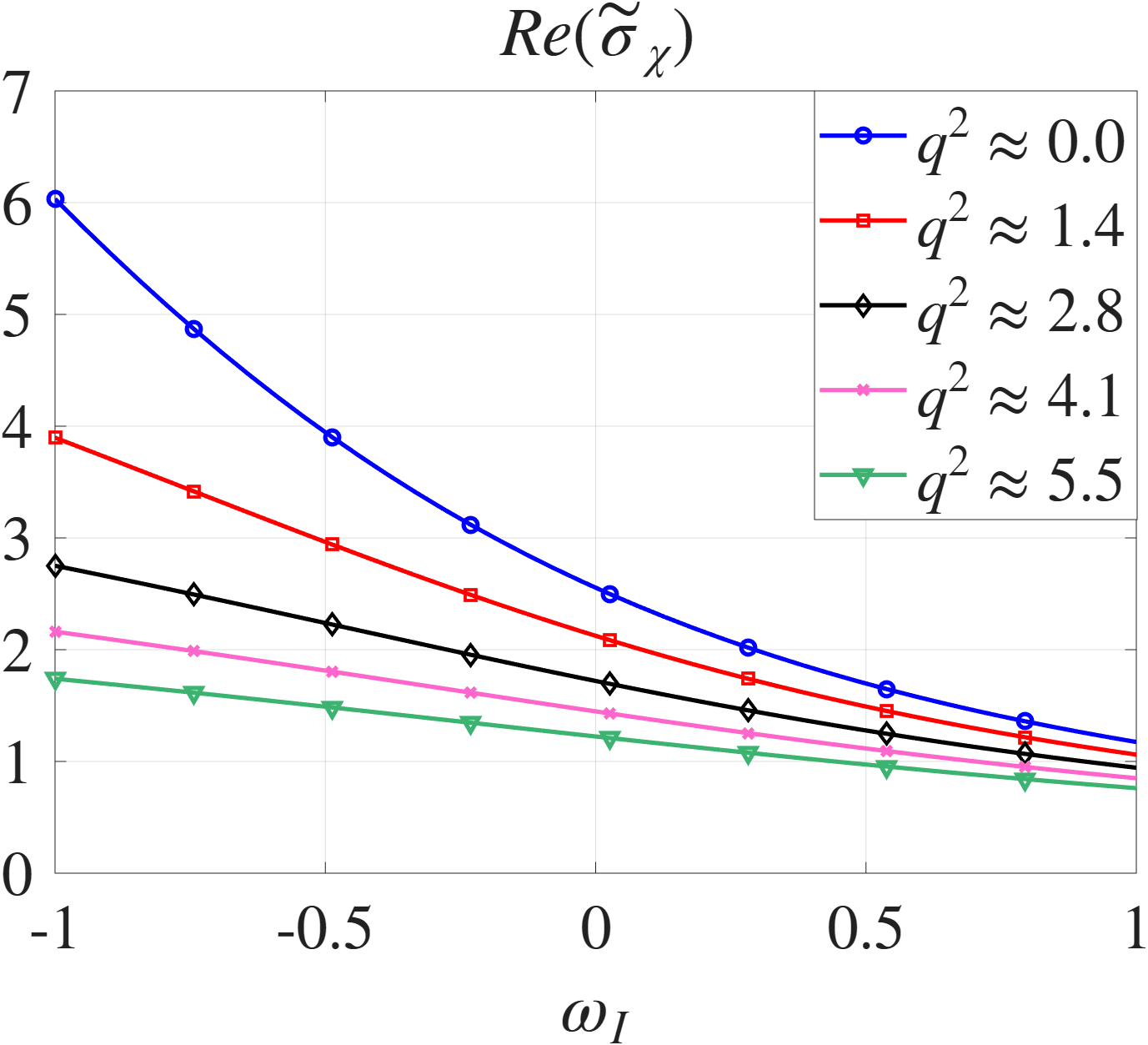}
        \caption{}
    \end{subfigure}
    \caption{Real part of $\tilde \sigma_\chi$ as a function of $\omega_I$ for: (a) $\omega_R = q^2 = 1$, (b) $\kappa B = 0.25$, $q^2 = 1$, and (c) $\kappa B = 0.25$, $\omega_R = 1$.}        \label{fig:re_sigmachi_wi}
\end{figure}

\begin{figure}[htbp]
    \centering
    \begin{subfigure}[b]{0.32\textwidth}
        \centering
        \includegraphics[width=\textwidth]{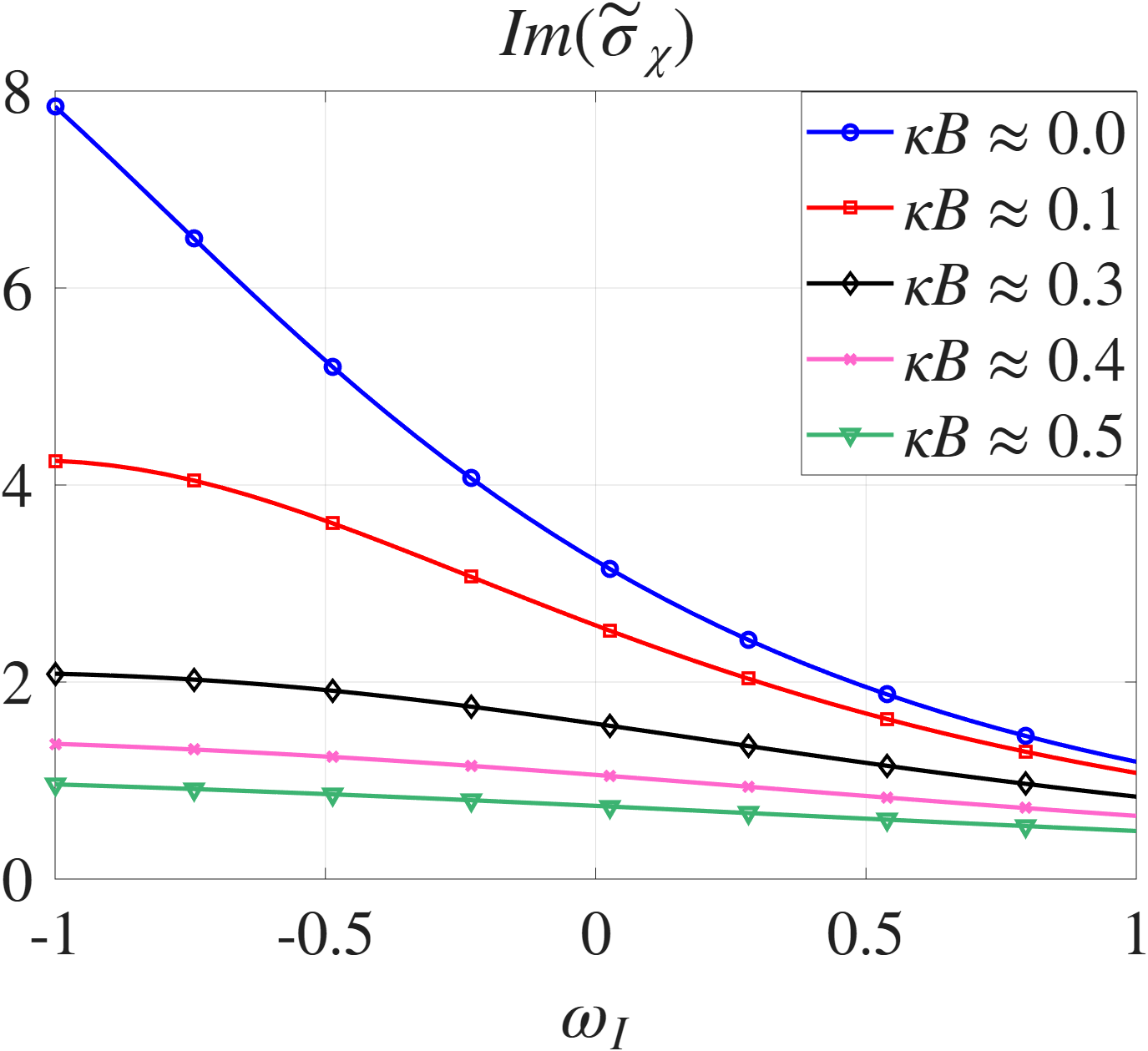}
        \caption{}
    \end{subfigure}
    \hfill
    \begin{subfigure}[b]{0.32\textwidth}
        \centering
        \includegraphics[width=\textwidth]{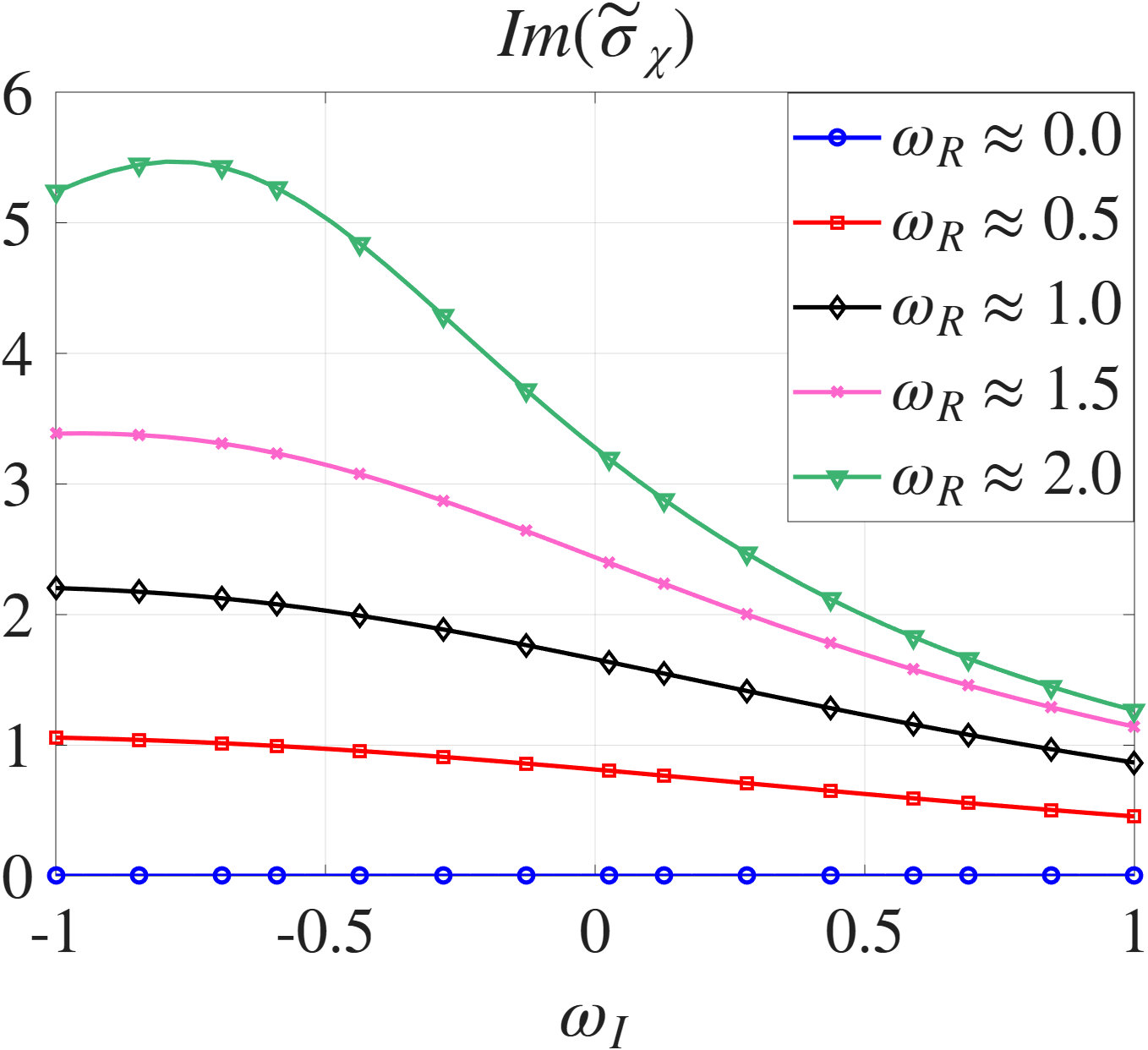}
        \caption{}
    \end{subfigure}
    \hfill
    \begin{subfigure}[b]{0.32\textwidth}
        \centering
        \includegraphics[width=\textwidth]{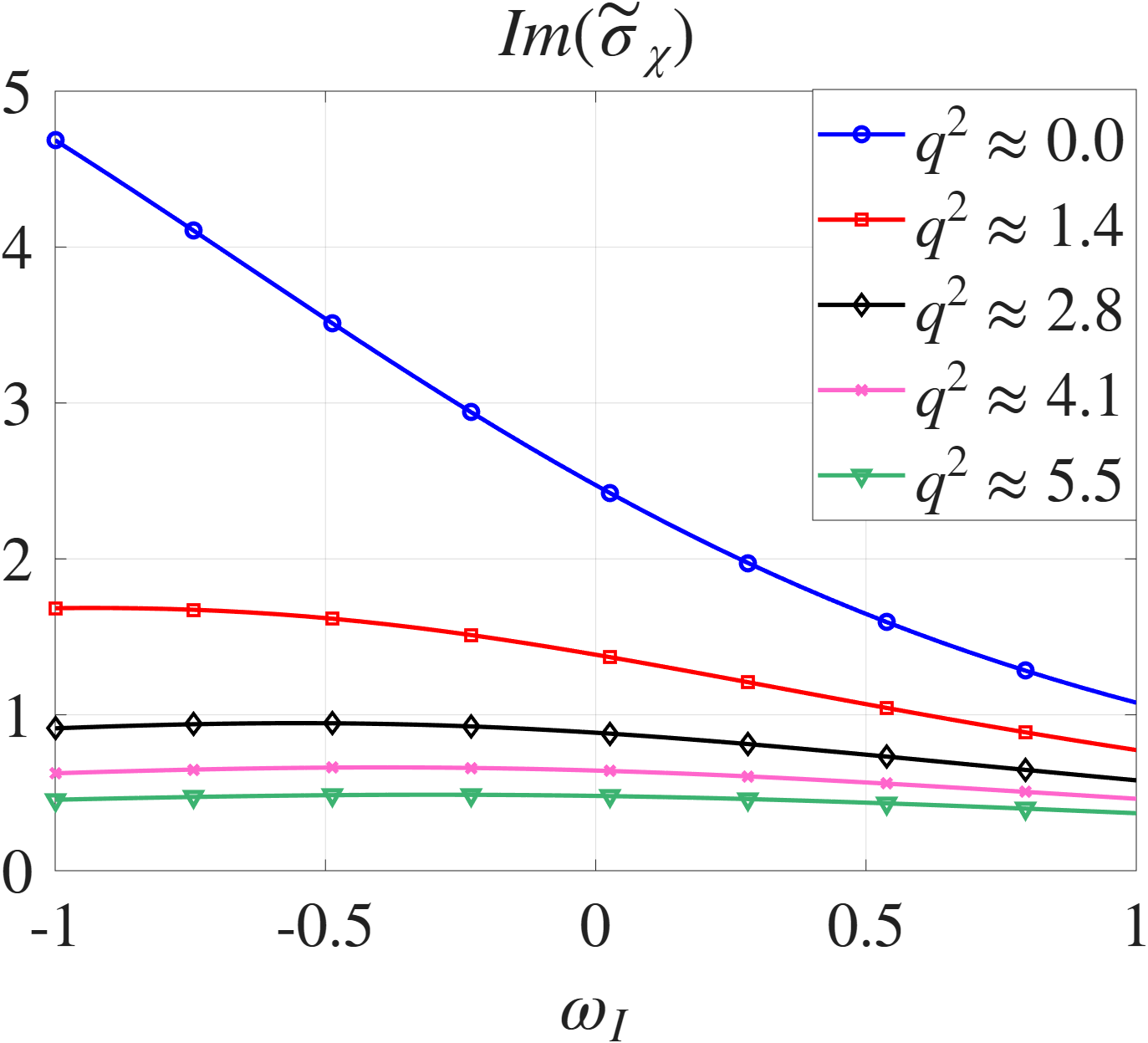}
        \caption{}
    \end{subfigure}
    \caption{Imaginary part of $\tilde \sigma_\chi$ as a function of $\omega_I$ for: (a) $\omega_R = q^2 = 1$, (b) $\kappa B = 0.25$, $q^2 = 1$, and (c) $\kappa B = 0.25$, $\omega_R = 1$.}        \label{fig:im_sigmachi_wi}
\end{figure}

\begin{figure}[htbp]
    \centering
    \begin{subfigure}[b]{0.32\textwidth}
        \centering
        \includegraphics[width=\textwidth]{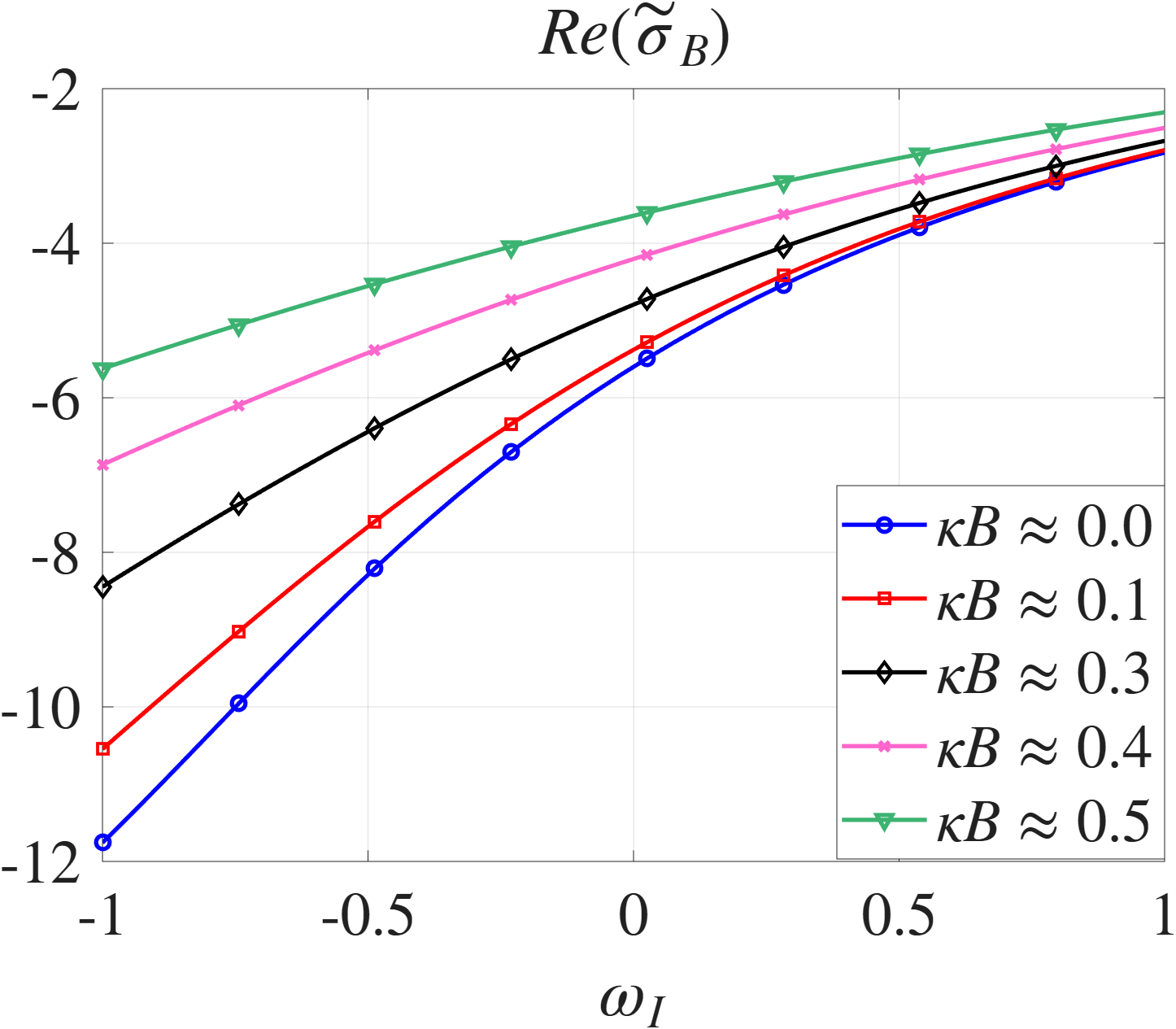}
        \caption{}
    \end{subfigure}
    \hfill
    \begin{subfigure}[b]{0.32\textwidth}
        \centering
        \includegraphics[width=\textwidth]{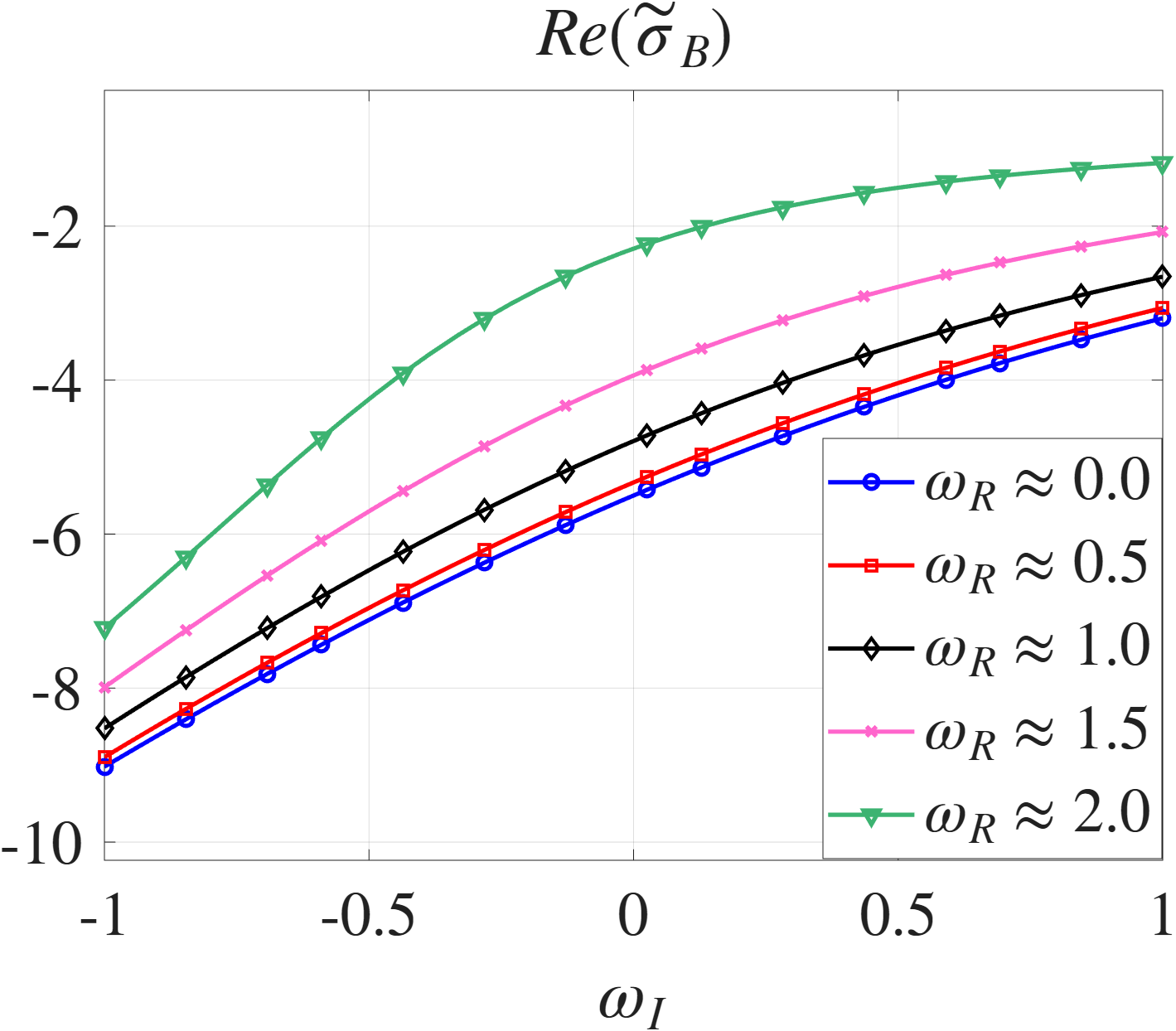}
        \caption{}
    \end{subfigure}
    \hfill
    \begin{subfigure}[b]{0.32\textwidth}
        \centering
        \includegraphics[width=\textwidth]{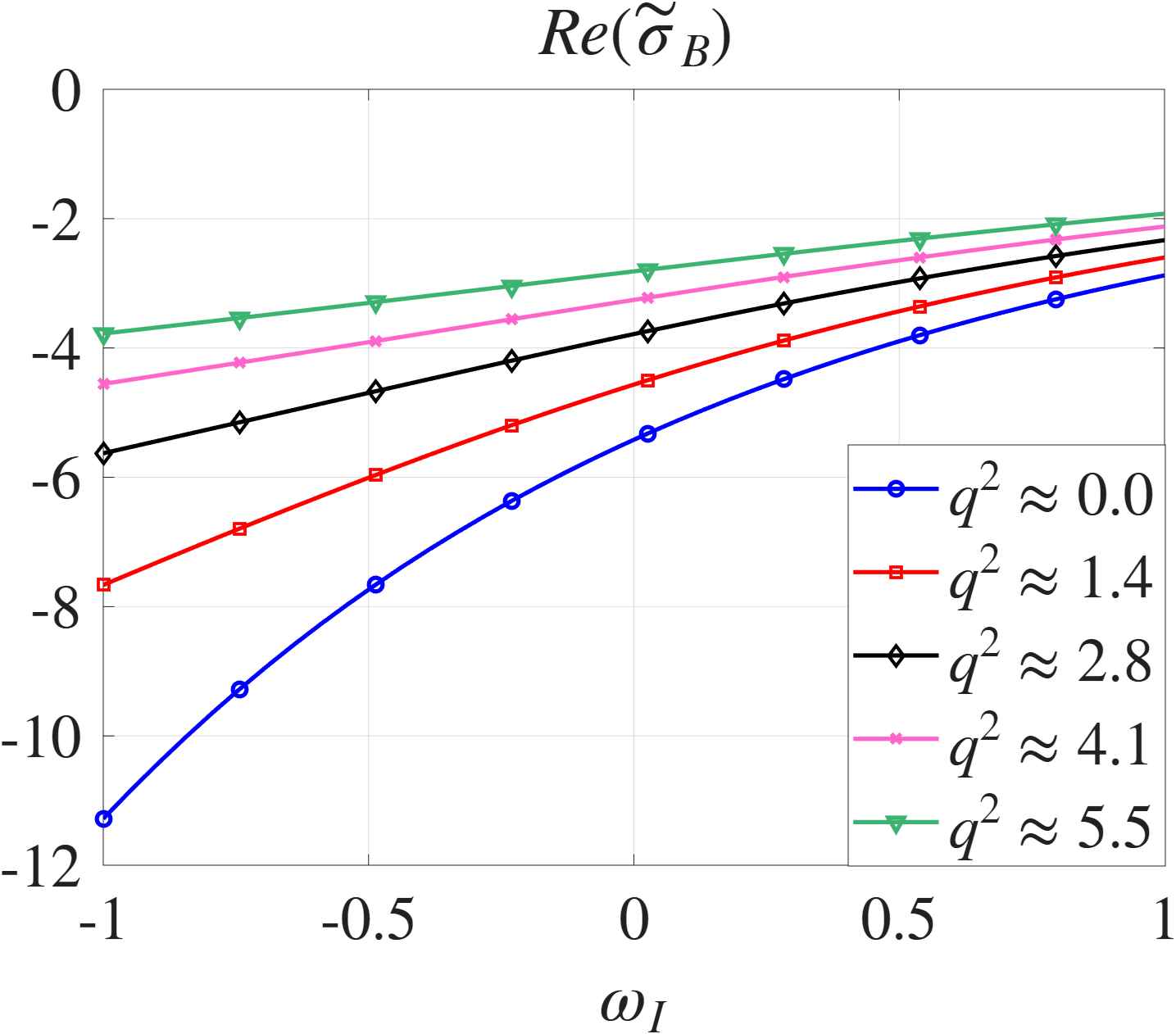}
        \caption{}
    \end{subfigure}
    \caption{Real part of $\tilde\sigma_B$ as a function of $\omega_I$ for: (a) $\omega_R = q^2 = 1$, (b) $\kappa B = 0.25$, $q^2 = 1$, and (c) $\kappa B = 0.25$, $\omega_R = 1$.}        \label{fig:re_sigmab_wi}
\end{figure}

\begin{figure}[htbp]
    \centering
    \begin{subfigure}[b]{0.32\textwidth}
        \centering
        \includegraphics[width=\textwidth]{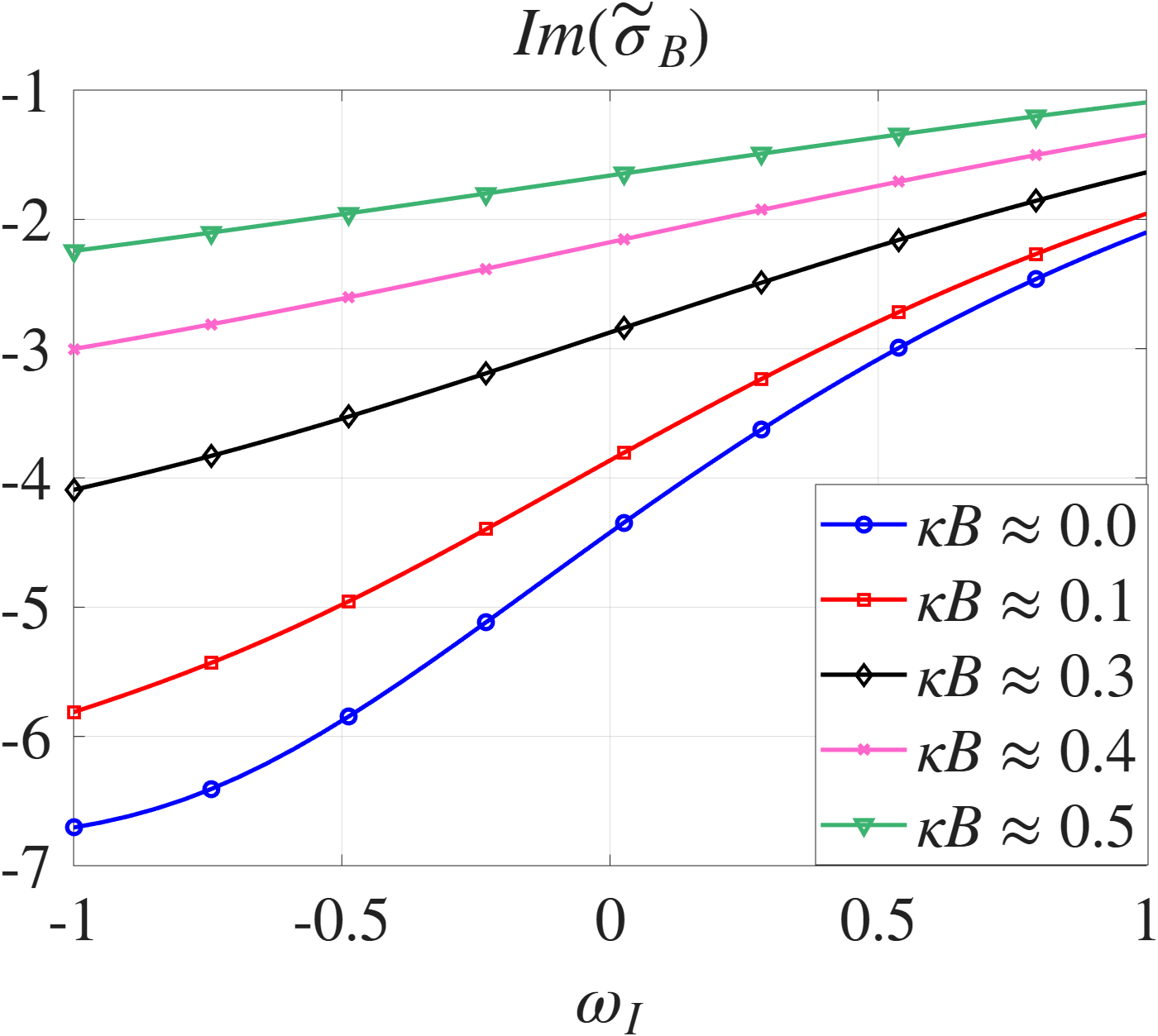}
        \caption{}
    \end{subfigure}
    \hfill
    \begin{subfigure}[b]{0.32\textwidth}
        \centering
        \includegraphics[width=\textwidth]{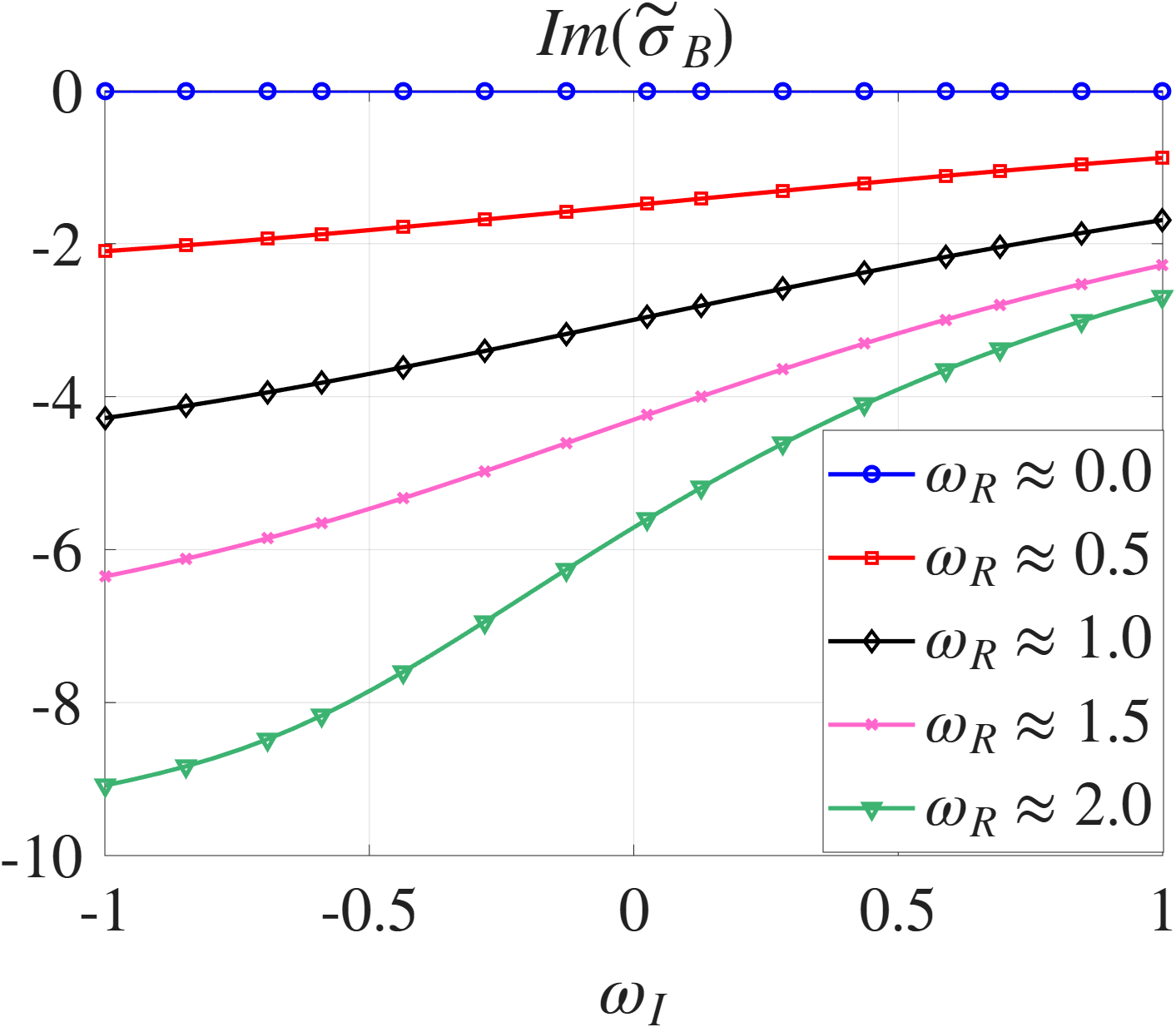}
        \caption{}
    \end{subfigure}
    \hfill
    \begin{subfigure}[b]{0.32\textwidth}
        \centering
        \includegraphics[width=\textwidth]{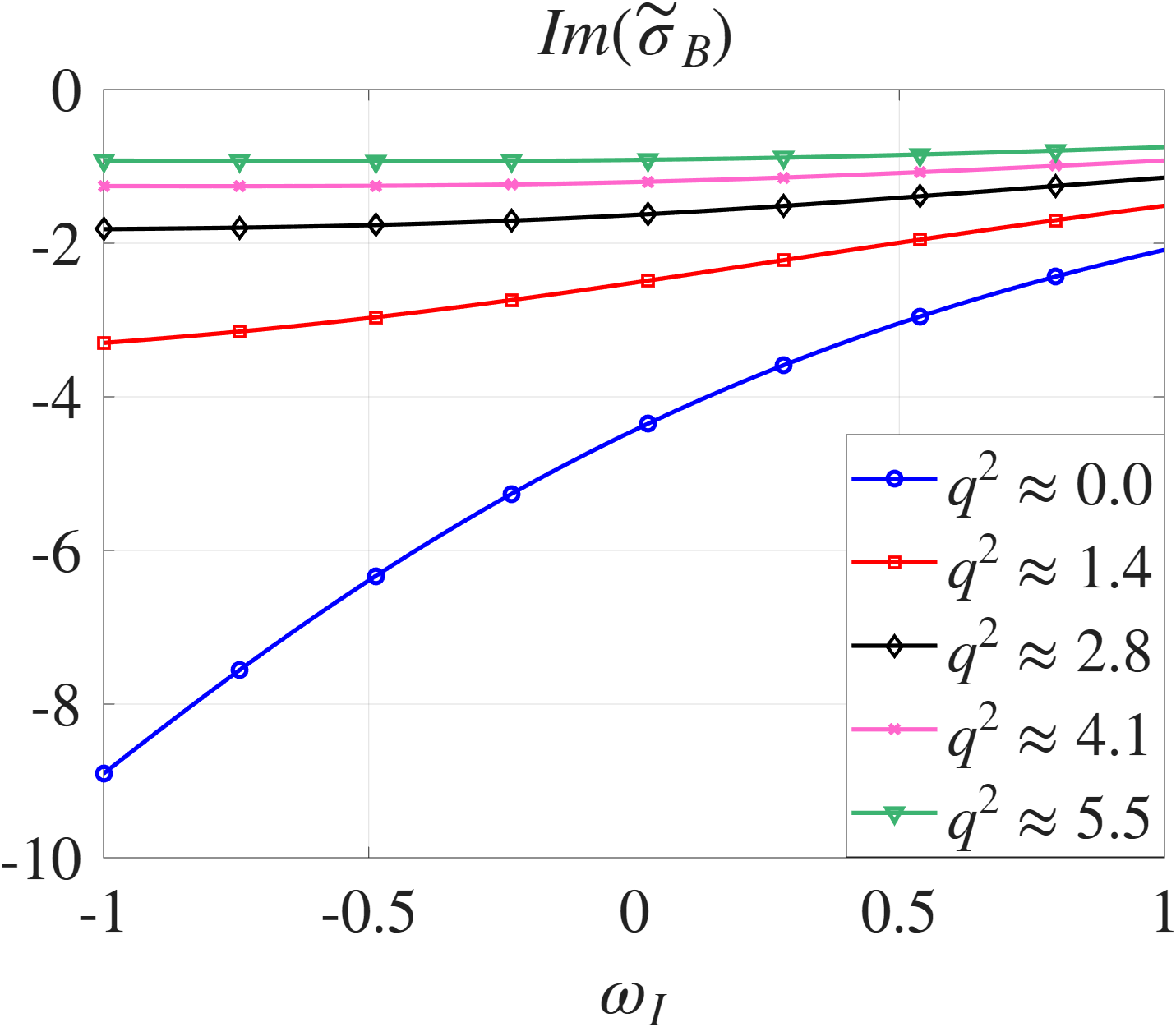}
        \caption{}
    \end{subfigure}
    \caption{Imaginary part of $\tilde\sigma_B$ as a function of $\omega_I$ for: (a) $\omega_R = q^2 = 1$, (b) $\kappa B = 0.25$, $q^2 = 1$, and (c) $\kappa B = 0.25$, $\omega_R = 1$.}         \label{fig:im_sigmab_wi}
\end{figure}

As a function of $\omega=\omega_R+i\omega_I$, all the TCFs have the following form:\footnote{In principle, one might study the analytical properties of the TCFs in the complex plane.}
\begin{equation}
\text{TCF}(\omega)=\text{TCF}_R(\omega^2)+i\omega\, \text{TCF}_I(\omega^2).
\end{equation}
Hence, under $\omega \rightarrow -\omega^*$
($\omega_R \rightarrow -\omega_R$),
\begin{equation}
    \Re[\text{TCF}(\omega)]=\Re[\text{TCF}(-\omega^*)]\,;\qquad
    \Im[\text{TCF}(\omega)]=-\Im[\text{TCF}(-\omega^*)].
\end{equation}
There is no such simple relation when $\omega\rightarrow\omega^*$. We analyze the TCFs
in the all-parallel configuration as a function of $\omega_I$. 
Figs. \ref{fig:re_diff_wi}--\ref{fig:im_sigmab_wi} for the TCFs reveal  strong asymmetries under the transformation $\omega_I~\rightarrow-\omega_I$. Most of the TCFs display  monotonic behavior
as a function of $\omega_I$.



\bibliographystyle{JHEP}
\bibliography{references}

\end{document}